\documentclass[twocolumn]{aastex62}
\pdfoutput=1 
\usepackage{amsmath,amstext}
\usepackage[T1]{fontenc}
\usepackage{apjfonts} 
\usepackage[figure,figure*]{hypcap}
\usepackage{url}

\shortauthors{Levy et al.}

\begin{document}
\renewcommand*{\sectionautorefname}{Section} 
\renewcommand*{\subsectionautorefname}{Section} 

\makeatletter
\newcommand{\manuallabel}[2]{\def\@currentlabel{#2}\label{#1}}
\makeatother

\newcommand{\D}			{$^\circ$}
\newcommand{\vrot}		{V_{\rm rot}}
\newcommand{\vrad}		{V_{\rm rad}}
\newcommand{\vsys}		{V_{\rm sys}}
\newcommand{\vflat}		{V_{\rm flat}}
\newcommand{\vbar}		{\bar{V}}
\newcommand{\rflat}		{R_{\rm flat}}
\newcommand{\rbeam}		{R_{\rm 2beam}}
\newcommand{\ha}		{\mbox{\rm{H}$\alpha$}}
\newcommand{\hb}		{\mbox{\rm{H}$\beta$}}
\newcommand{\hg}		{\mbox{\rm{H}$\gamma$}}
\newcommand{\hd}		{\mbox{\rm{H}$\delta$}}
\newcommand{\ttco}		{$^{13}$CO}
\newcommand{\kms}		{\mbox{km\,s$^{-1}$}}
\newcommand{\hi}        {\mbox{\rm H{\small I}}}
\newcommand{\HI}        {\hi}
\newcommand{\HII}       {\mbox{\rm H{\small II}}}
\newcommand{\SII}		{\mbox{\rm [S{\small II}]}}
\newcommand{\NII}		{\mbox{\rm [N{\small II}]}}
\newcommand{\OIII}		{\mbox{\rm [O{\small III}]}}
\newcommand{\OII}		{\mbox{\rm [O{\small II}]}}
\newcommand{\CII}		{\mbox{\rm [C{\small II}]}}
\newcommand{\HeI}		{\mbox{\rm [He{\small I}]}}
\newcommand{\htwo}      {\mbox{H$_{2}$}}
\newcommand{\Dv}		{\Delta V}
\newcommand{\eDv}		{\sigma_{\Delta V}}
\newcommand{\convol}	{{\fontfamily{cmtt}\selectfont convol}}
\newcommand{\regrid}	{{\fontfamily{cmtt}\selectfont regrid}}
\newcommand{\moment}	{{\fontfamily{cmtt}\selectfont moment}}
\newcommand{\miriad}	{{\fontfamily{cmtt}\selectfont Miriad}}
\newcommand{\ccdmom}	{{\fontfamily{cmtt}\selectfont ccdmom}}
\newcommand{\nmom}	{{\fontfamily{cmtt}\selectfont mom=32}}
\newcommand{\pipetd}	{\mbox{\rm{\small Pipe3D}}}
\newcommand{\chisq}		{$\chi_r^2$}
\renewcommand{\~}		{$\sim$}
\newcommand{\sigmaADC} {$\sigma_{\rm ADC}$}
\newcommand{\sigmaHg} {$\sigma_{\rm H\gamma}$}
\newcommand{\comment} {}
\newcommand{\comments} {}

\title{The EDGE-CALIFA Survey: Molecular and Ionized Gas Kinematics in Nearby Galaxies}

\author{Rebecca C. Levy}
\affiliation{Department of Astronomy, University of Maryland, College Park, MD 20742, USA}
\author{Alberto D. Bolatto}
\affiliation{Department of Astronomy, University of Maryland, College Park, MD 20742, USA}
\author{Peter Teuben}
\affiliation{Department of Astronomy, University of Maryland, College Park, MD 20742, USA}
\author{Sebasti\'{a}n F. S\'{a}nchez}
\affiliation{Instituto de Astronom\'{\i}a, Universidad Nacional Aut\'{o}noma de M\'{e}xico, A.P. 70-264, 04510 M\'{e}xico, D.F.,  Mexico}
\author{Jorge K. Barrera-Ballesteros}
\affiliation{Department of Physics \& Astronomy, Johns Hopkins University, Baltimore, MD 21218, USA}
\author{Leo Blitz}
\affiliation{Department of Astronomy, University of California, Berkeley, CA 94720, USA}
\author{Dario Colombo}
\affiliation{Max-Planck-Institut f\"{u}r Radioastronomie, D-53121, Bonn, Germany}
\author{Rub\'{e}n Garc\'{i}a-Benito}
\affiliation{Instituto de Astrof\'{\i}sica de Andaluc\'{\i}a, CSIC, E-18008 Granada, Spain}
\author{Rodrigo Herrera-Camus}
\affiliation{Max-Planck-Institut f\"{u}r Extraterrestrische Physik, D-85741 Garching bei M\"{u}chan, Germany}
\author{Bernd Husemann}
\affiliation{Max-Planck-Institut f\"{u}r Astronomie, K\"onigstuhl 17, D-69117 Heidelberg, Germany}
\author{Veselina Kalinova}
\affiliation{Max-Planck-Institut f\"{u}r Radioastronomie, D-53121, Bonn, Germany}
\author{Tian Lan}
\affiliation{Department of Astronomy, University of Illinois, Urbana, IL 61801, USA}
\affiliation{Department of Astronomy, Columbia University, New York, NY 10027, USA}
\author{Gigi Y. C. Leung}
\affiliation{Max-Planck-Institut f\"{u}r Astronomie, K\"onigstuhl 17, D-69117 Heidelberg, Germany}
\author{Dami\'{a}n Mast}
\affiliation{Universidad Nacional de C\'{o}rdoba, Observatorio Astron\'{o}mico de C\'{o}rdoba, C\'{o}rdoba, Argentina}
\affiliation{Consejo de Investigaciones Cient\'{\i}ficas y T\'{e}cnicas de la Rep\'{u}blica Argentina, C1033AAJ, CABA, Argentina}
\author{Dyas Utomo}
\affiliation{Department of Astronomy, University of California, Berkeley, CA 94720, USA}
\affiliation{Department of Astronomy, The Ohio State University, Columbus, OH 43210, USA}
\author{Glenn van de Ven}
\affiliation{Max-Planck-Institut f\"{u}r Astronomie, K\"onigstuhl 17, D-69117 Heidelberg, Germany}
\affiliation{European Southern Observatory (ESO), 85748 Garching bei M\"{u}nchen, Germany}
\author{Stuart N. Vogel}
\affiliation{Department of Astronomy, University of Maryland, College Park, MD 20742, USA}
\author{Tony Wong}
\affiliation{Department of Astronomy, University of Illinois, Urbana, IL 61801, USA}

\email{rlevy@astro.umd.edu}

\begin{abstract}
{\comment We present a comparative study of molecular and ionized gas kinematics in nearby galaxies.} These results are based on observations from the EDGE survey, which measured spatially resolved $^{12}$CO(J=1--0) in 126 nearby galaxies. Every galaxy in EDGE has corresponding resolved ionized gas measurements from CALIFA. Using a sub-sample of 17 rotation dominated{\comment, star-forming} galaxies where precise molecular gas rotation curves could be extracted, we derive CO and H$\alpha$ rotation curves using the same geometric parameters out to $\gtrsim$1 $R_e$. We find that $\sim$75\% of our sample galaxies have smaller ionized gas rotation velocities than the molecular gas in the outer part of the rotation curve. In no case is the molecular gas rotation velocity measurably lower than that of the ionized gas. We suggest that the lower ionized gas rotation velocity can be attributed to a significant contribution from extraplanar diffuse ionized gas in a thick, turbulence supported disk. Using observations of the H$\gamma$ transition also available from CALIFA, we measure ionized gas velocity dispersions and find that these galaxies have sufficiently large velocity dispersions to support a thick ionized gas disk. Kinematic simulations show that a thick disk with a vertical rotation velocity gradient can reproduce the observed differences between the CO and H$\alpha$ rotation velocities. Observed line ratios tracing diffuse ionized gas are elevated compared to typical values in the midplane of the Milky Way. In galaxies affected by this phenomenon, dynamical masses measured using ionized gas rotation curves will be systematically underestimated.
\end{abstract}


\section{Introduction}
\label{sec:intro}
Studying the molecular and ionized gas components of a galaxy gives powerful insights into various stages of star formation. The gas kinematics can reveal feedback mechanisms, such as inflows and outflows, and merger events which alter the star formation history (SFH) of the galaxy. The measurement of molecular kinematics of galaxies, as traced by $^{12}$CO, has vastly improved in recent years due to the advent of interferometers which allow for high spatial and spectral resolution measurements. Similar advances have been made in the optical regime through the use of integral field units (IFUs). Studying the multiwavelength kinematic properties of nearby galaxies provides information about their formation, SFH, and evolution. 

The multi-wavelength kinematics of disk galaxies have been compared in a number of case studies. \citet{wong04}, \citet{yim14}, and \citet{frank16} compare \hi\ and CO kinematics and generally find good agreement between the rotation velocities of the atomic and molecular components. However, comparisons with the ionized gas often lead to different results. Most notably is NGC\,891, which shows vertical gradients in the rotation velocity (``lags'') in \hi\ of -10 -- -20 \kms\,kpc$^{-1}$ \citep{swaters97,fraternali05} and in ionized gas of -15 \kms\,kpc$^{-1}$\citep{heald06b}. Similar lags in \ha\ and \hi\ are seen in NGC\,5775 \citep{lee01}, where the CO and \ha\ rotation velocities agree in the midplane \citep{heald06a}. However, lags between the \hi\ and \ha\ do not always agree \citep{fraternali04,fraternali05,zschaechner15a,zschaechner15b}. \citet{deblok16} study the CO, \hi, and \CII\ kinematics in ten nearby galaxies and find that the \CII\ velocity is systematically larger than that of the CO or \hi, although they attribute this is systematics in the data reduction and the low velocity resolution of the \CII\ data. \citet{simon05a} compare CO and \ha\ rotation curves in two disk galaxies: in NGC\,5963, the CO and \ha\ velocities agree to within 1 \kms, but the \ha\ in NGC\,4605 shows systematically slower rotation than the CO by 4.8 \kms. Clearly, comparisons among tracers of the different phases of the interstellar medium (ISM) are complicated. Large, homogeneous samples of galaxies in multiple tracers are needed to make headway towards understanding the causes and ubiquity of the kinematic differences between ISM phases.

\citet{davis13} studied the properties of 24 gas-rich early type galaxies from the ATLAS$^{3{\rm D}}$ survey by comparing the ionized, atomic, and molecular gas kinematics out to \~0.5 $R_e$. They find that 80\% of their sample show faster molecular gas rotation velocities than the ionized gas. The other 20\% have the same molecular and ionized gas rotation velocities. They attribute these rotation velocity differences to the velocity dispersion of the ionized gas. Therefore, the dynamically cold molecular gas is a better tracer of the circular velocity than the ionized gas. Such a study has yet to be carried out in a similar sample of star-forming disk galaxies.

One way to study the kinematics of a galaxy is through its rotation curve, the rotation velocity as a function of galactocentric radius. The velocity can be decomposed into rotational, radial, and higher order terms \citep[e.g.][]{begeman89,schoenmakers99,vandeVen10}. High spatial resolution data are needed to construct robust rotation curves using this method. This high resolution data on a large sample of galaxies has been lacking, particularly for the molecular gas tracers. The CALIFA IFU survey \citep{sanchez12a} measured optical spectra of 667 nearby galaxies, providing spatially and spectrally resolved \ha\ velocities, as well as intensities, velocities, and velocity dispersions for many other ionized gas lines. The EDGE-CALIFA survey \citep[EDGE,][]{bolatto17}, selected 126 galaxies from CALIFA and observed them in $^{12}$CO$(J=1-0)$ with the Combined Array for Millimeter Wave Astronomy (CARMA) at \~4.5" resolution. Together, these surveys allow for the molecular and ionized gas kinematics of a statistical sample of nearby, star-forming galaxies to be analyzed. {\comment Using a sub-sample of 17 EDGE-CALIFA galaxies, this} work constitutes the largest spatially resolved combined CO and \ha\ kinematic analysis to date for late-type galaxies.

Section \ref{sec:obs} presents the EDGE, CALIFA, and ancillary data used for this study. The rotation curve fitting routine, procedure to determine the kinematic parameters from the EDGE CO data, and the sub-sample of galaxies used in this work are discussed in Section \ref{sec:dataanalysis}. Section \ref{sec:results} presents comparisons of the CO and \ha\ rotation curves. Potential explanations and interpretations of the results are presented in Section \ref{sec:disc}, including the results of the kinematic simulations, velocity dispersions, and ionized gas line ratios. We present our conclusions and summary in Section \ref{sec:summary}. Throughout this paper, CO refers to $^{12}$C$^{16}$O$(J=1-0)$. 

\section{Observations and Data Reduction}
\label{sec:obs}
\subsection{The EDGE-CALIFA Survey}
\label{ssec:edgesurvey}
The EDGE-CALIFA survey \citep{bolatto17} measured CO in 126 nearby galaxies with CARMA in the D and E configurations. Full details of the survey, data reduction, and masking techniques are discussed in \citet{bolatto17}, and we present a brief overview here. The EDGE galaxies were selected from the CALIFA sample (discussed in the following section) based on their infrared (IR) brightness {\comment and are biased toward higher star formation rates (SFRs) \citep[see Figure 6 of][]{bolatto17}}. A pilot study of 177 galaxies was observed with the CARMA E-array. From this sample, 126 galaxies selected for CO brightness were re-observed in the D-array. These 126 galaxies with combined D and E array data constitute the main EDGE sample\footnote{The EDGE CO data cubes and moment maps for the main sample are publicly available and can be downloaded from \url{www.astro.umd.edu/EDGE}.}. The EDGE sample is the largest sample of galaxies with spatially resolved CO, with typical angular resolution of 4.5\arcsec\ (corresponding to \~1.5\,kpc at the mean distance of the sample). Data cubes were produced with 20 \kms\ velocity channels. At each pixel in the cube, a Gaussian is fit to the CO line. Velocity-integrated intensity, mean velocity, velocity dispersion, and associated error maps are created from the Gaussian fits. {\comment Pixels with velocities that differ from their nearest (non-blanked) neighbors by more than 40 \kms, generally caused by fitting failures in low signal-to-noise data, are replaced with the median value of the neighbors. This replacement is rare and occurs for \~0.5\% of pixels in a given galaxy.} Additional masking was applied to the CO maps where the Gaussian fitting introduced artifacts. This masking was based on signal to noise ratio (SNR) cut using the integrated intensity and associated error map. Pixels with SNR < 1 were blanked in the velocity field. {\comment Average CO velocity dispersions are derived and are listed in Table \ref{tab:KSSparams}. A beam smearing correction is applied and is discussed in Appendix \ref{app:BS}.} 

\subsection{The CALIFA Survey}
\label{ssec:califasurvey}
The CALIFA survey \citep{sanchez12a} observed 667 nearby (z = 0.005--0.03) galaxies. Full details of the CALIFA observations are presented in \citet{walcher14} and other CALIFA papers, but we present a brief overview for completeness. CALIFA used the PPAK IFU on the 3.5m Calar Alto observatory with two spectral gratings. The low resolution grating (V500) covered wavelengths from 3745--7500\,\AA\ with 6.0\,\AA\ (FWHM) spectral resolution, corresponding a FWHM velocity resolution of 275\,\kms\ at \ha. The moderate resolution grating (V1200) covered wavelengths from 3650--4840\,\AA\ with 2.3\,\AA\ (FWHM) spectral resolution, corresponding to a FWHM velocity resolution of 160\,\kms\ at \hg\ \citep{sanchez16}. The V500 grating includes many bright  emission lines, including \ha, \hb, \hg, \hd, the \NII\ doublet, and the \SII\ doublet. The V1200 grating contains many stellar absorption features used to derive the stellar kinematics as well as a few ionized gas emission lines, such as \hg\ and \hd. The typical spatial resolution of the CALIFA data are 2.5", corresponding to \~0.8\,kpc at the mean distance of the galaxies. The CALIFA galaxies were selected from the Sloan Digital Sky Survey (SDSS) DR7 to have angular isophotal diameters between 45" and 79.2" to make the best use of the PPAK field of view. The upper redshift limit was set so that all targeted emission lines were observable for all galaxies in both spectral set ups; the lower redshift limit was set so that the sample would not be dominated by dwarf galaxies. The data used for this study come from the final data release\footnote{The CALIFA data cubes are publicly available at \url{http://califa.caha.es}.} \citep{sanchez16} and data products come from \pipetd\ version 2.2 \citep{pipe3DI,pipe3DII} provided in the final form by the CALIFA Collaboration.

The wavelength calibration of the data is detailed in \citet{sanchez12a} and Appendix A.5 of \citet{husemann13} and is crucial to extract accurate line velocities. The wavelength calibration data are used to resample the spectra to a linear wavelength grid and to homogenize the spectral resolution across the band (6.0\,\AA\ FWHM for V500 and 2.3\,\AA\ FWHM for V1200). The calibration is done using HeHgCd lamp exposures before and after each pointing using 16 lines for the V500 data and 11 lines for the V1200 data. The resulting accuracy of the wavelength calibration is \~0.2--0.3\,\AA\ for the V500 data and \~0.1--0.2\,\AA\ for the V1200 data. However, in our analysis of the V1200 data, we found errors in the wavelength calibration resulting from a bad line choice used to anchor the wavelength scale. This has been remedied in the current version of the data used here.

Once the data are calibrated, \pipetd\ fits and removes the stellar continuum, measures emission line fluxes, and produces two-dimensional data products for each emission line. Full details of \pipetd\ and its application to the CALIFA data can be found in \citet{pipe3DI,pipe3DII}, and important details are reproduced here for completeness. The underlying stellar continuum is fit and subtracted to produce a continuum-subtracted or ``emission line only'' spectrum \citep[Section 2 of][]{pipe3DI}. A Monte Carlo method is used to first determine the non-linear stellar kinematic properties and dust attenuation at each pixel in the cube. Next, the results of this non-linear fitting are fixed and the properties of the underlying stellar population are determined from a linear combination of simple stellar population (SSP) templates \citep[see also Section 3.2 of][]{pipe3DII}. This model stellar spectrum is then subtracted from the CALIFA cube at each pixel to produce a continuum-subtracted cube. To determine the properties of the emission lines, \pipetd\ uses a {\comment nonparametric} fitting routine optimized for weaker emission lines (``flux\_elines") which extracts only the line flux intensity, velocity, velocity dispersion, and equivalent width (see Section 3.6 of \citet{pipe3DII} for full details). Each emission line of interest is fit using a moment analysis similar to optimal extraction. The line centroid is first guessed based on the rest-wavelength of the line, and a wavelength range is defined based on the input guess for the line FWHM. A set of 50 spectra in this range are generated using a Monte Carlo method and each is fit by a Gaussian. At each step in the Monte Carlo loop, the integrated flux of the line is determined by a weighted average, where the weights follow a Gaussian distribution centered on the observed line centroid and the input line FWHM. With the integrated flux fixed, the velocity of the line centroid is determined. {\comment The line fluxes are not corrected for extinction within \pipetd\ so the desired extinction correction can be applied in the analysis. We do not apply an extinction correction since this would have a minimal effect on the line centroid used here.} Additional masking was also applied to the CALIFA velocity fields using a SNR cut based on the integrated flux and error maps. Pixels with SNR < 3.5 were blanked. 

The linewidths of the V500 data (which covers \ha) are dominated by the instrumental linewidth (6.0\,\AA\ $\approx$ 275\,\kms\ at $\lambda({\rm \ha}) = 6562.68$\,\AA), and hence reliable velocity dispersions are not available for the V500 data. The instrumental linewidth can be removed from the V1200 data (2.3\,\AA\ $\approx$ 160\,\kms\ at $\lambda({\rm\hg})=4340.47$\,\AA). {\comment To determine the \hg\ linewidth, we start with the continuum-subtracted cube and isolate the \hg\ line. We fit the \hg\ line at each spaxel using a Gaussian, where the linewidth is given by the width of the Gaussian fit. Pixels with SNR < 3 are blanked. We convert the resulting maps from wavelength to velocity using the relativistic convention, producing maps of the velocity dispersions for each galaxy. Independently, \pipetd\ does provide velocity dispersion maps derived from non-parametric fitting. The values in these maps, however, are frequently lower than the instrumental velocity dispersion over extended regions (a problem we do not find in our Gaussian fitting), and it is known that the pipeline systematically finds dispersions lower than obtained from Gaussian fitting \citep[section 3.6 of][]{pipe3DII}. 
We compare the velocity dispersions extracted from \pipetd\ and our Gaussian fitting to non-parametric fitting done with NEMO \citep{nemo}. In this fitting we find the linewidth at each spaxel using the \ccdmom\ \nmom\ task. This finds the peak, locates the minima on either side of the peak, and takes a second moment over those channels. Velocity dispersions from this method agree much better with the Gaussian fitting results than with the \pipetd\ values, hence we adopt the Gaussian fitting results to determine the \hg\ velocity dispersion. 
Before using these velocity dispersion maps in our analysis (Section \ref{ssec:veldisp}), we remove the instrumental velocity dispersion, and and model and remove the beam smearing effects (the latter is a small effect in the regions were we are interested in measuring the gas velocity dispersion). This procedure is discussed in Appendix \ref{app:BS}, and caveats are discussed further in Section \ref{ssec:veldisp}. We regrid all CALIFA maps to the same grid as the corresponding EDGE map using the \miriad\ task \regrid\ \citep{miriad}.} 

{\comment CALIFA also derived effective radius ($R_e$) measurements for all EDGE galaxies as described in \citet{sanchez14}. These values are listed in Table \ref{tab:CALIFAparameters}.}

When comparing the velocity fields from the EDGE and CALIFA surveys, it is important to note that the velocities are derived using different velocity conventions: EDGE follows the radio convention, and CALIFA follows the optical convention. Because velocities in both surveys are referenced to zero, all velocities are converted to the relativistic velocity convention. In both the optical and radio conventions, the velocity scale is increasingly compressed at {\comment larger} redshifts; typical systemic velocities in the EDGE-CALIFA sample are \~4500 \kms. The relativistic convention does not suffer from this compression effect. Differences between these velocity conventions and conversions among them can be found in Appendix \ref{app:velconv}. All velocities presented here are in the relativistic convention, unless otherwise noted. 

\subsection{Convolving to a Common Spatial Resolution}
\label{ssec:convolution}
\begin{figure}
\label{fig:COHaVfields}
\centering
\includegraphics[width=\columnwidth]{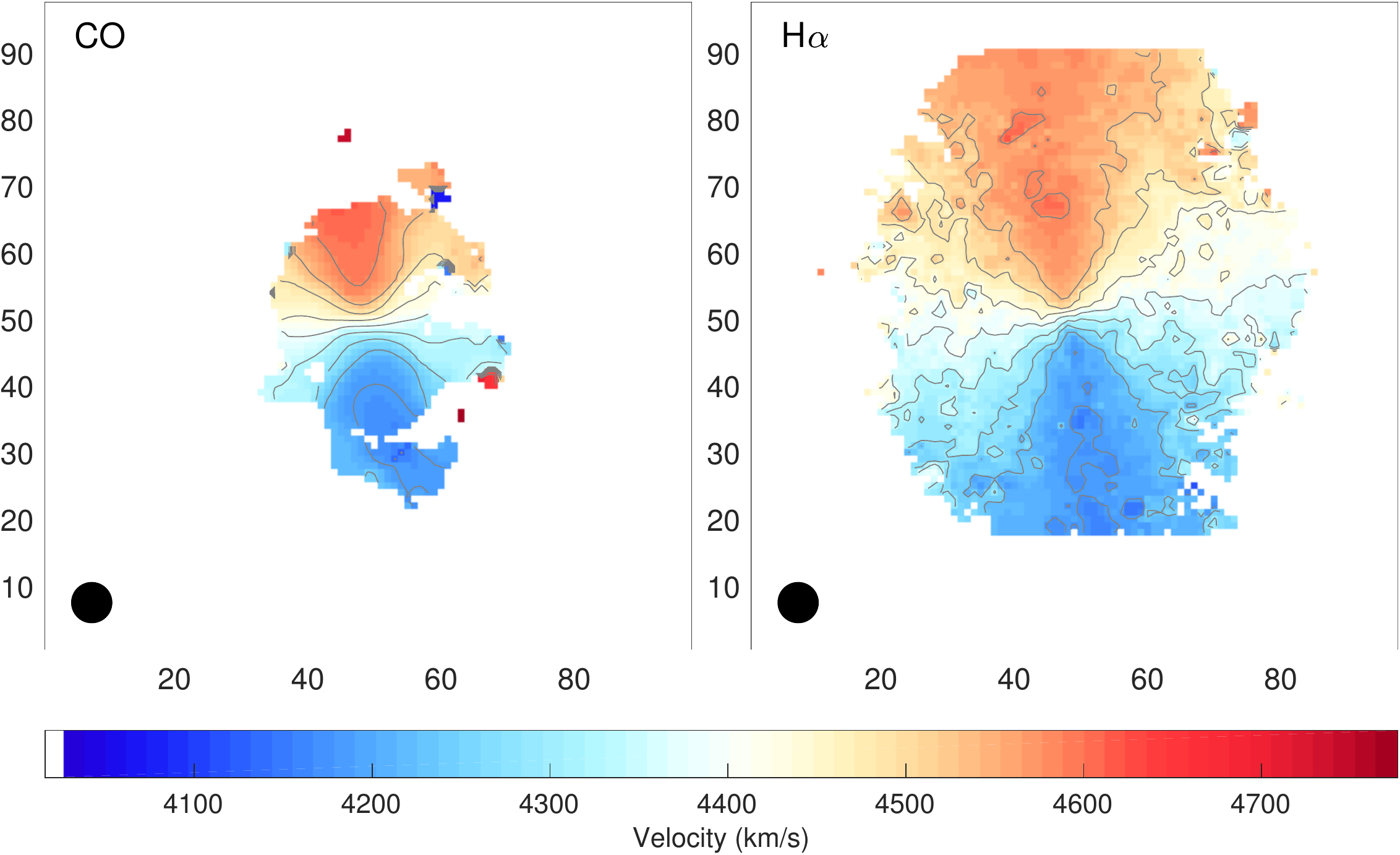}
\caption{\comment EDGE CO and CALIFA \ha\ velocity fields convolved to a 6" beam size for NGC\,2347. {\comments Isovelocity contours are shown in 50\,\kms\ increments out to $\pm250$\,\kms\ from the systemic velocity.} The circularized 6" beams are shown as the black circles.} 
\end{figure}

In order to accurately compare the CO and ionized gas velocity fields, the EDGE and CALIFA data cubes were convolved to the same angular resolution. The convolution was done using the \convol\ task in \miriad\ \citep{miriad}, which uses a Gaussian kernel. The EDGE beam was first circularized by convolving to a value 5\% larger than the beam major axis. The CALIFA point spread functions are circular \citep{sanchez16}. The EDGE and CALIFA cubes were convolved to a final 6" resolution, corresponding to \~2\,kpc at the mean distance of the galaxies. Data products were reproduced as outlined in Sections \ref{ssec:edgesurvey} and \ref{ssec:califasurvey}. {\comment The CO and \ha\ velocity fields for NGC\,2347 are shown in Figure \ref{fig:COHaVfields}. The rotation curves were derived as described in Section \ref{ssec:rcfitting}. There is excellent agreement between the native and convolved rotation curves for both CO and \ha, suggesting that while it is best to match physical resolution the convolution does not affect the results presented here. }

\subsection{GBT \hi\ Data}
\label{ssec:GBT}

\begin{figure}
\label{fig:HIspec}
\centering
\includegraphics[width=\columnwidth]{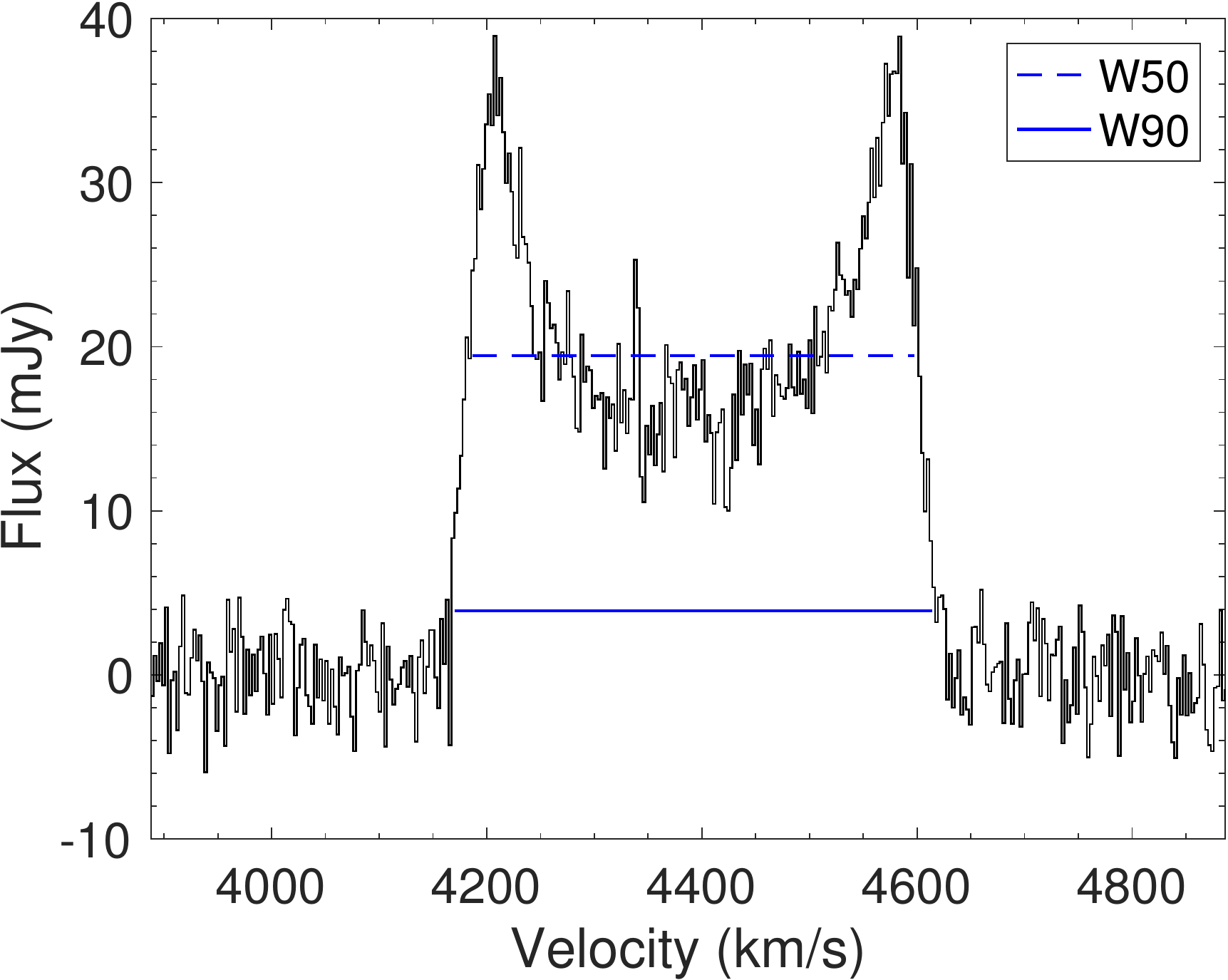}
\caption{The \hi\ spectrum from the GBT is shown for NGC\,2347. The spectrum has been clipped to $\pm 500$ \kms\ from the CO systemic velocity. The velocities here have been converted to the relativistic convention. Inclination corrected W50 and W90 values are indicated by the dashed and solid blue lines. For this work, the \hi\ data are used as a comparison to the molecular and ionized gas rotation velocities.} 
\end{figure}

The EDGE collaboration obtained \hi\ spectra for 112 EDGE galaxies from the Robert C. Byrd Green Bank Telescope (GBT) in the 2015B semester as part of GBT/15B-287 (PI: D. Utomo). We defer detailed discussion of these data for a future paper (Wong et al. 2018, in preparation) and present a brief overview. Observations were taken using the VEGAS spectrometer with a 100 MHz bandwidth, 3.1 kHz (0.65 \kms) spectral resolution, and a 3-$\sigma$ sensitivity of 0.51 mJy. On-source integration time was 15 minutes for each galaxy. The GBT primary beam FWHM was 9' compared to the average EDGE $D_{25}=1.6$', so the galaxies are spatially unresolved. Data were reduced using standard parameters in the observatory-provided GBTIDL package. A first or second order baseline was fit to a range of line-free channels spanning 300--500 \kms\ on either side of the signal range. The spectra were calibrated to a flux density scale assuming a gain of 2 K/Jy and a negligible coupling of the source size to the telescope beam. The widths containing 50\% and 90\% of the flux (W50 and W90 respectively) were derived from a Hanning smoothed spectrum to use as proxies for the maximum rotation velocity of the neutral atomic gas in these galaxies. These values are listed in Table \ref{tab:KSSparams}, if available. The \hi\ spectrum for NGC\,2347 is shown in Figure \ref{fig:HIspec}, with the inclination corrected W50 and W90 values marked.

If \hi\ data from GBT are not available, W50 values only were taken from \citet{springob05}. Specifically, we use their W$_C$ values which are W50 corrected for the instrumental and redshift effects. For this work, these data are used for only three galaxies and come from either the Green Bank 300 ft telescope (NGC\,5480 and NGC\,5633) or Arecibo (line feed system, UGC\,9067). These values are also listed in Table \ref{tab:KSSparams}. 

In this work, we use the \hi\ rotation velocities as point of comparison to the CO and \ha\ rotation velocities. We convert the W50 and W90 values to rotation velocities where $\vrot$=W$/(2\sin i)$, where $i$ is the galaxy's inclination as listed in Table \ref{tab:EDGEparameters}. 

\section{Data Analysis}
\label{sec:dataanalysis}
\subsection{Fitting CO Rotation Curves}
\label{ssec:rcfitting}

Rotation curves were determined for each galaxy using a tilted ring method \citep{rogstad74,begeman89}, which has previously been applied to \hi\ data \citep[e.g][]{begeman89,schoenmakers99,fraternali02,deblok08,iorio17}, ionized gas data \citep[e.g][]{vandeVen10,diteodoro16}, CO data \citep[e.g][]{wong04,frank16}, and recently ${\rm [C{\small II}]158\mu m}$ data in high redshift galaxies \citep{jones17}. Galaxies were deprojected (position angles and inclinations are listed in Table \ref{tab:EDGEparameters}) and divided into circular annuli. The radius of each annulus was determined such that the width was at least half a beam. The center position, inclination ($i$), and position angle (PA) are assumed the same for all annuli. The PA takes values between 0 and 360 degrees and increases counterclockwise, where ${\rm PA}=0$ indicates that the approaching side is oriented due north. The rotation ($\vrot$), radial ($\vrad$), and systemic ($\vsys$) velocity components were determined in each ring using a first order harmonic decomposition of the form
\begin{equation}
\label{eq:harmonicdecomp}
V(r) = \vrot(r)\cos\psi\sin i+\vrad(r)\sin\psi\sin i+\Delta\vsys(r)
\end{equation}
where $r$ is the galactocentric radius and $\psi$ is the azimuthal angle in the plane of the disk \citep{begeman89,schoenmakers99}. Before fitting, the central systemic velocity ($\vsys^{\rm cen}$) was subtracted from the entire map, so that the fitted systemic component is $\Delta\vsys(r)=\vsys(r)-\vsys^{\rm cen}$.

The initial values for the PAs and inclinations were chosen from photometric fits to outer optical isophotes \citep{falconbarroso17}. If values were not available from this method, they were taken from the HyperLeda database \citep{makarov14}. Initial central systemic velocity values ($\vsys^{\rm cen}$) and center coordinates (RA and Dec) were taken from HyperLeda. The kinematic PAs were determined from the results of the ring fitting by minimizing $\vrad$ at radii larger than twice the CO beam; an incorrect PA will produce a non-zero radial component. $\vsys^{\rm cen}$ values were refined by minimizing $\Delta\vsys$ at radii larger than twice the CO beam. The inclination is not as easily determined from kinematics; however, examining fits to individual annuli (rather than the rotation curve) can indicate whether the inclination is incorrect. Center offsets in RA and Dec (X$_{\rm off}$, Y$_{\rm off}$) were determined using a grid search method. At each point in the grid of X$_{\rm off}$ and Y$_{\rm off}$ values, a rotation curve was fit using that center. A constant was fit to the $\Delta\vsys$ component, and the combination of X$_{\rm off}$ and Y$_{\rm off}$ resulting in the best fit was selected as the center. The value of $\vsys^{\rm cen}$ was then adjusted as necessary to again minimize $\Delta\vsys$. The sign of the offset is such that the correct center is ($x_{\rm cen}$, $y_{\rm cen}$) = (RA--X$_{\rm off}$, Dec--Y$_{\rm off}$). If a rotation curve could not be fit, either because there is little or no detected CO or because the velocity field is very disturbed, the parameter values were unchanged from the initial values. The final values of the geometric parameters can be found in Table \ref{tab:EDGEparameters}, including whether the PA, inclination, and $\vsys^{\rm cen}$ values are derived from kinematics (this work), photometrically \citep{falconbarroso17}, or from HyperLeda.

\begin{figure*}
\label{fig:edgecalifasubplot}
\centering
    \gridline{\fig{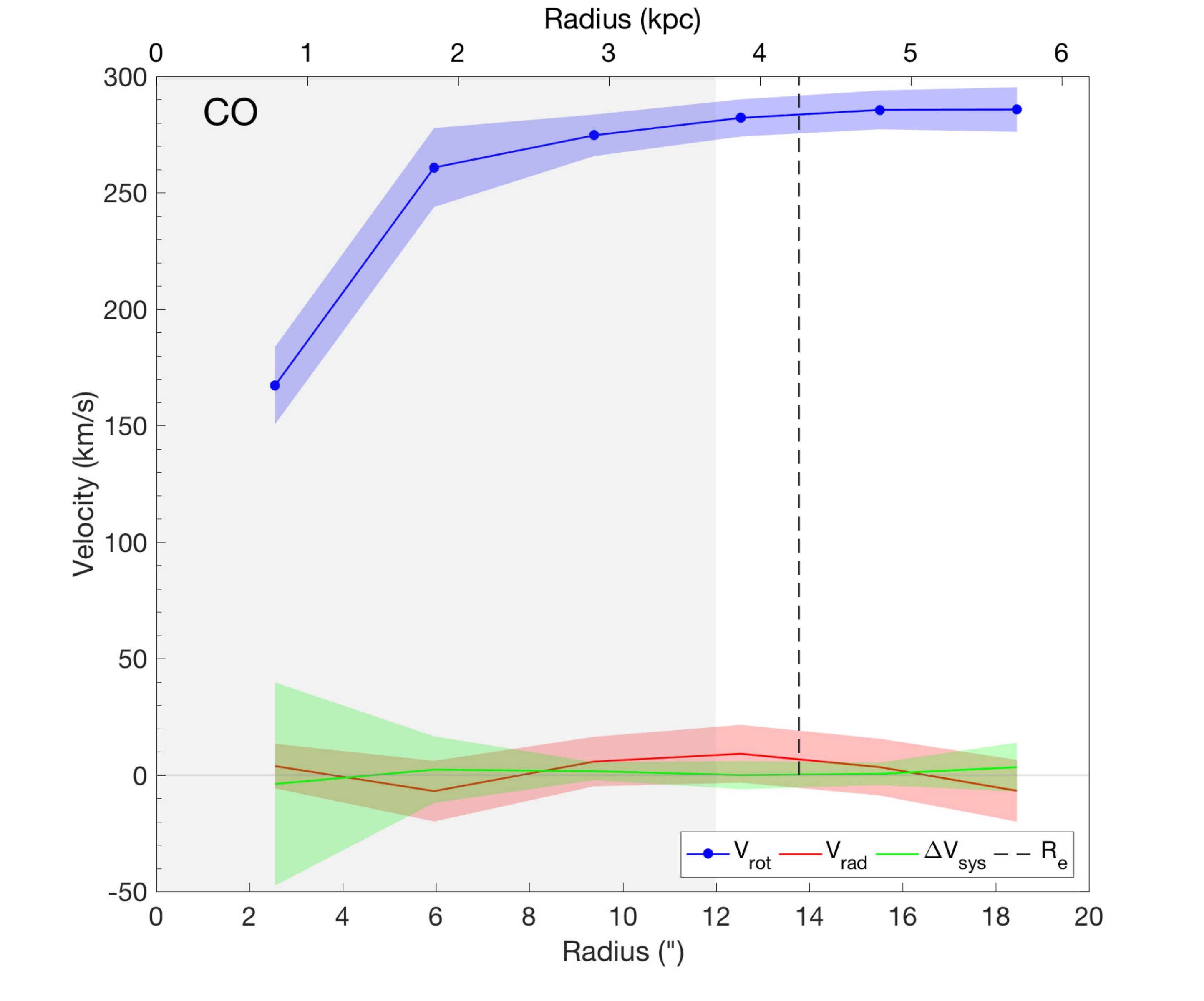}{1.1\columnwidth}{(a)}
    \fig{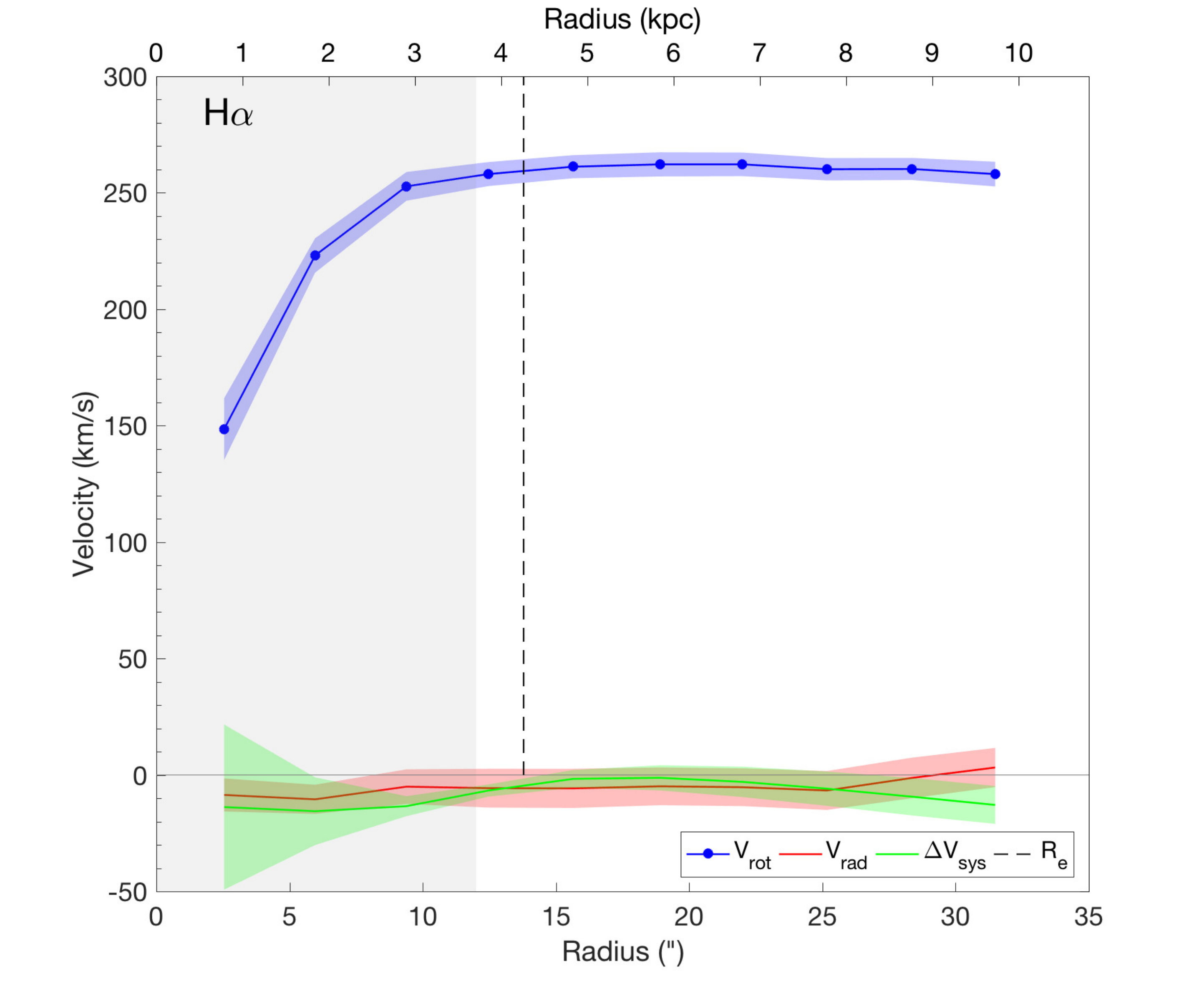}{1.1\columnwidth}{(b)}}
\caption{(a) The CO rotation curve for NGC\,2347, where $\vrot$ is shown in blue, $\vrad$ in red, and $\Delta\vsys$ in green. The colored shaded regions are the errors on the rotation curve from the Monte Carlo method. The gray shaded region shows the inner 2 beams where beam smearing can affect the rotation curve substantially. The black dashed line shows $R_e$ (Table \ref{tab:CALIFAparameters}). (b) The \ha\ rotation curve for NGC\,2347, where the colors of the curves are the same as (a). In both cases, the $\vrad$ and $\Delta\vsys$ components are small and consistent with zero within the error ranges. The $\vrot$ components flatten at larger radii. Interestingly, $\vrot$(\ha) is noticeably smaller than $\vrot$(CO).} 
\end{figure*}

Errors on the rotation curve were determined using a Monte Carlo method in which the geometrical parameters were drawn randomly from a uniform distribution. The center position was allowed to vary by 1" in either direction, since over the whole EDGE sample, the average change in the CO (or \ha) center position from the original value is 0.7". The inclination is varied by 2\D, which is the average difference between the final and initial inclinations over the whole EDGE sample. The PA was also allowed to vary by 2\D, which is the median difference between the final and initial PAs over the whole EDGE sample. This allows typical uncertainties in the kinematic parameters to be reflected in the rotation curves. The shaded error regions shows the standard deviation of 1000 such rotation curves. The CO rotation curve showing $\vrot, \vrad, {\rm\ and \ } \Delta\vsys$ for one galaxy is shown in Figure \ref{fig:edgecalifasubplot}a.

{\comments We note this method to determine errors on the rotation curve differs from methods which use the differences between the approaching and receding sides of the galaxy, assuming that those differences are at the 2-$\sigma$ level \citep[e.g.][]{swaters99,deblok08}. Typical uncertainties on the CO rotation velocity are $\sim$3--10 km/s (1-$\sigma$). For the uncertainties stemming from the difference in rotation velocity between the approaching and receding sides to exceed the typical 1-$\sigma$ uncertainties we find in our Monte Carlo method, rotation velocities of the approaching and receding sides would have to differ by 12--40 km/s. This seems unlikely, especially given that uncertainties on the rotation velocities derived from the differences between the approaching and receding sides presented by \citet{deblok08} are generally \~10\,\kms. Therefore, we conclude that the major source of uncertainty in deriving our rotation curves are uncertainties in the geometric parameters.}

Due to the beam size of the EDGE data, the observed velocities are affected by beam smearing, especially in the centers of the galaxies \citep{bosma78,begeman87}. \citet{leung18} analyzed the effect of beam smearing in the EDGE sample and found that it is only significant in the inner portions of the galaxy ($\lesssim$0.5 $R_e$) where the velocity gradient is steep. For this study, we do not correct for beam smearing but rather exclude from the analysis points in the rotation curve within 2 beams from the center. The radius corresponding to twice the CO beam is referred to as $\rbeam$ throughout. The excluded central region is shown in gray in Figure \ref{fig:edgecalifasubplot}. Excluding the center of the galaxy also minimizes any effects from a bulge or an active galactic nucleus (AGN).

\rotate
\begin{deluxetable*}{cccccccccccccccc}
\tablecaption{Parameters for the KSS \label{tab:KSSparams}}
\tabletypesize\footnotesize
\tablehead{
\colhead{Name} & \colhead{Morph (Type)} & \colhead{log(M$_*$)} & \colhead{log(SFR)} & \colhead{$d$} & \colhead{D$_{25}$} &  \colhead{CO V$_{\rm{max}}$} & \colhead{$\Dv$} & \colhead{HI W50} & \colhead{HI W90} & \colhead{$\Sigma_*$} & \colhead{$\sigma_{\rm CO}$} & \colhead{$\sigma_{\rm ADC}$} & \colhead{$\sigma_{\rm H\gamma}$}& \colhead{[SII]/H$\alpha$} & \colhead{[NII]/H$\alpha$} \\
& & \colhead{(M$_\odot$)} & \colhead{(M$_\odot$\,yr$^{-1}$)} & \colhead{(Mpc)} & \colhead{(')}  & \colhead{(km\,s$^{-1}$)} & \colhead{(km\,s$^{-1}$)}& \colhead{(km\,s$^{-1}$)}& \colhead{(km\,s$^{-1}$)} & \colhead{(M$_\odot$\,pc$^{-2}$)} & \colhead{(km\,s$^{-1}$)}& \colhead{(km\,s$^{-1}$)}& \colhead{(km\,s$^{-1}$)} & & }
\startdata
IC1199 & Sbc (3.7) & 10.8 & 0.2 & 68.3 & 1.2 & 199.0$\pm$5.7 & 2.6$\pm$4.3 & 141.8 & 229.0 & 189.6 & 11.1$\pm$5.5 & 52.1$\pm$26.5 & 34.6$\pm$9.2 & 0.19$\pm$0.01 & 0.39$\pm$0.01 \\
NGC2253 & Sc (5.8) & 10.8 & 0.5 & 51.2 & 1.4 & 174.2$\pm$7.9 & 1.1$\pm$7.7 & 211.0 & 243.9 & 205.6 & 9.2$\pm$2.5 & 24.9$\pm$37.2 & 28.0$\pm$3.4 & 0.17$\pm$0.01 & 0.37$\pm$0.01 \\
NGC2347 & Sb (3.1) & 11.0 & 0.5 & 63.7 & 1.6 & 285.6$\pm$2.0 & 24.1$\pm$0.4 & 267.7 & 289.2 & 336.5 & 10.1$\pm$4.0 & 85.9$\pm$5.0 & 36.6$\pm$5.1 & 0.18$\pm$0.01 & 0.38$\pm$0.01 \\
NGC2410 & Sb (3.0) & 11.0 & 0.5 & 67.5 & 2.2 & 227.2$\pm$5.9 & 15.3$\pm$5.1 & 247.3 & 267.3 & 198.6 & 17.6$\pm$7.5 & 67.7$\pm$16.5 & 38.2$\pm$12.1 & 0.20$\pm$0.01 & 0.47$\pm$0.04 \\
NGC3815 & Sab (2.0) & 10.5 & 0.0 & 53.6 & 1.4 & 185.8$\pm$5.2 & 18.8$\pm$3.6 & 176.5 & 223.5 & 215.4 & 15.1$\pm$5.1 & 52.9$\pm$8.0 & 26.9$\pm$3.3 & 0.20$\pm$0.02 & 0.38$\pm$0.01 \\
NGC4047 & Sb (3.2) & 10.9 & 0.6 & 49.1 & 1.5 & 216.3$\pm$2.2 & 14.3$\pm$0.5 & 220.3 & 243.2 & 296.6 & 9.5$\pm$2.4 & 63.2$\pm$9.5 & 29.5$\pm$3.7 & 0.16$\pm$0.01 & 0.35$\pm$0.01 \\
NGC4644 & Sb (3.1) & 10.7 & 0.1 & 71.6 & 1.5 & 186.1$\pm$5.8 & 13.5$\pm$2.8 & 171.0 & 214.7 & 132.6 & 16.6$\pm$8.1 & 53.2$\pm$10.9 & 31.3$\pm$7.7 & 0.16$\pm$0.01 & 0.40$\pm$0.01 \\
NGC4711 & SBb (3.2) & 10.6 & 0.1 & 58.8 & 1.2 & 142.8$\pm$10.5 & 5.8$\pm$0.9 & 167.3 & 185.2 & 121.1 & 10.0$\pm$4.4 & 47.9$\pm$37.2 & 32.9$\pm$7.7 & 0.18$\pm$0.01 & 0.36$\pm$0.01 \\
NGC5016 & SABb (4.4) & 10.5 & -0.0 & 36.9 & 1.6 & 178.4$\pm$0.8 & 17.7$\pm$8.7 & 187.2 & 199.1 & 248.9 & 11.3$\pm$5.6 & 56.4$\pm$11.0 & 26.6$\pm$3.8 & 0.16$\pm$0.01 & 0.38$\pm$0.01 \\
NGC5480 & Sc (5.0) & 10.2 & 0.2 & 27.0 & 1.7 & 111.4$\pm$6.6 & 4.8$\pm$5.1 & 147.7 & - & 180.0 & 11.2$\pm$3.5 & 37.1$\pm$41.5 & 24.0$\pm$2.9 & 0.20$\pm$0.01 & 0.33$\pm$0.01 \\
NGC5520 & Sb (3.1) & 10.1 & -0.1 & 26.7 & 1.6 & 162.3$\pm$0.2 & 15.3$\pm$1.4 & 158.1 & 170.1 & 202.3 & 13.4$\pm$6.8 & 60.5$\pm$0.4 & 26.5$\pm$2.5 & 0.20$\pm$0.01 & 0.39$\pm$0.01 \\
NGC5633 & Sb (3.2) & 10.4 & 0.2 & 33.4 & 1.1 & 187.5$\pm$9.4 & 9.1$\pm$2.9 & 200.6 & - & 396.0 & 11.3$\pm$2.5 & 53.3$\pm$23.4 & 25.1$\pm$1.8 & 0.17$\pm$0.01 & 0.37$\pm$0.01 \\
NGC5980 & Sbc (4.4) & 10.8 & 0.7 & 59.4 & 1.6 & 216.3$\pm$4.4 & 8.2$\pm$4.8 & 219.4 & 248.6 & 214.8 & 12.3$\pm$2.3 & 52.5$\pm$24.0 & 34.3$\pm$3.8 & 0.17$\pm$0.01 & 0.41$\pm$0.01 \\
UGC04132 & Sbc (4.0) & 10.9 & 1.0 & 75.4 & 1.2 & 238.5$\pm$11.4 & 16.3$\pm$8.0 & 255.6 & 291.5 & 206.5 & 15.8$\pm$3.9 & 79.4$\pm$40.0 & 33.1$\pm$4.3 & 0.20$\pm$0.01 & 0.40$\pm$0.01 \\
UGC05111 & Sbc (4.0) & 10.8 & 0.6 & 98.2 & 1.5 & 216.2$\pm$11.9 & 9.1$\pm$4.4 & - & - & 154.4 & 15.4$\pm$3.6 & 52.2$\pm$33.5 & - & 0.23$\pm$0.02 & 0.41$\pm$0.02 \\
UGC09067 & Sab (2.0) & 11.0 & 0.7 & 114.5 & 0.8 & 211.7$\pm$4.2 & 1.0$\pm$2.1 & 212.7 & - & 110.7 & 14.1$\pm$7.6 & 12.5$\pm$40.7 & 29.0$\pm$4.9 & 0.21$\pm$0.01 & 0.35$\pm$0.01 \\
UGC10384 & Sab (1.6) & 10.3 & 0.7 & 71.8 & 1.2 & 187.4$\pm$8.1 & 14.2$\pm$7.0 & 180.7 & 200.9 & 74.5 & 18.9$\pm$6.2 & 45.0$\pm$12.9 & 35.4$\pm$2.7 & 0.21$\pm$0.01 & 0.35$\pm$0.02 \\
\enddata
\tablecomments{The table lists relevant parameters for the KSS galaxies not already listed in Table \ref{tab:EDGEparameters}. The morphology and types are from HyperLeda and are listed in \citet{bolatto17}. Values for M$_*$, SFR, distances ($d$), and the diameter of the 25th magnitude isophote (D$_{25}$) are taken from CALIFA and are also found in \citet{bolatto17}. Errors on log(M$_*$) and log(SFR) are $\pm$0.1 in the corresponding units. CO V$_{\rm max}$ is the maximum CO rotation velocity, determined by the median of the CO RC at radii larger than twice the CO beam; the error is the standard deviation. $\Dv$ is the median difference between CO and \ha\ RCs, as described in Section \ref{sec:results}. \hi\ W50 and W90 listed here are uncorrected for inclination. All values are taken from the GBT except for NGC 5480, NGC 5633, and UGC 9067 (see Section \ref{ssec:GBT}). $\Sigma_*$ are averages of stellar surface density radial profiles from \citep{utomo17}. $\sigma_{\rm CO}$ are the velocity dispersions estimated from the beam-smearing corrected CO maps (Section \ref{ssec:edgesurvey}). $\sigma_{\rm ADC}$ are the velocity dispersions estimated from the ADC (Section \ref{ssec:veldispest}). $\sigma_{\rm H\gamma}$ are the median velocity dispersions measured from the H$\gamma$ line; errors are the weighted standard deviations (Section \ref{ssec:veldisp}). [SII]/H$\alpha$ and [NII]/H$\alpha$ are the median intensity ratios, where is error is the standard error Section (\ref{ssec:SII}).}
\end{deluxetable*}

\subsection{Fitting Ionized Gas Rotation Curves}
\label{ssec:califafitting}
The CALIFA data were fit using the methods described in in previous section and the same PA and inclination as the CO listed in Table \ref{tab:EDGEparameters}. In some cases, the \ha\ velocity contours are noticeably offset from the CO contours. CALIFA provides refinements to their astrometry in the headers of the data; however, these refinements are not large enough to account for some observed offsets. The CALIFA pipeline registers the RA and Dec for the center of the PPAK IFU to the corresponding center of the SDSS DR7 image \citep{garciabenito15}. In DR2, 7\% of the galaxies have registration offsets from SDSS >3" \citep{garciabenito15}. However, this registration process is known to fail in some cases. Indeed, in many of the galaxies for which we find offsets, this registration process has failed. Therefore, CALIFA centers were re-fit in the same way as the EDGE data, as described in Section \ref{ssec:rcfitting}. Because the V500 and V1200 data were taken on different days, the centers of the V500 and V1200 data were re-fit independently. For both the V500 and V1200 data, the average magnitude of the center offset is 0.9". The center offsets and $\vsys^{\rm cen}$ values for the CALIFA data are presented in Table \ref{tab:CALIFAparameters}. The \ha\ rotation curve for NGC\,2347 is shown in Figure \ref{fig:edgecalifasubplot}b.

\subsection{The Kinematic Sub-Sample}
\label{ssec:KSS}
Of the 126 EDGE galaxies, $\approx$100 have peak brightness temperatures $\ge 5\sigma$ \citep{bolatto17}. Reliable CO and \ha\ rotation curves could not, however, be derived for every detected EDGE galaxy. To best compare the CO and \ha\ rotation curves, a sub-sample of galaxies for which reliable CO and \ha\ rotation curves could be derived is used for the remainder of the analysis (the Kinematic Sub-Sample or KSS). A reliable rotation curve has small $\vrad$ and $\Delta\vsys$ components at radii larger than $\rbeam$ (as in Figure \ref{fig:edgecalifasubplot}). In the centers of galaxies, there may be radial motions due to bars and other effects, but these should not affect the larger radii we consider here. Ensuring that both the CO and \ha\ have small $\vrad$ components validates our assumptions that the CO and \ha\ have the same PA and inclination and that the PA and inclination do not change much over the disk (i.e. there are no twists or warps). In addition to the criteria on the rotation curves, there are four galaxies (NGC\,4676A, NGC\,6314, UGC\,3973, and UGC\,10205) for which the observed CO velocity width may not be fully contained in the band \citep{bolatto17}. One galaxy (UGC\,10043) has a known \ha\ outflow \citep{lopezcoba17}. These galaxies are also excluded from the subsample. Finally, we exclude galaxies with inclinations larger than 75\D. At large inclinations, the line profiles can become skewed and a Gaussian fit to the line profiles is not appropriate and can lead to systematic biases in the mean velocities. We do, however, plan to analyze the highly inclined galaxies in a forthcoming paper (Levy et al. 2018, in preparation). Under these criteria, our sample size is reduced to 17 galaxies. Figure \ref{fig:KSS} shows CO and \ha\ velocity fields and rotation curves for all galaxies in the KSS. Specific notes on each galaxy in the KSS can be found in Appendix \ref{app:galbygal}. Table \ref{tab:KSSparams} lists global quantities for the KSS not listed in Tables \ref{tab:EDGEparameters} or \ref{tab:CALIFAparameters}.


\section{Results}
\label{sec:results}

\begin{figure*}
\label{fig:RCcomp}
\centering
	\gridline{\fig{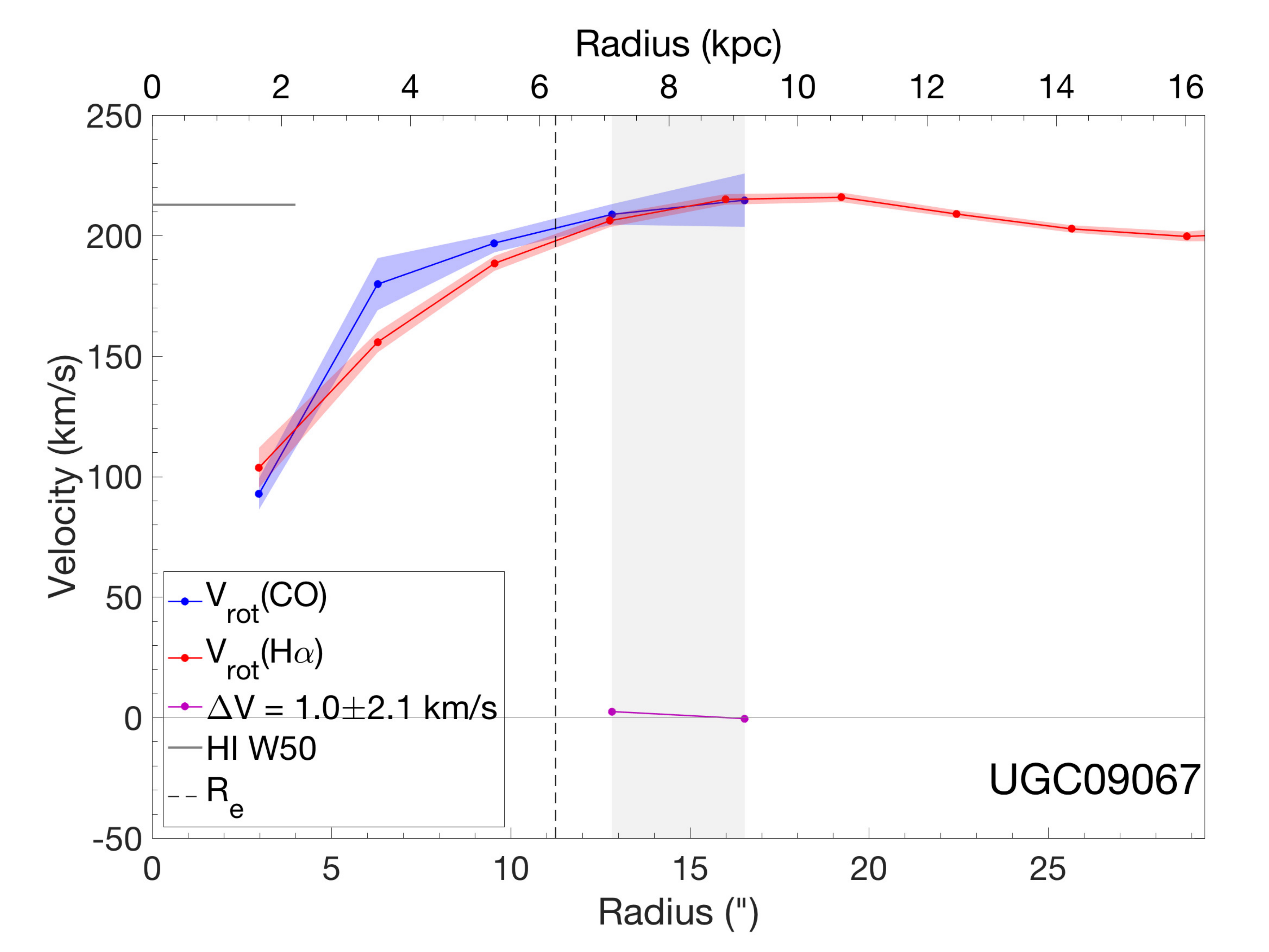}{1.05\columnwidth}{(a)}
    \fig{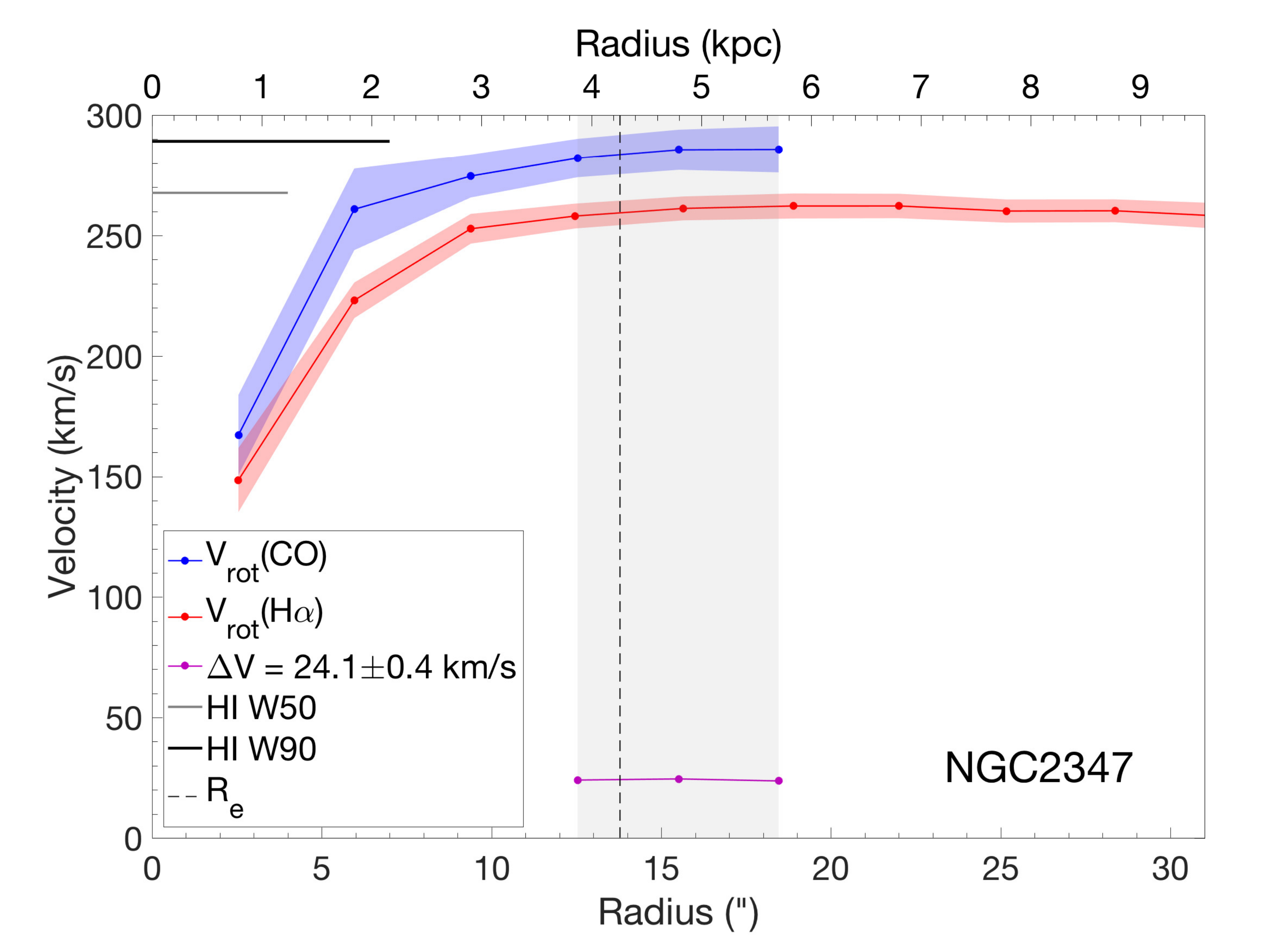}{1.05\columnwidth}{(b)}}
\caption{CO and \ha\ rotation curve comparisons for two galaxies. In both plots, $\vrot$(CO) is in blue and $\vrot$(\ha) is in red. The purple curves show the difference between the rotation curves at radii greater than $\rbeam$ to the furthest CO extent. The median difference between $\vrot$(CO) and $\vrot$(\ha) ($\Dv$) is quoted and the error is the standard deviation. The solid gray and black lines show the inclination corrected $\vrot$(\hi) values from W50 and W90 for comparison. The black dashed lines show $R_e$ (Table \ref{tab:CALIFAparameters}). (a) UGC\,9067 has CO and \ha\ rotation curves which are consistent within the error bars. (b) NGC\,2347 shows a difference of 24 \kms\ between the CO and \ha\ rotation curves. The \hi\ rotation velocities tend to agree better with the CO rotation curve. Many galaxies in the KSS show $\Dv$ which are larger than the errors on the rotation curves.} 
\end{figure*}

{\comment Previous comparisons of molecular and ionized gas rotation velocities for individual galaxies show variations in agreement \citep[e.g.][]{wong04,simon05a,heald06b,deblok16}.  \citet{davis13}, for example, found that for 80\% of their sample of 24 gas-rich early-type galaxies (ETGs) the ionized gas rotation velocities were lower than for the molecular gas. For a few of the star-forming disk galaxies in our KSS, the molecular and ionized gas rotation velocities agree within the errors, such as UGC\,9067 shown in Figure \ref{fig:RCcomp}a.} The majority of our galaxies, however, have CO rotation velocities which are measurably higher than the \ha\ rotation velocities (such as for NGC\,2347, shown in Figure \ref{fig:RCcomp}b). In no case is the \ha\ rotation velocity measurably higher than the CO. To quantify the differences between the CO and \ha\ rotation curves, the rotational component of the \ha\ rotation curve was linearly interpolated and resampled at the same radii as the CO rotation curve. $\vrot$(CO) and $\vrot$(\ha) are compared at radii larger than twice the convolved beam ($\rbeam$) to ensure that beam smearing is not affecting the results; the gray shaded regions in Figure \ref{fig:RCcomp} show the radii over which the rotation curves are compared. The differences between $\vrot$(CO) and $\vrot$(\ha) are shown in Figure \ref{fig:RCcomp} (purple points). The median of these differences ($\Dv$) was taken to determine an average velocity difference between the CO and \ha\ rotational velocities. The standard deviation of the difference at each radius ($\eDv$) is quoted as an error on $\Dv$. Galaxies have measurably different CO and \ha\ rotation velocities if $|\Dv| > \eDv$ and are consistent if $|\Dv| \le \eDv$. Of the 17 galaxies in the KSS, $77\%^{+23\%}_{-0\%}$ ($13^{+4}_{-0}$) show measurably higher CO rotation velocities than \ha, and the other $23\%^{+0\%}_{-23\%}$ ($4^{+0}_{-4}$) show consistent CO and \ha\ rotation velocities. This is remarkably similar to the ETG results of \citet{davis13}.

\begin{figure}
\label{fig:gausshistdiffs}
\centering
\includegraphics[width=\columnwidth]{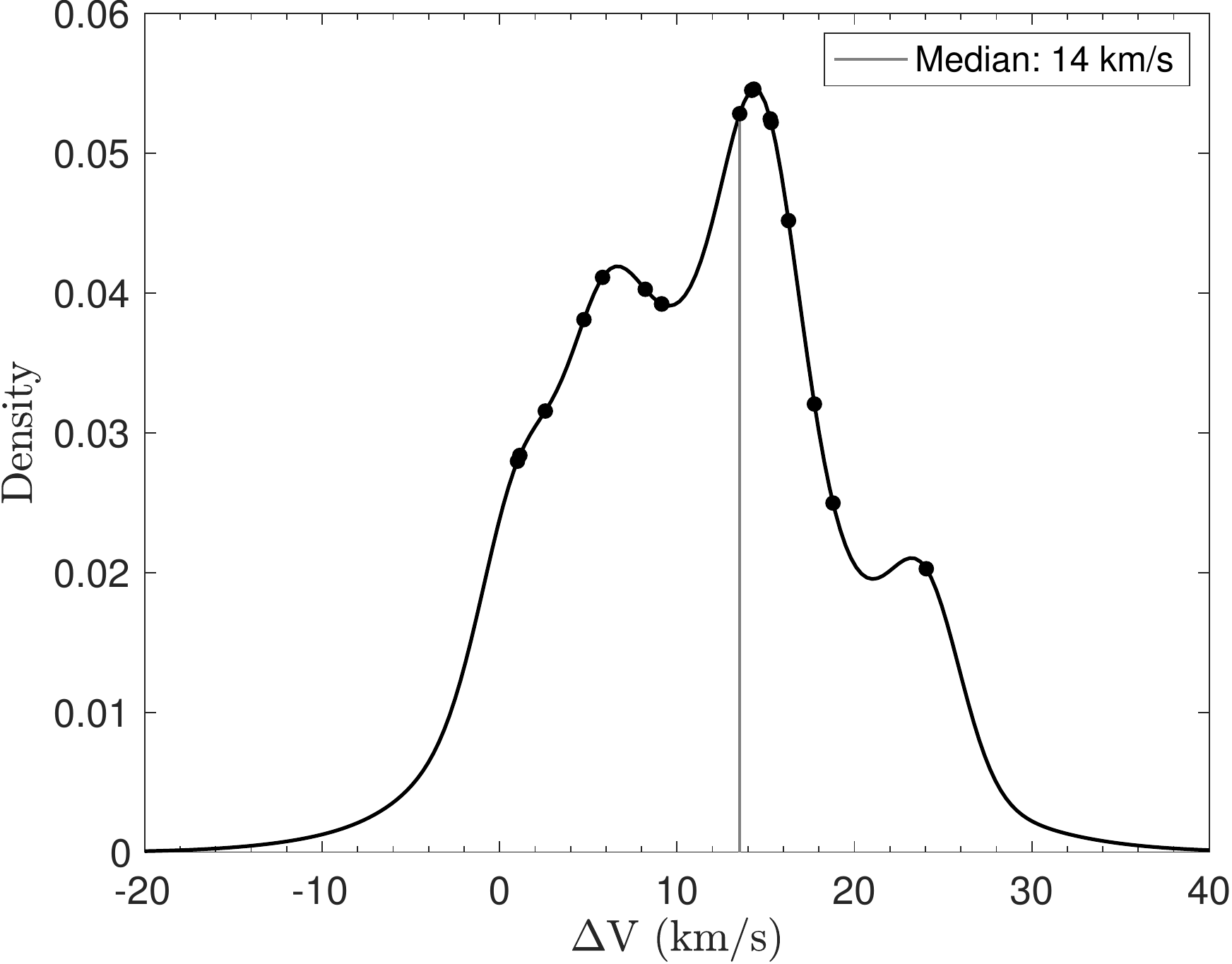} 
\caption{Kernel density estimator showing the distribution of $\Dv$ (the median $\vrot$(CO)-$\vrot$(\ha)) in the KSS. Each galaxy is represented as a Gaussian with centroid $\mu=\Dv$, $\sigma=\eDv$, and unit area. A minimum $\eDv$ of 2\,\kms\ is imposed. The Gaussians are summed to produce a histogram and normalized to unit area. The circles indicate the peak of the Gaussian for each galaxy. All galaxies in the KSS have $\Dv>0$. The median $\Dv$ is 14\,\kms.}
\end{figure}

To better understand the distribution of $\Dv$ in the KSS, a kernel density estimator (KDE) was formed, where each galaxy is represented as a Gaussian with centroid $\mu=\Dv$, $\sigma=\eDv$, and unit area. These Gaussians were summed and re-normalized to unit area. The resulting distribution is shown in Figure \ref{fig:gausshistdiffs}, showing that all galaxies in the KSS have $\Dv>0$. The median $\Dv$ of the sample is 14 \kms.

We find no strong radial trends in $\Dv$, likely because the range of radii probed is relatively small. Over the 17 KSS galaxies, the median gradient in $\Dv$ with radius is $-0.2\pm6.4$\,\kms\,kpc$^{-1}$.

\begin{figure}
\label{fig:ionRC}
\centering
\includegraphics[width=\columnwidth]{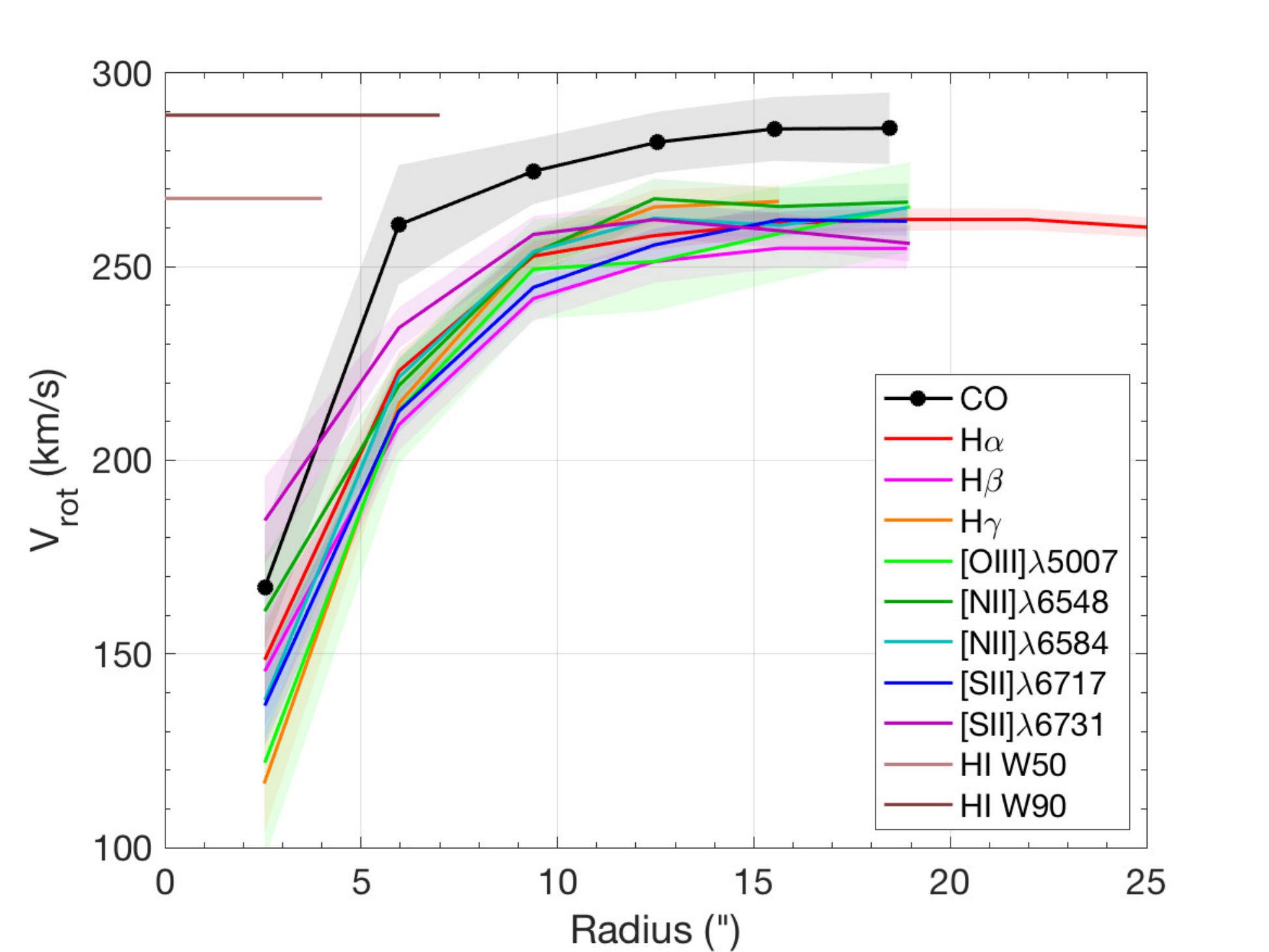} 
\caption{Rotation curves in several ionized gas lines are consistent with each other (using fixed geometric parameters listed in Table \ref{tab:CALIFAparameters}). The figure shows an example for NGC\,2347. For CO, the shaded region shows the error on the rotation curves from variations in the kinematic parameters using the Monte Carlo method described in Section \ref{ssec:rcfitting}. For all other curves, the shaded regions indicate the formal errors from the rotation curve fitting. The rotation curves from the ionized gas are consistent with one another and are all below the CO rotation curve. Ionized rotation curves other than \ha\ are truncated at the same radius as CO. The tan and brown horizontal lines show $\vrot$(\hi) from W50 and W90 measurements, which tend to agree with the CO rotation velocity. Note that variations among the rotation curves are enhanced as the y-axis does not extend down to zero.}
\end{figure}

In addition to \ha, rotation curves were derived for other ionized lines available from CALIFA using the same method and parameters described in Section \ref{ssec:califafitting}. These lines include \hb, \OIII$\lambda$5007, \NII$\lambda$6548, \NII$\lambda$6583, \SII$\lambda$6717, and \SII$\lambda$6731 from the V500 grating and \hg\ from the V1200 grating. Rotation curves from these lines (as well as CO and \ha) are shown for NGC\,2347 in Figure \ref{fig:ionRC}, where the colored shading indicates the errors on the rotation curves. Within these errors, the ionized gas rotation curves are consistent with one another, and below the CO rotation curve. Also shown are the W50 and W90 measurements from the HI data. These values straddle the CO rotation curve and both are larger than the ionized gas rotation velocities. 

\begin{figure}
\label{fig:DvEW}
\centering
\includegraphics[width=\columnwidth]{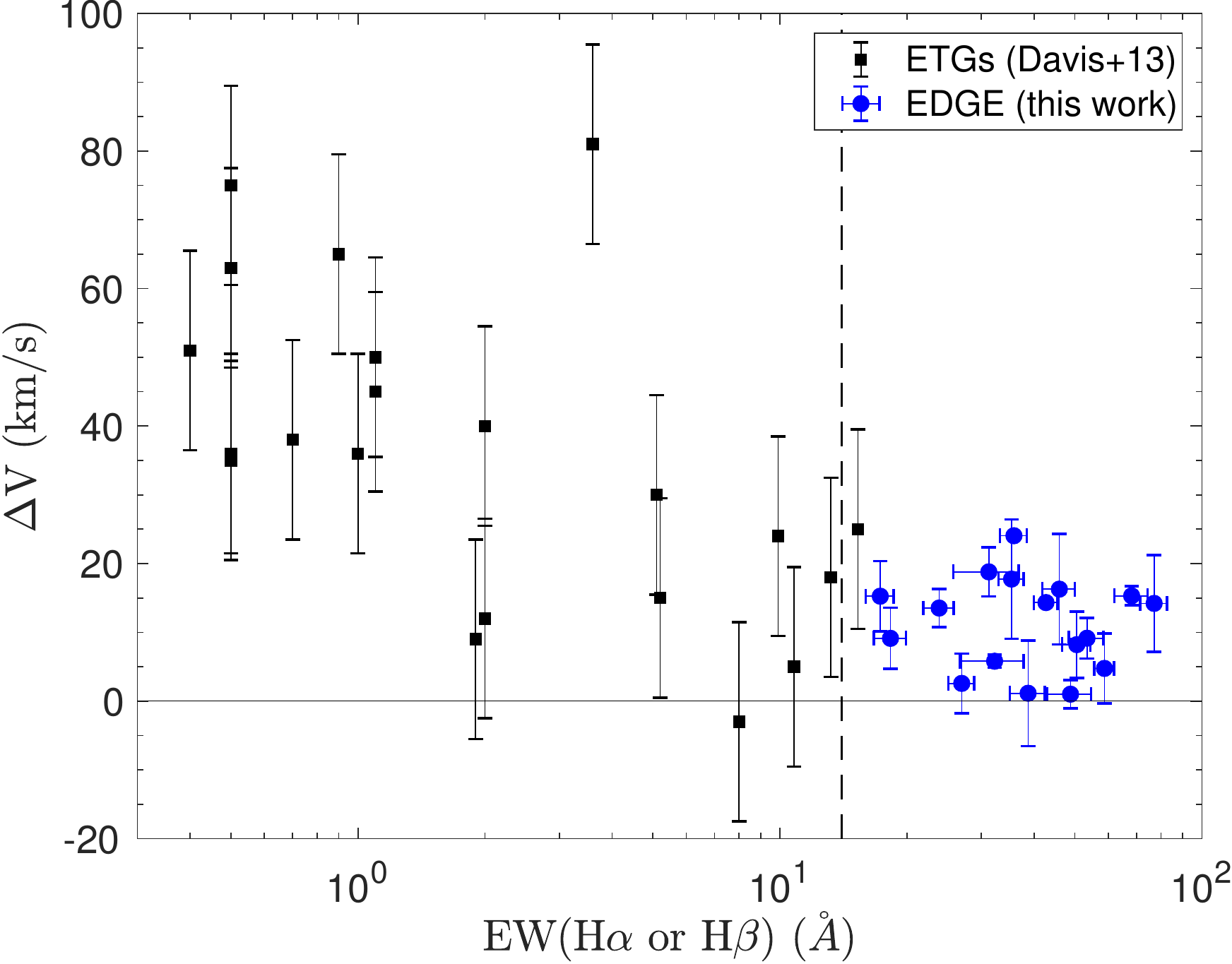}
\caption{Equivalent width of the \ha\ or \hb\ emission versus rotation velocity difference between the molecular and ionized gas, $\Dv$. The blue points are the KSS galaxies and use EW(\ha), and the black points are ETGs from \citet{davis13} who use the luminosity weighted mean EW(\hb), excluding those which are counter-rotating. Error bars are not provided for the EW(\hb) data. Our points are consistent with the low $\Dv$ end of the ETGs. EWs > 14\,\AA\ (vertical dashed line) trace star formation \citep{lacerda18}, so it is not surprising that the star-forming disk galaxies used in this study have larger EWs than the ETGs use by \citet{davis13}.}
\end{figure}

The low end of the $\Dv$ values measured by \citet{davis13} are comparable to those we measure (Figure \ref{fig:DvEW}). \citet{davis13} also measure the luminosity weighted mean equivalent width (EW) of \hb\ (a measure of the dominance of star formation). CALIFA provides maps of the EW(\ha), and we find the median EW(\ha) in the same region as where $\Dv$ is calculated (excluding the inner 12" out to the furtherest CO extent). The error is the standard deviation of EWs divided by the square root of the number of beams over the region. {\comment As shown in Figure \ref{fig:DvEW}, there is a trend between the EW and $\Dv$. The EW(\ha) values we measure are larger than those measured by \citet{davis13}. EWs > 14\,\AA\ trace {\comment star-forming complexes}, and galaxies where the ionization is dominated by \HII\ regions in the midplane tend to have larger EWs \citep{lacerda18}. This implies that the bulk of the ionized gas emission in our objects comes from the \HII\ regions in midplane, which naturally rotate at the same velocity as the molecular gas (since they represent recent episodes of star formation). \citet{lacerda18} also find that EWs < 3\,\AA\ trace regions of diffuse gas ionized by low-mass, evolved stars. These are prevalent in elliptical galaxies and bulges and can also be present above or below the midplane in spirals. EWs between these values are likely produced by a mixture of ionization processes. 

This suggests a scenario where ionized gas caused by recent star formation (such as gas associated with \HII\ regions), which is close to the galaxy midplane and has a small scale height, shares the rotation of the molecular gas from which the star formation arose. Ionized gas associated with older stellar populations or produced by cosmic rays (which typically have much larger scale heights), or possibly gas that has been shock-ionized (experiencing an injection of momentum that may drive it to large scale heights) or otherwise vertically transported may rotate at lower speeds.

This scenario is in agreement with the trend seen in Figure \ref{fig:DvEW}, in which the ETGs have lower EW(\hb) and higher $\Dv$ than the star-forming spirals studied here. It also agrees with studies that find vertical gradients in the rotation velocity of the ionized gas in some galaxies \citep[e.g.,][]{rand97,rand00}. As pointed out by a number of authors, however, the steady-state solution for a homogeneous barotropic fluid immersed in an axisymmetric potential does not allow for such vertical rotation velocity gradients \citep[e.g.,][]{barnabe06,marinacci10}. Having the ionized disks in equilibrium while maintaining such gradients may require an anisotropic velocity dispersion, similar to what may be expected for a galactic fountain \citep{marinacci10}.}

\section{Discussion}
\label{sec:disc}
The ubiquity and magnitude of the differences between the CO and \ha\ rotation velocities are striking. We propose that these differences could be due to the presence of significant extraplanar diffuse ionized gas (eDIG) in our KSS galaxies. In the following subsections, we give some background on previous eDIG detections, rule out scenarios other than eDIG that could produce this effect, and give support for eDIG in these systems from velocity dispersions and ionized gas line ratios. We also suggest that this thick, pressure supported disk would have a vertical gradient in the rotation velocity, with gas at higher latitudes rotating more slowly than gas in the midplane. 

\subsection{Previous Detections of Extraplanar Diffuse Ionized Gas}
\label{ssec:eDIG}
eDIG has been observed and discussed in the literature, and we highlight some results here for context. The importance of the warm ionized medium (WIM) as a significant fraction of the ISM in the Milky Way (MW) has been known for over four decades \citep[e.g.][]{reynolds71,reynolds73,kulkarni87,cox89,mckee90} and for over two decades in other galaxies \citep[e.g.][]{dettmar90,rand90,rand96,hoopes99,rossa03a,rossa03b}. In particular, diffuse \ha\ can contribute >50\% of the total \ha\ luminosity with large variations \citep[][and references therein]{haffner09}. In the MW, half of the \HII\ is found more than 600 pc from the midplane \citep{reynolds93}. How such a large fraction of  diffuse gas can be ionized at these large scale heights is debated, but it is widely believed that leaky \HII\ regions containing O-star clusters can produce WIM-like conditions out to large distances from the cluster and the midplane {\comment\citep[by taking advantage of chimneys and lines-of-sight with little neutral gas created by past feedback,][]{reynolds01a,madsen06}, although ionization sources with large penetration depths (such as cosmic rays) or re-accretion from the halo may also play a role.} 

Extraplanar \HI\ exhibiting differential rotation has been detected in the MW \citep[e.g.][]{levine08} and in studies of individual galaxies \citep[e.g.][]{swaters97,schaap00,chaves01,fraternali02,fraternali05,zschaechner15a,zschaechner15b,vargas17}. Velocity gradients between the high latitude gas and the dynamically cold midplane are generally -10 -- -30\,\kms\,kpc$^{-1}$ extending out to a few kpc, but there are large variations among and within individual galaxies.

Extraplanar \ha\ (i.e. eDIG) has also been found and studied in galaxies, primarily from photometry. In a recent study tracing the WIM in the spiral arms of the MW, \citet{krishnarao17} find an offset between the CO and \ha\ velocity centroids, although they do not interpret this offset as eDIG. Outside the MW, NGC\,891 is the prototypical galaxy with bright eDIG extending up to 5.5\,kpc from the midplane \citep{rand90,rand97} and a vertical velocity gradient of -15\,\kms\,kpc$^{-1}$ \citep{heald06b}, in agreement with measurements of its extraplanar \HI\ \citep{swaters97,fraternali05}. \citet{boettcher16} find that the thermal and turbulent velocity dispersions (11\,\kms\ and 25\,\kms\ respectively) are insufficient to support eDIG in hydrostatic equilibrium with a 1\,kpc scale height. NGC\,5775 has observed \HI\ loops and filaments with their rotation lagging the midplane \citep{lee01} as well as \ha\ lags of -8\,\kms\,kpc$^{-1}$ detected up to 6--9\,kpc from the midplane \citep{heald06b,rand00}. NGC\,2403 has eDIG which lags the midplane by 80\,\kms\ extending a few kpc above the midplane \citep{fraternali04}, in rough agreement with the lags observed in \HI\ which extend 1--3\,kpc from the midplane \citep{fraternali02}. Finally, eDIG has been observed in the face-on galaxy M\,83 with a lag relative to the midplane of 70\,\kms\ with a vertical scale height of 1\,kpc \citep{boettcher17}. There is a range of eDIG velocity gradients and scale heights, and, moreover, \HI\ and \ha\ vertical velocity gradients are not always similar or present \citep[e.g.][]{zschaechner15a}.

Apart from these case studies, there are several large photometric studies of eDIG independent of HI. Following the work of \citet{rand96}, \citet{miller03I} and \citet{rossa03a,rossa03b} performed larger photometric surveys of nearby edge-on spiral galaxies. \citet{rossa03a,rossa03b} had a sample of 74 edge-on disks and found that 40.5\% of the sample had eDIG extending 1--2\,kpc from the midplane. In their sample of 17 galaxies, \citet{miller03I} observe eDIG in all but one galaxy. \citet{miller03II} did a spectroscopic follow up study of nine edge-on galaxies with observed eDIG and found vertical velocity gradients ranging from $-30$ to $-70$\,\kms\,kpc$^{-1}$. 

\subsection{Comparison with Stellar Dynamical Modeling}
\label{ssec:stellarcomp}
The stellar circular velocity curve, which accounts for stellar velocity dispersion, should agree with the CO rotation curve if CO is a dynamically cold tracer. \citet{leung18} test three different dynamical models of the galaxy's potential to determine stellar circular velocity curves and compare these to CO rotation curves of 54 EDGE galaxies. Overall, they find agreement between the CO rotation curves and the three models to within 10\% at 1 $R_e$. We defer to \citet{leung18} for a complete discussion of the details of the stellar modeling. The agreement between the stellar dynamical modeling and the CO rotation curves verifies that CO is indeed a dynamically cold tracer, indicating that the \ha\ is exhibiting anomalous behavior rather than the CO. {\comment Moreover, our measured CO velocity dispersions (Table \ref{tab:KSSparams}) are small (\~10\,\kms), further indicating that the CO is dynamically cold.}

\subsection{An Inclined Disk}
\label{ssec:incdisk}
\begin{figure}
\label{fig:IncTrend}
\centering
\includegraphics[width=\columnwidth]{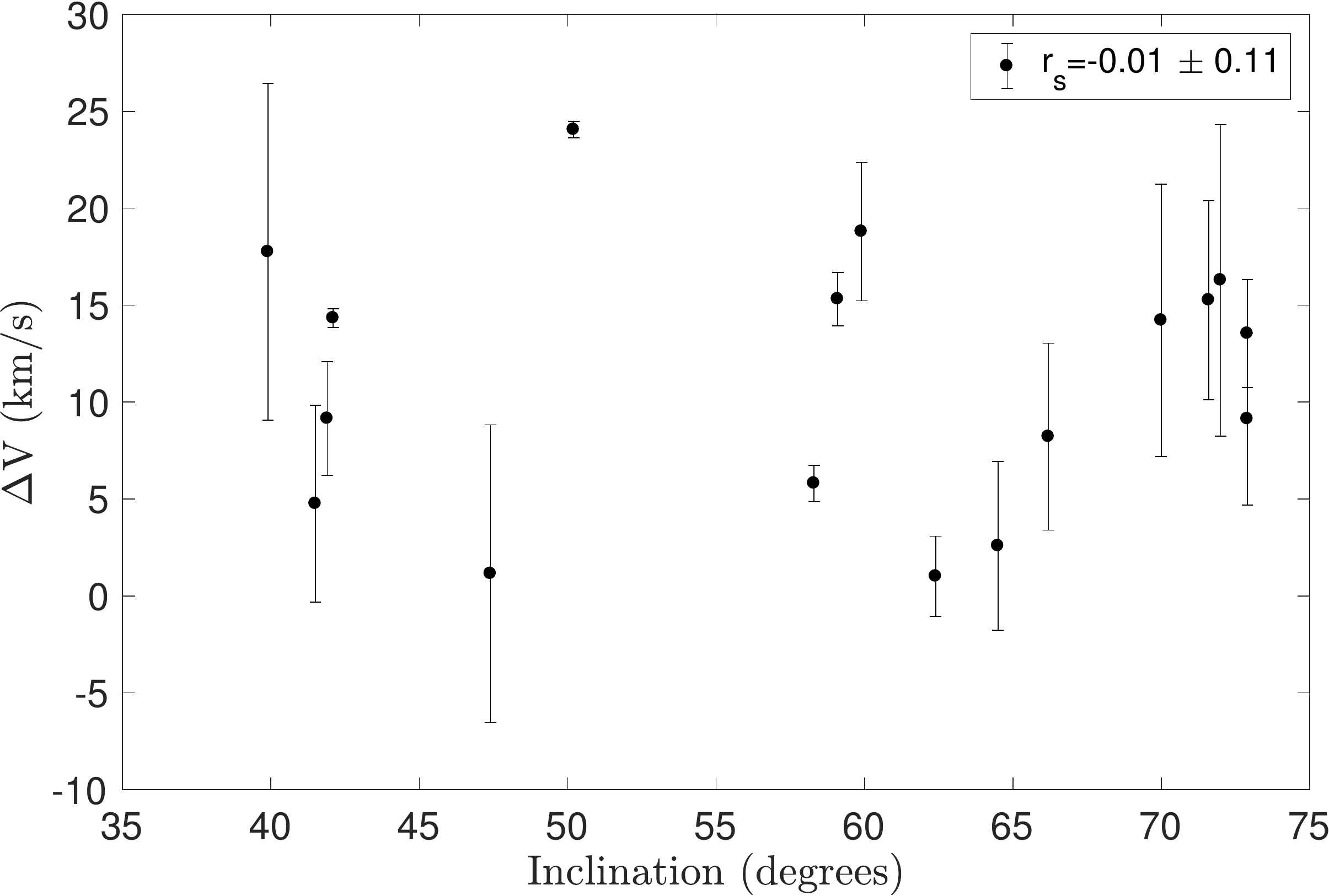} 
\caption{Lack of a trend between galaxy inclination and $\Dv$. The cause of the observed rotation velocity difference $\Dv$ cannot be purely an inclination effect. The Spearman rank correlation coefficient ($r_s$) is consistent with 0.}
\end{figure}

It is possible that the observed difference between the CO and ionized gas rotation velocities could be produced by the inclination of the disk{\comment; however,} we find no correlation between $\Dv$ and inclination, as shown in Figure \ref{fig:IncTrend}. To determine a correlation, we calculate the Spearman rank correlation coefficient ($r_s$), which quantifies how well the relationship between the variables can be described by any monotonic function. Variables which are perfectly monotonically correlated will have $r_s = \pm 1$, assuming that there are no repeated values of either variable. The Spearman rank correlation coefficient does not, however, take the errors on the data into consideration. The errors on $\Dv$ are especially important here. Therefore, we use a Monte Carlo method to determine the correlation coefficient over 1000 samples drawn from a uniform random distribution within the error ranges on each point. The error reported on $r_s$ is the standard deviation of all 1000 $r_s$ values. As shown in Figure \ref{fig:IncTrend}, the difference between the CO and \ha\ rotation velocities is not a result of the inclination of the galaxy ($r_s=-0.01\pm0.11$).

{\comment We note that if the molecular and ionized gas disks had different inclinations, our assumption that they are the same could produce a $\Dv$. However, to produce only $\Dv > 0$ would require that all ionized gas disks are less highly inclined (with respect to us) than the molecular gas disks, which for a sample of 17 galaxies is extremely unlikely. We can, therefore, rule out that the inclination affects the results in this way.}

$\Dv$ was also plotted against other global parameters of the galaxies, such as stellar mass (M$_*$), SFR, specific SFR (sSFR $\equiv$ SFR/M$_*$), morphology, physical resolution, and CO V$_{\rm{max}}$. There are no trends with any of these global parameters, and they are shown for completeness in Figure \ref{fig:Dvtrends} in Appendix \ref{app:galbygal}. As mentioned in Section \ref{ssec:convolution}, the lack of trend with physical resolution (Figure \ref{fig:Dvtrends}e) justifies our choice to convolve to a common angular resolution rather than to a common physical resolution. These values and their sources are listed in Table \ref{tab:KSSparams}. The lack of trends with these parameters agrees with \citet{rossa03b}, who also found no trends in the presence of eDIG with such global parameters.

\subsection{Star Formation Rate Surface Density Threshold}
\label{ssec:SigmaSFR}
\begin{figure}
\label{fig:SFRarea}
\centering
\includegraphics[width=\columnwidth]{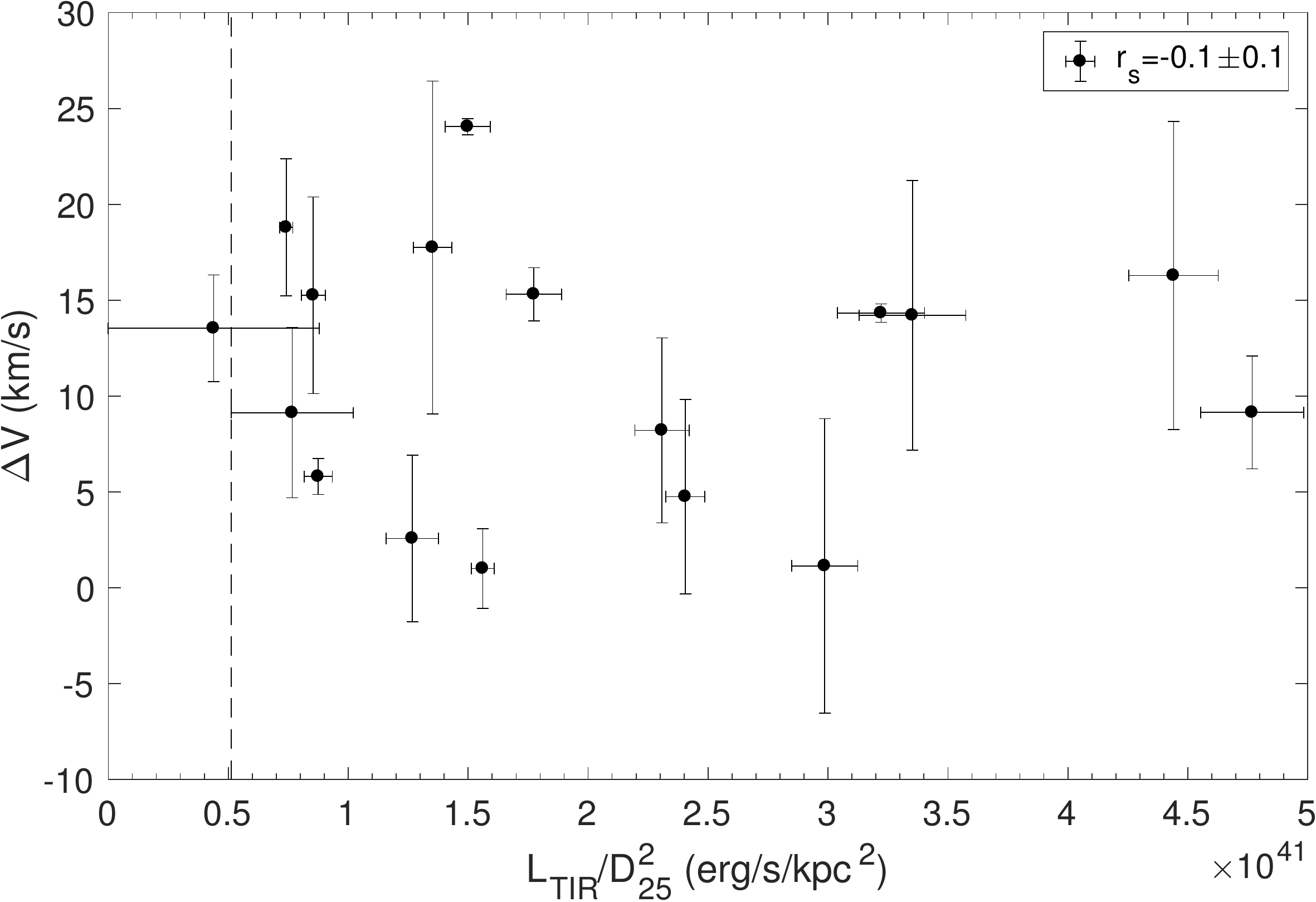}
\caption{$L_{\rm TIR}/D_{25}^2$ versus $\Dv$ for the KSS. The vertical dashed line shows the eDIG threshold from \citet{rossa03a} of  $L_{\rm TIR}/D_{25}^2 = 5.1\times 10^{40} {\rm\ erg\ s}^{-1}$ (Equation \ref{eq:thresTIR}). We find that > 90\% of our sample exceed this threshold, indicating the likely presence of eDIG in these systems. There is not, however, a trend between $L_{\rm TIR}/D_{25}^2$ and $\Dv$.}
\end{figure}

In previous studies of eDIG, \citet{rand96} and \citep{rossa03a} find a possible trend in the amount of eDIG with the SFR per unit area ($\Sigma_{\rm SFR}$) as traced by the far infrared (FIR) luminosity ($L_{\rm FIR}$). The physical picture is that a minimum level of widespread star formation is needed to sustain a thick disk that covers the entire plane of the galaxy. \citet{rossa03a} determine a threshold $\Sigma_{\rm SFR}$ above which they claim that an eDIG will be ubiquitous. This does not guarantee, however, that galaxies above this threshold will always have an eDIG or that galaxies below it cannot have eDIG. \citet{rossa03a} define this threshold $\Sigma_{\rm SFR}$ as
\begin{equation}
\label{eq:thres}
\frac{L_{\rm FIR}}{D_{25}^2} = (3.2\pm0.5)\times 10^{40} {\rm\ erg\ s}^{-1}{\rm\,kpc}^{-2}
\end{equation}
where $D_{25}$ is diameter of the 25th magnitude isophote. \citet{catalantorrecilla15} measure total IR (TIR, 8--1000 $\mu$m) luminosities ($L_{\rm TIR}$) for 272 CALIFA galaxies, and $L_{\rm TIR}$ measurements are available for 15/17 KSS galaxies. The threshold defined by \citet{rossa03a} can be converted to TIR by multiplying by 1.6 \citep{sanders96} so 
\begin{equation}
\label{eq:thresTIR}
\frac{L_{\rm TIR}}{D_{25}^2} = (5.1\pm0.8)\times 10^{40} {\rm\ erg\ s}^{-1}{\rm\,kpc}^{-2}
\end{equation}
For the two galaxies without $L_{\rm TIR}$ measurements, we can estimate $L_{\rm TIR}$ from the SFR measured by CALIFA from extinction corrected \ha\ where
\begin{equation}
L_{\rm TIR} = \frac{1.6}{4.5\times10^{-44}}\Big[\frac{\rm SFR}{M_\odot\ {\rm yr}^{-1}}\Big] {\rm erg\ s^{-1}}
\end{equation}
\citep{kennicutt98b} and the factor of 1.6 comes from converting from $L_{\rm FIR}$ to $L_{\rm TIR}$ \citep{sanders96}. Using measurements of $D_{25}$ from HyperLeda (values are listed in Table \ref{tab:KSSparams}), we compare the values of $L_{\rm TIR}/D_{25}^2$ for all KSS galaxies to the eDIG threshold (Equation \ref{eq:thresTIR}) in Figure \ref{fig:SFRarea}. We find that $94^{+6}_{-0}\%$ of galaxies in the KSS have $L_{\rm TIR}/D_{25}^2$ greater than this threshold and should have eDIG based on this criterion. 

If many galaxies have a thick ionized disk, this could underestimate the dynamical mass of the galaxy derived from the ionized gas rotation velocity. Because we find that the ionized gas rotates more slowly than the molecular gas in galaxies with large $\Sigma_{\rm SFR}$, this effect could be significant in local star-forming galaxies and even more so at higher redshifts where there is more star formation occurring on average.

\subsection{Ionized Gas Velocity Dispersion}
{\comment  The trends between the eDIG and the star-formation rate surface density discussed in Section \ref{ssec:SigmaSFR} \citep[e.g.][]{rand96,rossa03a} are suggestive of star-formation feedback playing an important role in forming the eDIG. In order for ionized gas to remain above or below the disk midplane in a long-lived configuration, it must have sufficient velocity dispersion (or at least a vertical bulk motion). This effectively acts as as additional pressure term, allowing the gas to remain at larger scale heights. Therefore, we expect that galaxies with larger ionized gas velocity dispersions should have larger eDIG scale heights. Measuring the ionized gas velocity dispersion is, therefore, an important way to test these ideas.}

\subsubsection{\hg\ Velocity Dispersion Measurements}
\label{ssec:veldisp}

\begin{figure*}
\label{fig:DvVelDisp}
\centering
\gridline{\fig{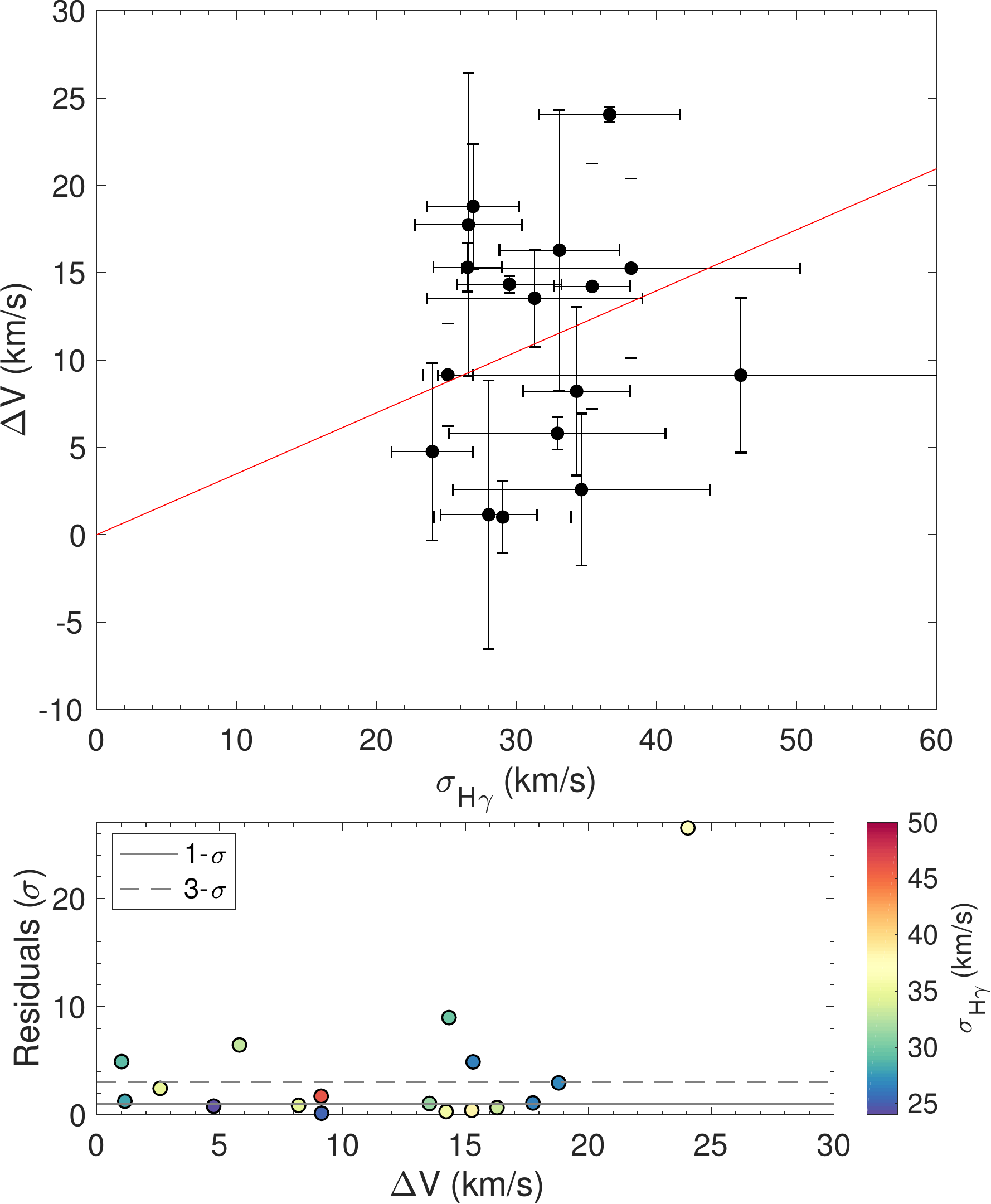}{0.9\columnwidth}{(a)}
	\fig{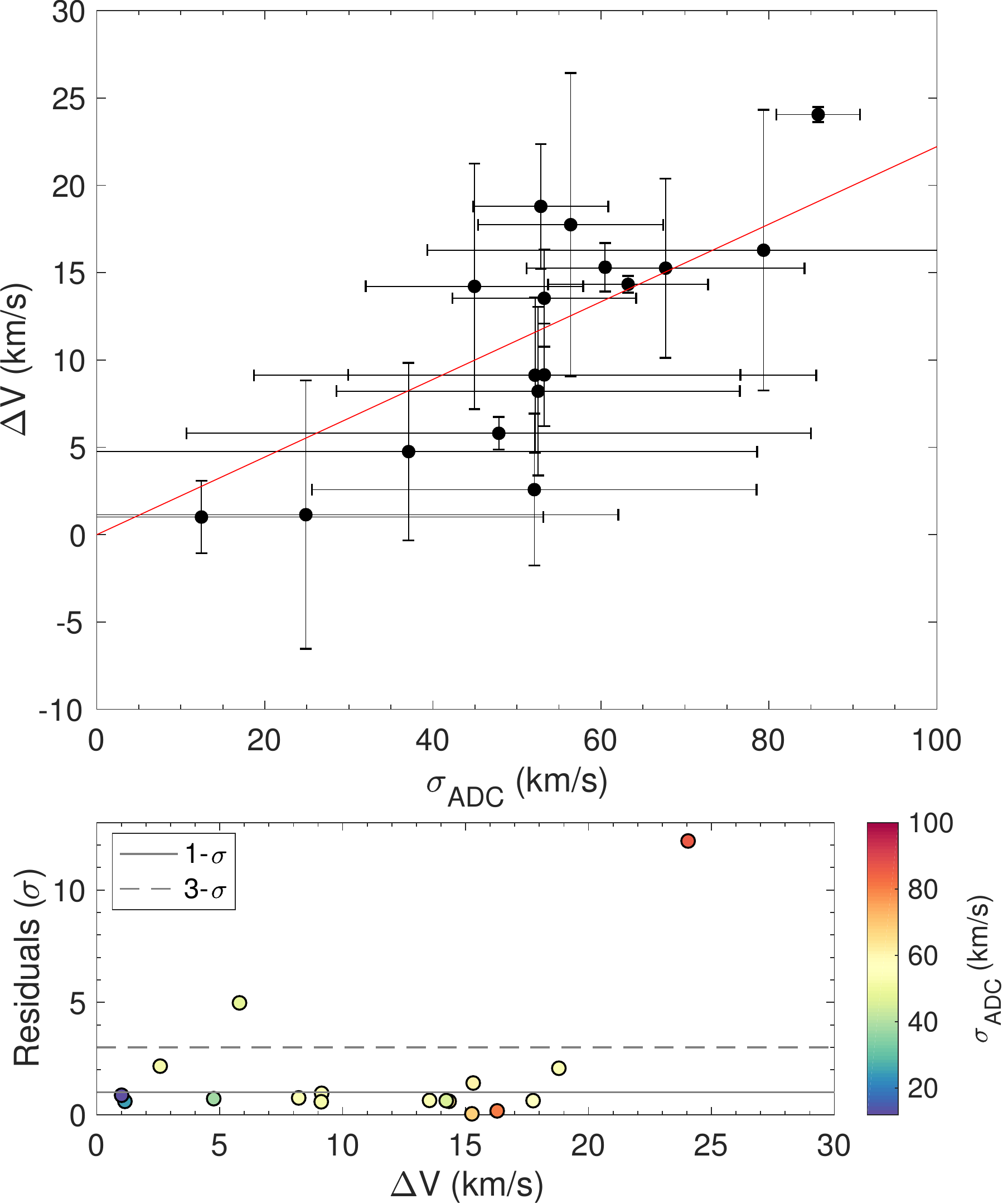}{0.9\columnwidth}{(b)}}
\caption{(a) The top panel shows the trend between the velocity dispersion measured from the \hg\ linewidth (\sigmaHg) and $\Dv$. Error bars reflect the propagated error.  The red line is a linear fit to the data points. The lower panel shows the perpendicular distance of each point from the line in units of standard deviations plotted against $\Dv$. Points are color-coded by \sigmaADC. The solid and gray dashed lines show 1- and 3-$\sigma$. {\comment 35\%} of the galaxies are within 1-$\sigma$ of the best-fit line and {\comment 71\%} are within 3-$\sigma$. If there is an underlying correlation between \sigmaHg\ and $\Dv$, it is weak. (b) The same as (a) but using the velocity dispersions inferred from the ADC (\sigmaADC). The trend results from the form of the ADC used here (Equation \ref{eq:ADC}). {\comment 71\%} of the galaxies are consistent with the best-fit line within 1-$\sigma$ and 88\% are consistent within 3-$\sigma$. In both bottom panels, the galaxy many $\sigma$ away from the best-fit line is NGC\,2347.}
\end{figure*}

{\comment CALIFA data cannot, unfortunately, be used to accurately measure the \ha\ linewidth due to the low spectral resolution of the V500 grating employed (6.0\,\AA\ FWHM\,$\approx275$\,\kms\ at \ha). CALIFA observes \hg\ with the moderate-resolution V1200 grating (2.3\,\AA\ FWHM\,$\approx160$\,\kms\ at \hg), however, and those data can be used in principle to establish ionized gas velocity dispersions. Starting with the continuum-subtracted cubes, we fit the \hg\ line with a Gaussian at each spaxel where the width of the Gaussian corresponds to the velocity dispersion, as discussed in Section \ref{ssec:califasurvey}. The measured linewidth is the convolution of the instrumental response with the actual gas velocity dispersion, and has a small contribution from rotation smearing caused by the finite angular resolution.
We apply a beam smearing correction which also accounts for the instrumental velocity dispersion. This method is described in detail in Appendix \ref{app:BS}. The accuracy of the resulting ionized gas velocity dispersion depends critically on the exact knowledge of the spectral resolution and response of the grating, because the instrumental velocity dispersion ($\sigma_{\rm inst}\approx 68$\,\kms) is of the same order as the observed \sigmaHg\ before removal: in other words, the spectral resolution of the V1200 observation is marginal for the purposes of measuring the velocity dispersion in these galaxies, and our results should be considered tentative.
Inspection of the maps suggests that it is likely that the beam smearing-corrected \sigmaHg\ values reported here are lower limits to the real ionized gas velocity dispersion, and we caution against over-interpretation of these values. We do not correct our \sigmaHg\ for inclination and hence assume that the velocity dispersion is isotropic. As discussed in Section \ref{sec:results}, anisotropic velocity dispersions may be required to maintain ionized gas disks with vertical gradients in the rotation velocity \citep{marinacci10}. With these caveats in mind, we calculate the average beam smearing-corrected \hg\ velocity dispersion over the same region where $\Dv$ is calculated. Velocity dispersions range from \~25--45\,\kms\ (Figure \ref{fig:DvVelDisp}a). The scale height of the disk ($h$) corresponding to a given isotropic velocity dispersion ($\sigma$) is
\begin{equation}
\label{eq:scaleheight}
h = \frac{\sigma^2}{\pi G\Sigma_*(r)}
\end{equation}
\citep{vanderkruit88,burkert10}. We use azimuthally averaged radial profiles of the stellar surface density ($\Sigma_*$) from \citet{utomo17} to find $\Sigma_*$ over the same range of radii where $\Dv$ is calculated (listed in Table \ref{tab:KSSparams}). The scale heights corresponding to the observed \sigmaHg\ are \~0.1--1.3\,kpc. Previous measurements of eDIG scale heights range from 1--2\,kpc \citep{rossa03a,miller03II,fraternali04} up to a few kpc above the disk \citep{rand00,rand97,miller03I}. Because our \sigmaHg\ are likely lower limits, the scale heights may indeed be larger than we report here.}

\subsubsection{\comment Velocity Dispersion Estimates from an Asymmetric Drift Correction}
\label{ssec:veldispest}
\begin{figure}
\label{fig:SigmaHgADC}
\centering
\includegraphics[width=\columnwidth]{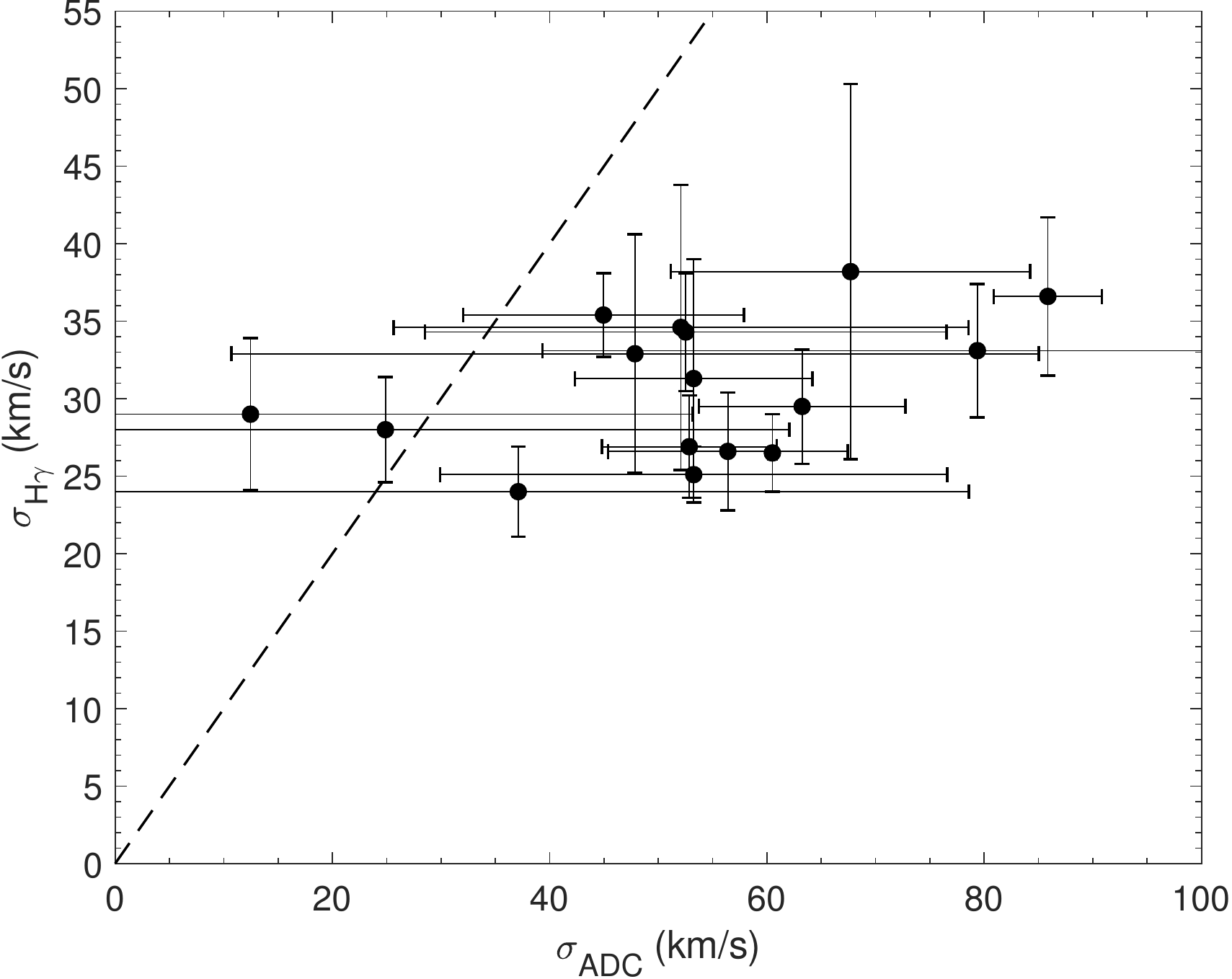}
\caption{{\comment The velocity dispersions inferred from the ADC compared to those measured for the \hg\ emission in each galaxy. The error bars reflect the statistical errors, measured from the standard deviation of the measurements in the annulus. The dashed line is one-to-one. The velocity dispersion \sigmaHg\ is roughly comparable to those inferred from the \sigmaADC; although \sigmaHg\ tends to be smaller than \sigmaADC this should not be over-interpreted given the difficulty of the measurement discussed in Section \ref{ssec:veldisp}.}}
\end{figure}

It is possible to infer the velocity dispersion needed to produce the observed $\Dv$ using an asymmetric drift correction (ADC). Generally, an ADC is used to find $V_{\rm circ}$ given $V_{\rm rot}$, $\sigma$, and $\Sigma_*(r)$; however, since $\vrot({\rm CO})$ traces $V_{\rm circ}$ (Section \ref{ssec:stellarcomp}), we can invert the ADC to find $\sigma$ instead, with $\vrot=\vrot$(\ha). If we assume that {\comment the velocity dispersion is isotropic ($\sigma_r=\sigma_z=\sigma_\phi$)}, $\sigma(r)=$ constant, and $\Sigma_*(r)=2\rho(r,z)h(z)$ \citep{binney08}, then
\begin{equation}
\label{eq:ADC}
\sigma^2=\frac{V_{\rm circ}^2-V_{\rm rot}^2}{-d\ln\Sigma_*/d\ln r}.
\end{equation}
We use azimuthally averaged radial profiles for $\Sigma_*(r)$ from \citet{utomo17} to find $d\ln\Sigma_*/d\ln r$. {\comment $V_{\rm circ}$, $V_{\rm rot}$, and $\Sigma_*(r)$} are averaged over the same radii as where $\Dv$ is calculated; this excludes the central two beams (12"\~4\,kpc) where beam smearing or a bulge can affect the rotation curve. Velocity dispersions from the ADC method (\sigmaADC, Equation \ref{eq:ADC}) range from {\comment \~15--85\,\kms} in the KSS (Figure \ref{fig:DvVelDisp}b). There is an apparent trend with $\Dv$ resulting from Equation \ref{eq:ADC}. Using Equation \ref{eq:scaleheight}, we find scale heights ranging from {\comment\~0.1--2.0\,kpc, again in rough agreement with previous measurements. For individual galaxies, the velocity dispersions and scale heights predicted from the ADC tend to be larger than those measured in the \hg\ (Figure \ref{fig:SigmaHgADC}), but given the difficulty in the measurement the agreement is reasonable.}

\subsubsection{Trends between $\Dv$ and the Velocity Dispersion}
\label{ssec:dvveldisp}
To explain the difference in rotation velocities observed between the molecular and ionized gas as resulting from the presence of eDIG, we would expect that galaxies with a larger $\Dv$ should have larger velocity dispersions as well. There is a trend between $\Dv$ and \sigmaADC\ (Figure \ref{fig:DvVelDisp}b) stemming directly from the form of the ADC used (Equation \ref{eq:ADC}). There is not, however, an immediately apparent relation between $\Dv$ and \sigmaHg\ (Figure \ref{fig:DvVelDisp}a). Because the errors on both $\Dv$ and \sigmaHg\ are large, however, there could be an underlying correlation. To assess whether an underlying correlation could exist, we fit a line to the data points (top panels of Figure \ref{fig:DvVelDisp}). We then calculate the perpendicular distance of each point from the line as well as the error on that distance accounting for the error bars on both quantities. From this, we determine the distance from the best-fit line in standard deviations (bottom panels of Figure \ref{fig:DvVelDisp}). For the ADC, {\comment 71\%} of the galaxies are consistent with the best-fit line within 1-$\sigma$ and 88\% are consistent within 3-$\sigma$ (Figure \ref{fig:DvVelDisp}a). This tight correlation again stems from the form of the ADC used, since \sigmaADC\ depends explicitly on $\sqrt{\vrot{\rm (CO)}^2-\vrot{\rm (\ha)}^2}$ which is $\sim\sqrt{\vrot\Dv}$ (Equation \ref{eq:ADC}). For \sigmaHg, however, only {\comment 35\% }of the galaxies are within 1-$\sigma$ of the best-fit line and {\comment 71\%} are within 3-$\sigma$ (Figure \ref{fig:DvVelDisp}a), so any underlying correlation between \sigmaHg\ and $\Dv$ is weak. Nonetheless, most of our galaxies have high enough velocity dispersions to support a thick ionized gas disk, and nearly all of our subsample have sufficient $\Sigma_{\rm SFR}$ (Figure \ref{fig:SFRarea}). 

\subsection{\SII/\ha\ and \NII/\ha\ Ratios}
\label{ssec:SII}
The velocity dispersion is not the only tracer of eDIG. The ratios of \NII$\lambda$6583/\ha\ (\NII/\ha) and \SII$\lambda$6717/\ha\ (\SII/\ha) increase with distance from the midplane and are used to probe the ionization conditions of the WIM \citep[e.g][]{miller03I,miller03II,fraternali04,haffner09}. \SII/\ha\ varies only slightly with temperature, whereas \NII/\ha\ is used to trace variations in the excitation temperature of the gas \citep{haffner09}. From observations of the MW and a few other galaxies, \SII/\ha\,$=0.11\pm0.03$ and \NII/\ha \,$\sim0.25$ in the midplane \citep{madsen04,madsen06}, whereas \SII/\ha\,$=0.34\pm0.13$ and \NII/\ha\,$\gtrsim 0.5$ in the eDIG \citep{blanc09,madsen04}. Observations of these ratios indicate that there must be additional heating sources aside from photoionization from leaky \HII\ regions to produce the WIM \citep[][and references therein]{haffner09}.

\begin{figure}
\centering
\label{fig:SIINII}
\gridline{\fig{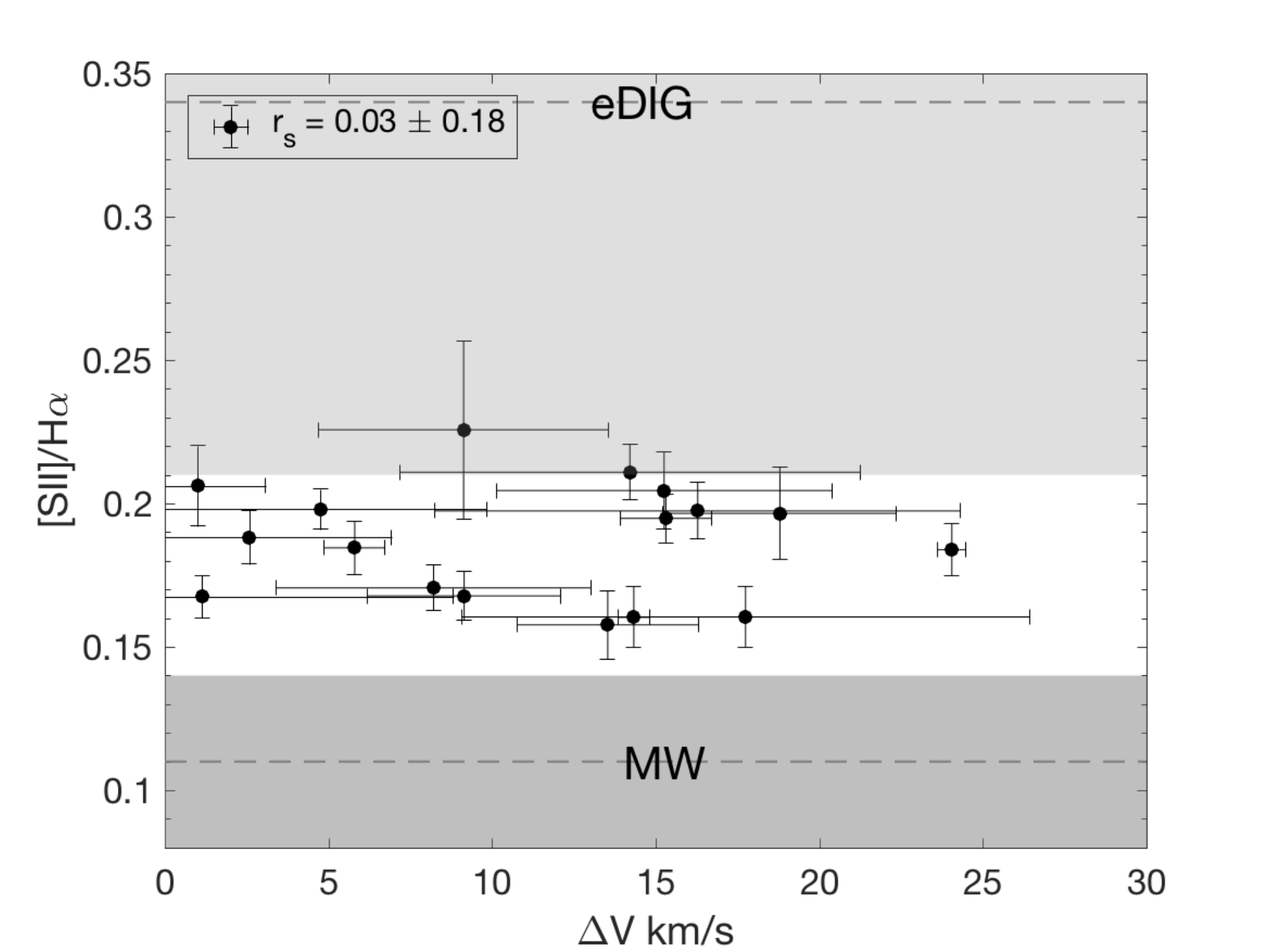}{\columnwidth}{(a)}}
\gridline{\fig{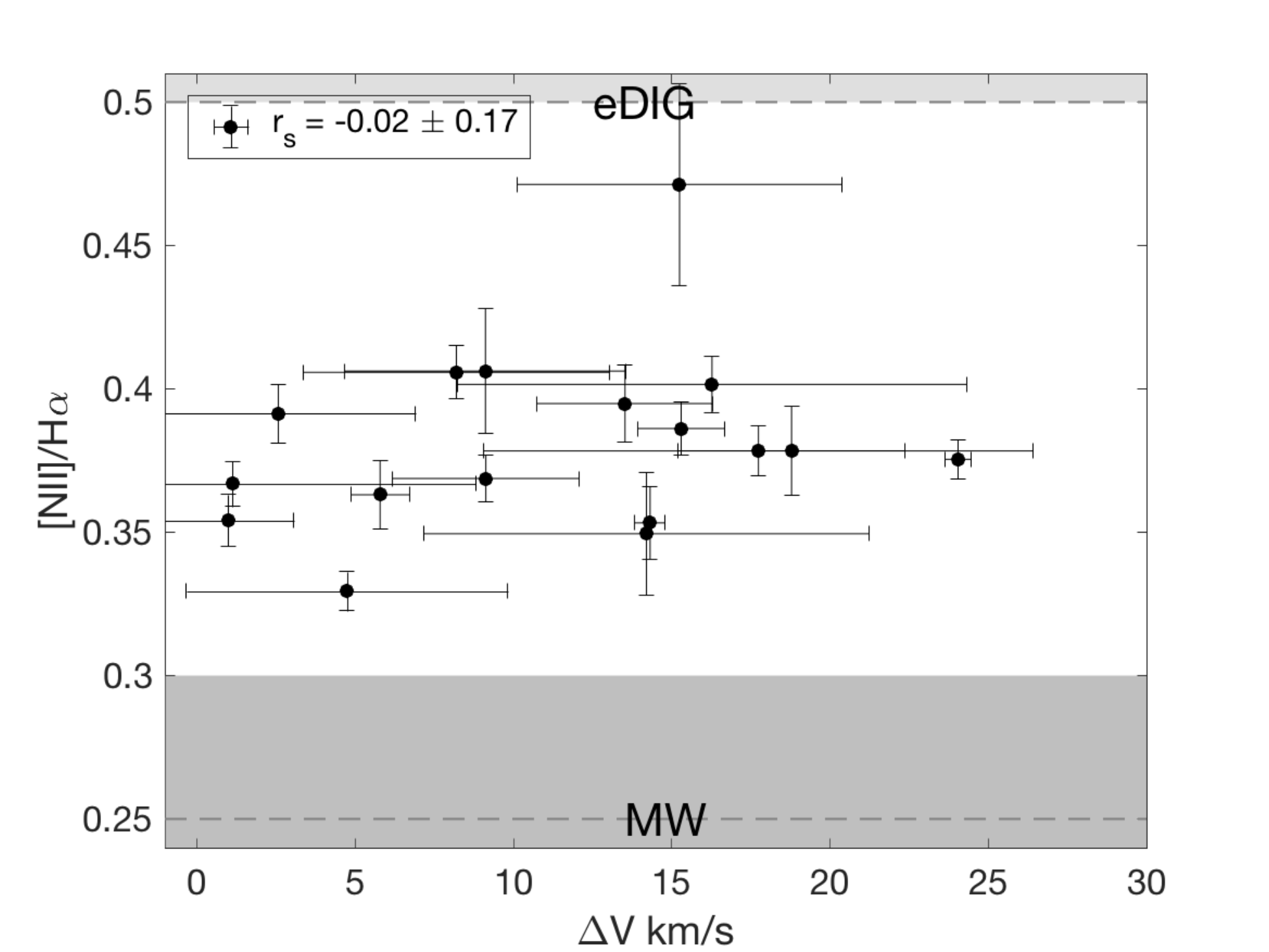}{\columnwidth}{(b)}}
\caption{(a) There is no trend between $\Dv$ and \SII/\ha, although the errors bars are large. All of our \SII/\ha\ ratios are larger than for the MW (dark gray shaded region, \SII/\ha\,$=0.11\pm0.03$) but only a few fall within the observed range for the eDIG (light gray shaded region, \SII/\ha\,$=0.34\pm0.13$). The dashed gray lines show the median values for the MW and eDIG. (b) There is no trend between \NII/\ha\ and $\Dv$. As with \SII/\ha, all of our measured ratios exceed those found in the plane of the MW (dark gray shaded region, \NII/\ha\,$\sim0.25$ where we have adopted a 20\% error range) but are not as high as is generally observed in the eDIG (light gray shaded region, \NII/\ha\,$\gtrsim0.5$). The dashed gray lines show the median values for the MW and eDIG.}
\end{figure}

CALIFA provides \ha, \SII, and \NII\ intensity maps for all galaxies. These were masked to cover the same radii as where $\Dv$ is calculated. As shown in Figure \ref{fig:SIINII}, there are no trends between \SII/\ha\ or \NII/\ha\ with $\Dv$ ($r_s=0.03\pm0.18$ and $r_s=-0.02\pm0.17$ respectively). For both \SII/\ha\ and \NII/\ha, our ratios are all larger than for the plane of the MW but only a few fall within the observed range for the eDIG. This is perhaps not unexpected, since emission from the plane is mixed with emission from the eDIG which would systematically lower the observed ratios. This is an encouraging hint that the observed $\Dv$ could be due to eDIG in a thick disk. 

\subsection{Kinematic Simulations}
\label{ssec:sims}
To further investigate how the disk's geometry affects the observed rotation curve, we perform a suite of kinematic simulations using NEMO. Disks are given different scale heights and vertical rotation velocity distributions as described in the follow subsections. The particle velocities are given by an input rotation curve which rises linearly from $r=0-1$ units and is constant at $V_0=200$\,\kms\ from $r=1-6$ units. The disk is then inclined and ``observed" with a 1 unit beam. The velocity is derived by fitting the peak of the line at each point in the simulated data cube. $\vrot(r)$ is averaged between $r=1$ and $r=5$ to give $\bar{V}$ in the flat part of the rotation curve. The error on $\bar{V}$ ($\sigma_{\bar{V}}$) is the standard deviation of $\vrot(r)$. A simulated $\Dv$ is computed by $V_0-\bar{V}$, which is analogous to the $\Dv$ defined previously ($V_0$ corresponds to $\vrot$(CO) and $\bar{V}$ corresponds to the median $\vrot(\ha)$ in the outer part of the galaxy). We test four disk configurations, which are described below. To convert the scale heights to physical units (i.e. kpc), we use the turn-over radius ($R_0$) of the rotation curve (1 unit in the simulations) and find the average $R_0$ of the KSS \ha\ data. We fit the KSS \ha\ rotation curves with $V_{\rm model}=V_0(1-e^{-R/R_0})$ \citep[e.g.][]{boissier03,leroy08} and fix $V_0$ to the value determined for that galaxy. The average $R_0$ over the sample is found, and the scaling is 1 unit = $1.77\pm0.25$\,kpc. We note that the assumed beam in the simulations is 1 unit, which is nearly identical to the beam size of the observations (6" = 1.73\,kpc at the average distance of the KSS galaxies).

\subsubsection{Thin Disks}
\label{ssec:thindisk}
\begin{figure*}
\label{fig:SimTrends}
\centering
\gridline{\fig{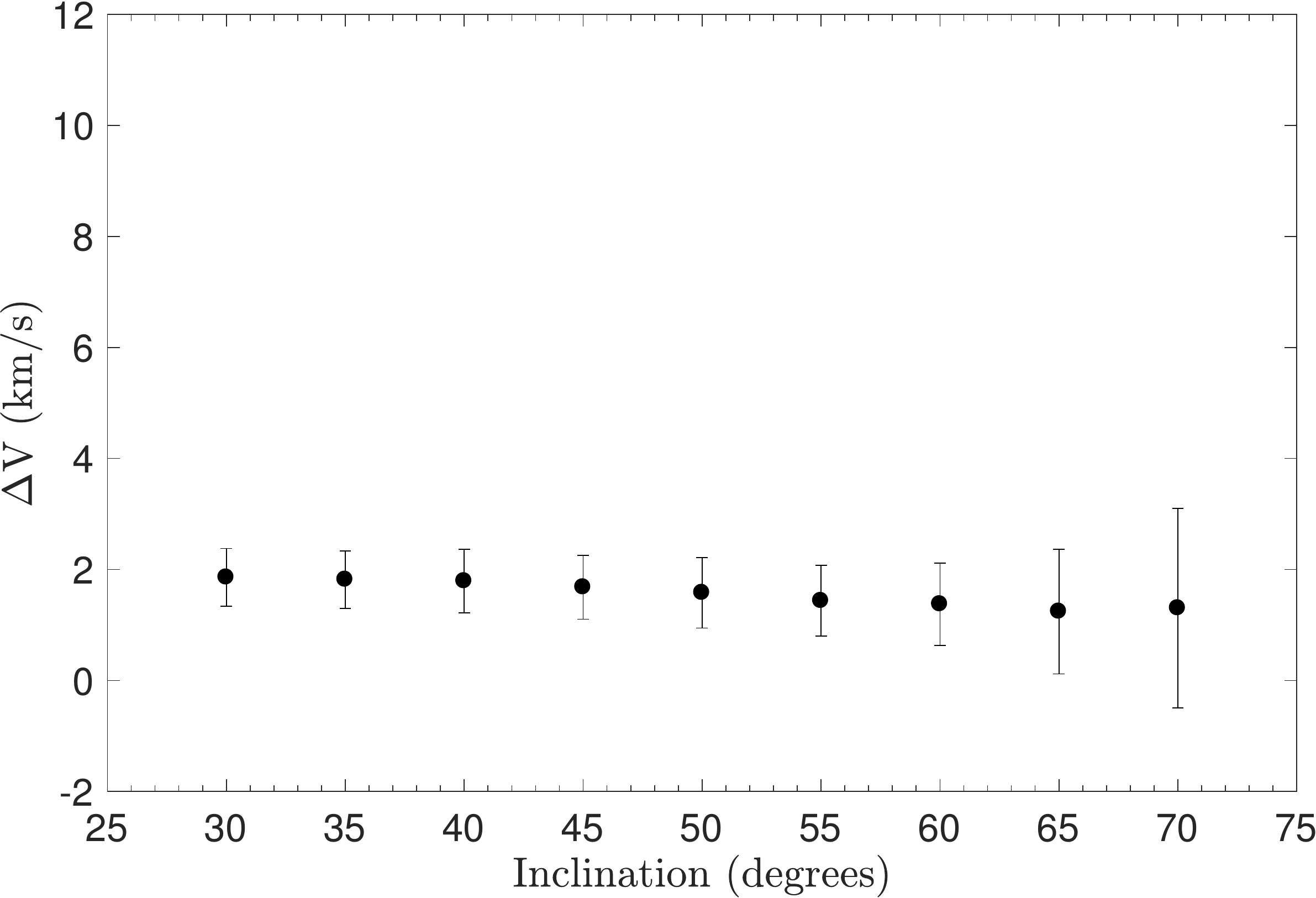}{0.85\columnwidth}{(a)}
		\fig{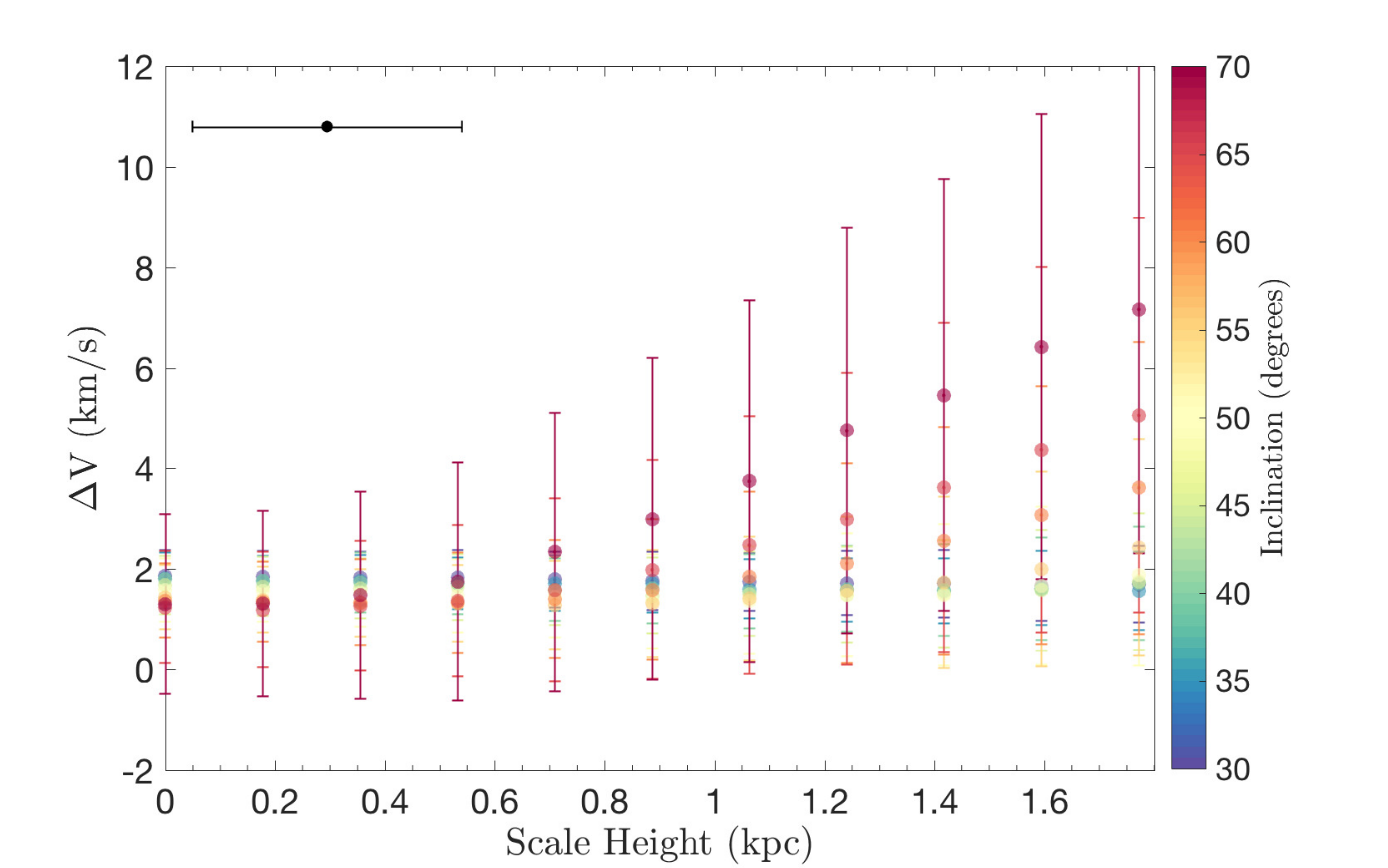}{\columnwidth}{(b)}}
\gridline{\fig{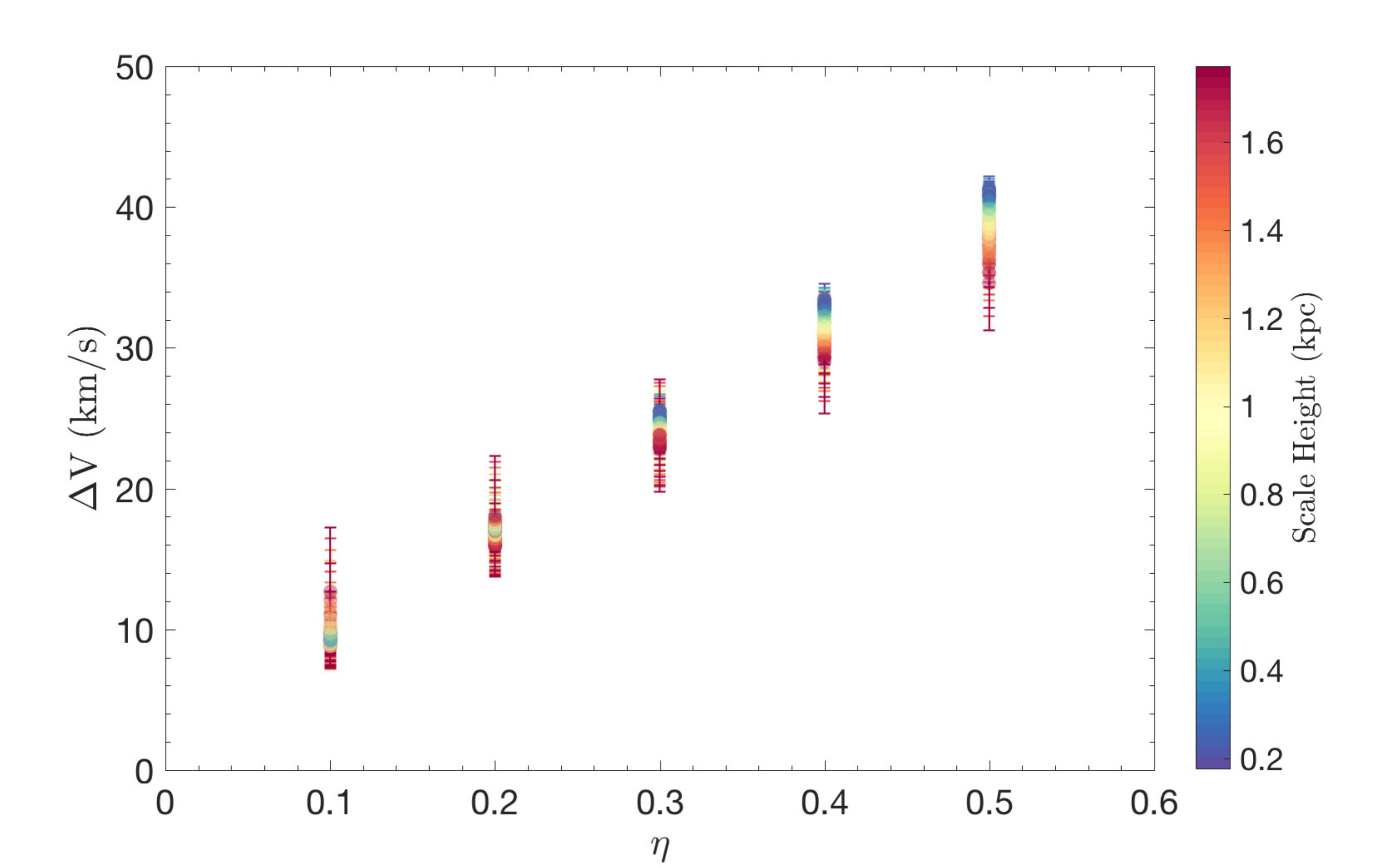}{\columnwidth}{(c)}
		\fig{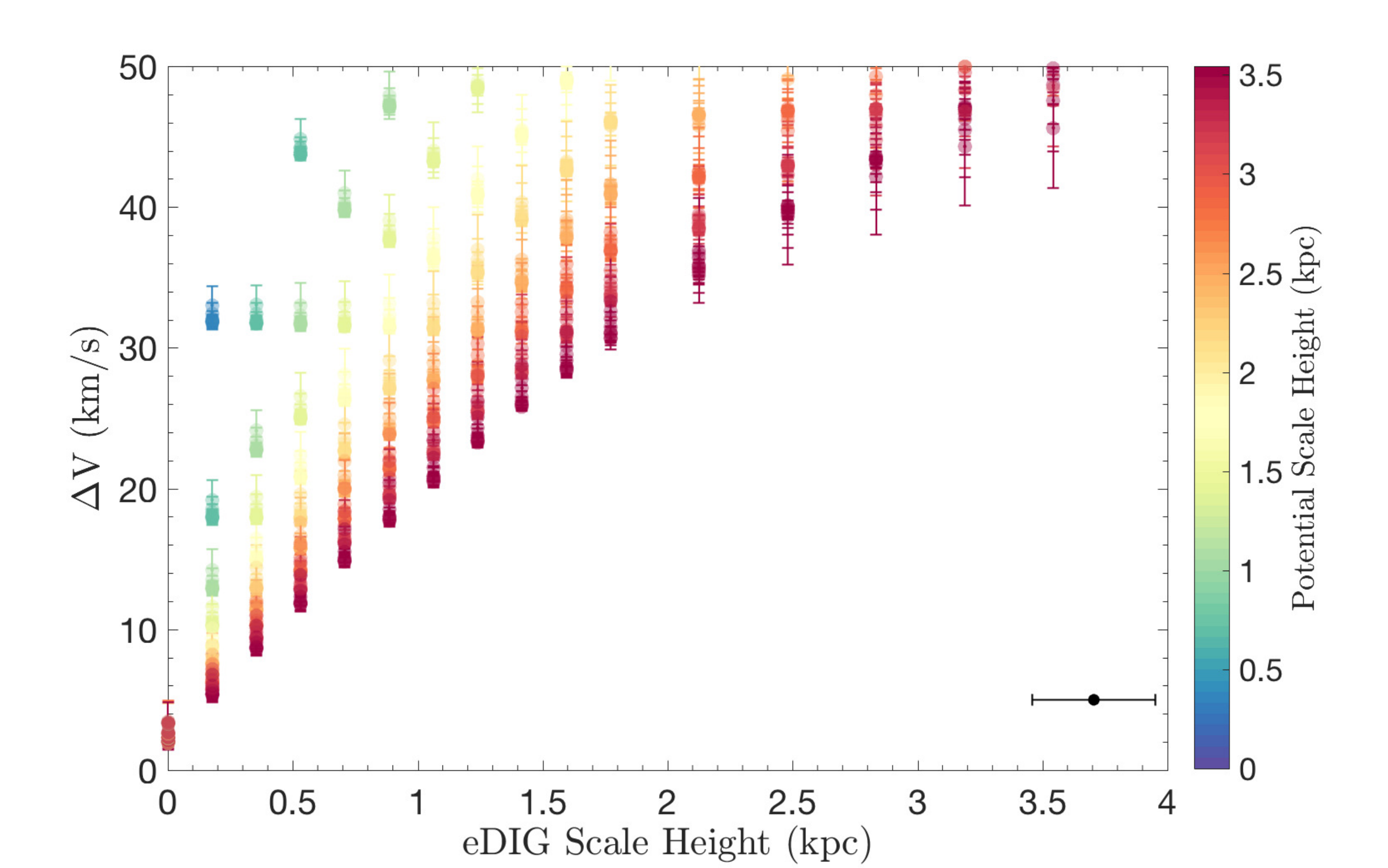}{\columnwidth}{(d)}}
\caption{(a) From the simulations of thin inclined disks, the $\Dv$ values are all very small, so inclination cannot explain the observed $\Dv$. The \~2\,\kms\ offset away from $\Dv=0$ is due entirely to beam smearing. (b) From the simulations of thick disks with no vertical rotation velocity gradient (i.e. $\vrot(z)=V_0$), there is no trend between $\Dv$ and the scale height. There is only a trend for the most highly inclined galaxies (and the errors are large). So a thick disk without a vertical rotation velocity gradient cannot explain the $\Dv$ measured in all systems. Again, the constant offset away from $\Dv=0$ is due entirely to beam smearing. The black point in the upper left corner shows the error on the scale height from the scaling between units in the simulation to kpc. Points are color-coded based on their inclination. (c) For a thick disk with a linear vertical rotation velocity gradient, there is a strong trend between the magnitude of the gradient (parameterized by $\eta$, see Equation \ref{eq:linearVrotz}) and $\Dv$, producing $\Dv$ values comparable to those observed for $\eta\lesssim0.3$. Points are color-coded based on the scale height. (d) Using the more realistic model with $\vrot(r,z)$ given by Equation \ref{eq:Vrotz} and $\rho_{\rm eDIG}(z)$ given by Equation \ref{eq:rhoeDIG}, we find that there is a strong trend between the eDIG scale height ($h_{\rm eDIG}$) and $\Dv$. It is possible to produce $\Dv$ in the range we observe (5--25\,\kms) with $h_{\rm eDIG} \lesssim 1.5$\,kpc. There is also a trend with the scale height of the underlying potential ($h_p$, shown in the color-coding), but this trend is weaker. The black point in the lower right corner shows the error on the scale height from the scaling between units in the simulation to kpc.}
\end{figure*}

First, we create a thin disk of particles with scale height $h=0$. The simulated $\Dv$ as a function of inclination is shown in Figure \ref{fig:SimTrends}a. The resulting $\Dv$ values are all very small and are insufficient to explain the offsets seen in Figure \ref{fig:gausshistdiffs}. We recover the input rotation curve to within \~2\,\kms. This \~2\,\kms\ offset is due to beam smearing. If a smaller beam is used, this offset disappears. So neither the inclination nor beam smearing of a thin disk can produce the observed $\Dv$.

\subsubsection{Thick Disks}
\label{ssec:thickdisk}
Using the same simulation set up described above, the initial disk can be given a scale height. The same input rotation curve is used at all heights, $z$, such that $\vrot(r,z)=\vrot(r,z=0)=V_0$. Particles are distributed vertically using a Gaussian distribution where the FWHM is $2\sqrt{2\ln2}h$ (so the scale height above the midplane is $h$). $\Dv$ is then calculated from these simulations as described previously. Figure \ref{fig:SimTrends}b shows $\Dv$ as a function of $h$, color-coded by inclination. There is only a trend for the highest inclinations, and even at these high inclinations, $\Dv$ is not as large as many galaxies in the KSS. Therefore, a thick disk with $\vrot(z)=V_0$ cannot cause the observed $\Dv$, except perhaps for very highly inclined galaxies. For intermediate inclinations, we recover the input rotation curve again to within the \~2\,\kms\ from beam smearing.

\subsubsection{Thick Disks with Vertical Rotation Velocity Gradients}
\label{ssec:simgrad}
We also test a thick disk with a vertical gradient in the rotation velocity. As mentioned in Section \ref{ssec:eDIG}, vertical gradients in the rotation velocity have been observed in the extraplanar \HI\ and eDIG of several galaxies. The physical rationale behind a vertical gradient in the rotation velocity is related to turbulent pressure support. Gas with larger velocity dispersions can be in pressure equilibrium at larger distances from the disk midplane. Material off the plane should have an orbit inclined with respect to the main disk enclosing the galactic center (like the stars); however, pressure support forces the gas to orbit parallel to the main disk. The gas further off the plane rotates more slowly than gas closer to the plane, creating the vertical gradient in the rotation velocity. First, we impose a linear vertical rotation velocity gradient parameterized by $\eta$ where 
\begin{equation}
\label{eq:linearVrotz}
\vrot(z)=V_0\Big(1-\eta\frac{z}{h}\Big).
\end{equation}
As shown in Figure \ref{fig:SimTrends}c, there is a strong trend between $\eta$ and $\Dv$. A linear vertical gradient in the rotation velocity can produce the observed values of $\Dv$ for $\eta\lesssim0.3$, meaning that at $z=h$, $\vrot\gtrsim0.7V_0$. For a given value of $\eta$, larger scale heights have less of an effect on $\Dv$, as seen from their shallower slope in Figure \ref{fig:SimTrends}c. 

Next, we test a more realistic model for $\vrot(z)$ than the linear vertical velocity gradient. The rotation velocity of material above the disk is governed by the potential. Here we assume that the rotation velocity in the disk midplane is constant ($V_0$) and, hence, the radial surface density profile, $\Sigma(r)$, is described by a Mestel disk \citep{mestel63,binney08}. We also assume that the vertical density distribution is exponential. Therefore, the total density distribution of material that dominates the potential has the form
\begin{equation}
\rho(r,z)=\frac{V_0^2}{2\pi Grh_p}e^{-|z|/h_p}
\end{equation}
where $h_p$ is the vertical scale height of the material that dominates the potential. The potential of a thin Mestel disk is 
\begin{equation}
\label{eq:MestPot}
\phi(r)=-V_0^2\ln\Big(\frac{r}{r_{\rm max}}\Big)
\end{equation}
where $r_{\rm max}$ is the maximum extent of the disk \citep{binney08}. Therefore, the total potential is
\begin{equation}
\begin{split}
\label{eq:intpot}
\phi(r,z)=\int\phi(r,z-z')\xi(z')dz'\\
=-V_0^2\ln\Big(\frac{r}{r_{\rm max}}\Big)\frac{1}{h_p}\int e^{-|z'|/h_p}dz'\\
=V_0^2\ln\Big(\frac{r}{r_{\rm max}}\Big)e^{-|z|/h_p}+c\\
\end{split}
\end{equation}
where the constant of integration ($c$) can be found by demanding that $\phi(r,z\rightarrow\infty)\rightarrow 0$. Therefore, 
\begin{equation}
\label{eq:potential}
\phi(r,z)=V_0^2\ln\Big(\frac{r}{r_{\rm max}}\Big)e^{-|z|/h_p}.
\end{equation}
The rotation velocity from a potential is given by $V_{\rm rot}^2=r\frac{\partial}{\partial r}\phi(r,z)$ \citep{binney08} so from Equation \ref{eq:potential}
\begin{equation}
\label{eq:Vrotz}
V_{\rm rot}(r,z)=V_0\sqrt{e^{-|z|/h_p}}
\end{equation}
\noindent where the absolute value preserves the symmetry above and below the disk. The eDIG has its own density distribution and scale height ($h_{\rm eDIG}$) which are largely independent of the potential and determined mostly by the star formation activity. Although large ratios of $h_{\rm eDIG}/h_p$ are physically unlikely, here we treat these two scale heights as independent quantities. We can find the eDIG density as a function of $z$ where the vertical density distribution is described by the hydrostatic Spitzer solution \citep{spitzer42,binney08,burkert10}:
\begin{equation}
\label{eq:rhoeDIG}
\rho_{\rm eDIG}(r,z)=\rho_0{\rm sech}^2\Big(\frac{z}{h_{\rm eDIG}}\Big)
\end{equation}
where $\rho_0$ is the density in the midplane. We repeat the NEMO simulations as before, but using $\vrot(r,z)$ given by Equation \ref{eq:Vrotz} (rather than Equation \ref{eq:linearVrotz}) and an eDIG density distribution give by Equation \ref{eq:rhoeDIG} (rather than a Gaussian). The results of this more realistic model are shown in Figure \ref{fig:SimTrends}d. There is a strong trend between $h_{\rm eDIG}$ and $\Dv$. There is also a secondary trend between $h_{p}$ and $\Dv$. With this model, it is possible to reproduce the observed range of $\Dv$ with $h_{\rm eDIG} \lesssim 1.5$\,kpc. {\comment Using Equation \ref{eq:scaleheight} and the median $\Sigma_*$ of the KSS at the radius where the measurements are done ($\approx200$\,M$_\odot$\,pc$^{-2}$), this implies a velocity dispersion $\lesssim60$\,\kms, which agrees with the range of velocity dispersions inferred from the ADC and the measurements from \hg.}

\subsection{Other Possible Explanations}
\label{ssec:bulge}

\begin{figure}
\centering
\label{fig:bulge}
\gridline{\fig{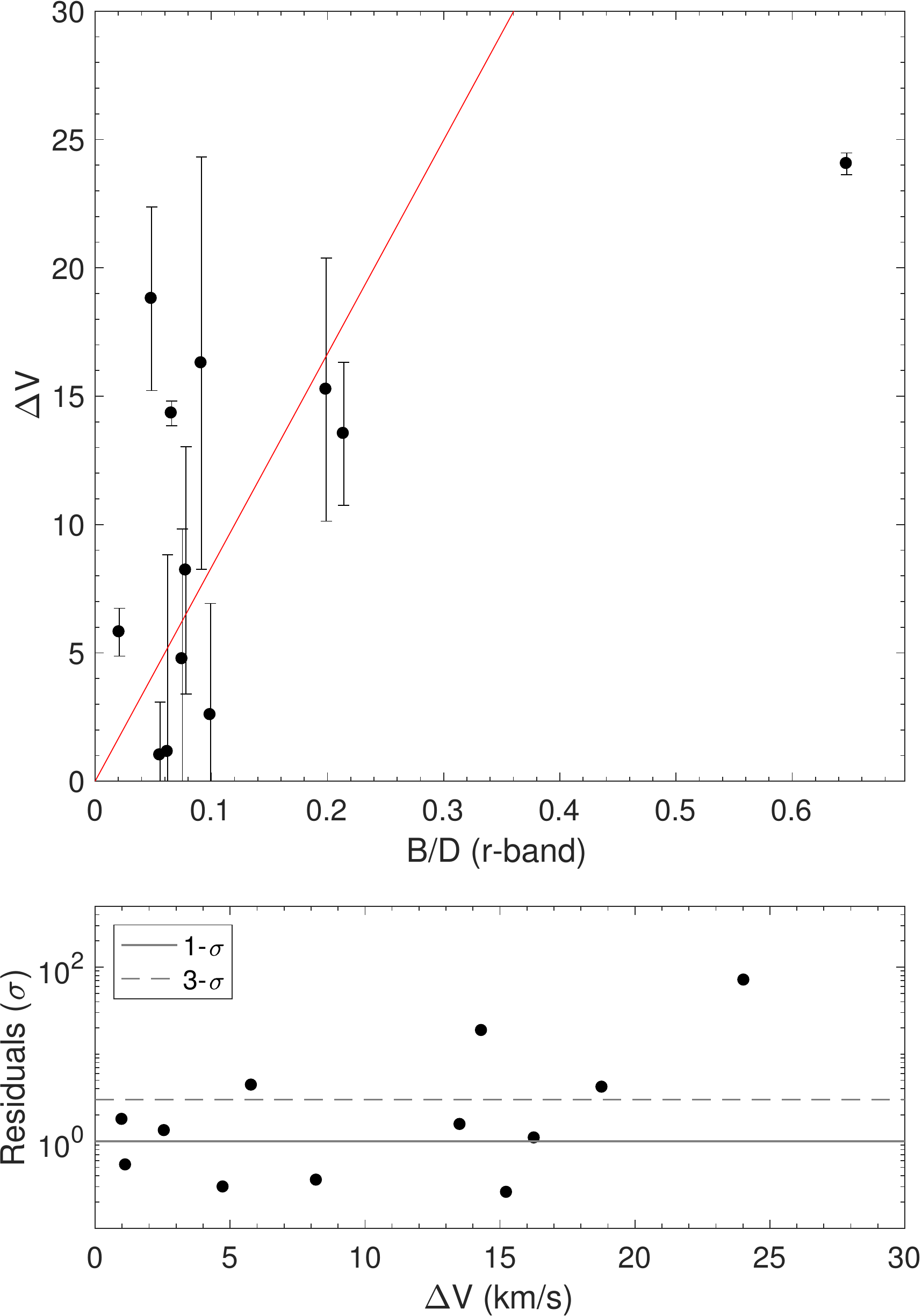}{0.8\columnwidth}{(a)}}
\gridline{\fig{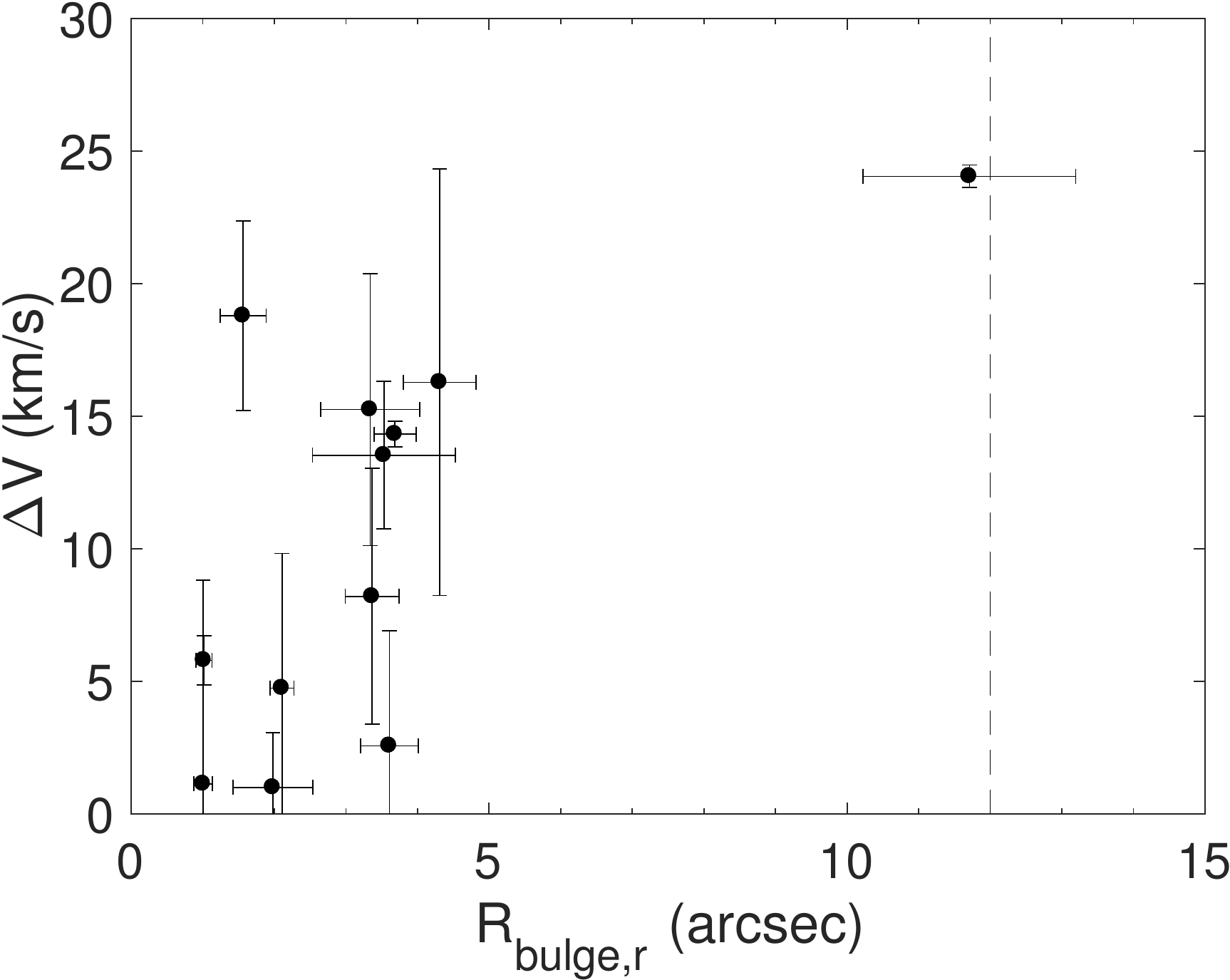}{0.75\columnwidth}{(b)}}
\caption{(a) The trend between the B/D ratio and $\Dv$. B/D ratios are derived from \citet{mendezabreu17}, who do not report errors on those values. The red line is a linear fit to the data points, excluding NGC\,2347 whose B/D ratio is an outlier compared to the other galaxies used here. The bottom panel shows the perpendicular distance from the line for each point in units of standard deviations plotted against $\Dv$. The solid and gray dashed lines show 1- and 3-$\sigma$. For the r-band, 33\% of the galaxies are consistent with the best-fit line within 1-$\sigma$ and 67\% are consistent within 3-$\sigma$. (b) We also investigate the relationship between the effective radius of the bulge ($R_{\rm bulge}$) from \citet{mendezabreu17} and $\Dv$. All bulges (except NGC\,2347) are much smaller than $\rbeam$ {\comment(vertical dashed line)}, so contamination from the bulge is likely small in these galaxies.}
\end{figure}

In the previous subsections, we motivated our hypothesis that the observed difference between the molecular and ionized gas rotation velocities is due to eDIG in a thick disk with a vertical gradient in the rotation velocity. That is not, however, the only explanation. It is possible that we are instead measuring ionized gas velocities and velocity dispersions in the galactic bulge. To test this, we explore potential correlations between $\Dv$ and measured the bulge-to-disk (B/D) luminosity ratios \citep{mendezabreu17}. Here, we use only results from the r-band since that overlaps with \ha. This is shown in the top panel of Figure \ref{fig:bulge}a. To determine if there are any underlying correlations, we follow the same methodology as was done for the velocity dispersions discussed in Section \ref{ssec:dvveldisp}. There is one galaxy with a large B/D ratio which is a clear outlier from the rest; this galaxy is NGC\,2347 which has the largest $\Dv$ in the subsample. It is excluded from the fitting of the linear regression. The bottom panel of Figure \ref{fig:bulge}a shows the distance of each galaxy from the best-fit line. For r-band, 33\% of galaxies are consistent with the best-fit line within 1-$\sigma$ and 67\% are consistent within 3-$\sigma$. It is possible that the bulge could be affecting the measured rotation velocities and contributing to the measured $\Dv$. This affect is likely minimized, however, as $\Dv$ is only measured at radii larger than $\rbeam$ (12"\~4\,kpc). \citet{mendezabreu17} also measure the effective bulge radius ($R_{\rm bulge}$). We plot $R_{\rm bulge}$ versus $\Dv$ in Figure \ref{fig:bulge}b. The dashed vertical line marks $\rbeam$, which is the smallest radius included when measuring $\Dv$. Bulge contamination is likely for NGC\,2347 since $R_{\rm bulge}\approx\rbeam$, but the other galaxies in the sample have much smaller bulges so the likelihood of contamination from the bulge is lessened. The exact contributions of the bulge and eDIG to $\Dv$ are difficult to disentangle in detail, however, and it is likely that both contribute to the measured $\Dv$ at some level.

\section{Summary}
\label{sec:summary}
We present a kinematic analysis of the EDGE-CALIFA survey, combining high resolution CO maps from EDGE with CALIFA optical IFU data. Together, the CO and \ha\ kinematics can be compared in a statistical sample. We summarize our results as follows, indicating the relevant figures and/or tables:
\begin{enumerate}
\itemsep0em 
\item Using a sub-sample of 17 galaxies {\comment from the EDGE-CALIFA survey} where precise molecular gas rotation curves could be derived, we fit CO and \ha\ rotation curves using the same geometric parameters out to $\gtrsim 1\ R_e$ (Figure \ref{fig:KSS}). 

\item In our sub-sample, we find that most galaxies (\~75\%) have CO rotation velocities which are measurably higher than the \ha\ rotation velocity in the outer part of the rotation curves. We refer to the median difference between the CO and \ha\ rotation velocity as $\Dv$. Measurable differences between CO and \ha\ rotation velocities range from 5--25\,\kms, with a median value of 14\,\kms\ (Figure \ref{fig:gausshistdiffs}).  

\item The rotation velocity differences between CO and \ha\ are not driven by inclination effects since we find no significant trend between the inclination and $\Dv$ (Figure \ref{fig:IncTrend}).

\item We suggest that these differences are caused by extraplanar diffuse ionized gas (eDIG) in these galaxies, which may constitute a thick, turbulent disk of ionized gas. Extraplanar ionized gas would be caused by stellar feedback, so we expect that galaxies with sufficient SFR per unit area ($\Sigma_{\rm SFR}$) would have extended thick ionized gas disks and that galaxies with smaller $\Sigma_{\rm SFR}$ might have patchy extraplanar ionized gas above \HII\ regions \citep{rand96,rossa03a}. Indeed, the majority of the galaxies in the high-quality rotation curve sub-sample (\~95\%) have sufficient $\Sigma_{\rm SFR}$ to harbor eDIG (Figure \ref{fig:SFRarea}). Because EDGE galaxies were selected from CALIFA based on their FIR brightness \citep{bolatto17}, it is not surprising that the majority of them are star-forming disk galaxies with large $\Sigma_{\rm SFR}$.

\item If galaxies frequently feature thick ionized gas disks, the effect described above would cause a systematic underestimate of galaxy dynamical masses derived from ionized gas rotation velocity. Because we find that the ionized gas rotates more slowly than the molecular gas in galaxies with large $\Sigma_{\rm SFR}$, this effect could be significant in local active star-forming galaxies and even more so at higher redshifts where star formation rates are much higher on average.

\item  {\comment We measure ionized gas velocity dispersion using the \hg\ line (\sigmaHg) as a proxy for \ha, and compare them to predictions for the  asymmetric drift correction (ADC), assuming velocity dispersion explains the observed difference in rotation velocities (Figure \ref{fig:DvVelDisp}). For the ADC, we infer velocity dispersion which would support a thick ionized gas disk with scale height ranging from \~0.1--2.0\,kpc. The velocity dispersion measured from the \hg\ is comparable but somewhat smaller than predicted from the ADC (Figure \ref{fig:SigmaHgADC}). The low spectral resolution of the data makes this measurement very difficult, so these results are tentative.}

\item We find that \SII/\ha\ and \NII/\ha, which are tracers of the WIM, are elevated in these galaxies compared to typical values in the plane of the MW \citep[\SII/\ha=0.11, \NII/\ha\~0.25;][]{madsen04,madsen06}, but are not as large as typically found in the eDIG \citep[\SII/\ha=0.34, \NII/\ha$\gtrsim$0.5;][]{blanc09,madsen04}. This is likely because emission from the midplane and eDIG are mixed together in these measurements (Figure \ref{fig:SIINII}).

\item We investigate the effect of disk geometry by performing a suite of kinematic simulations with NEMO. We find that neither a thin disk nor a thick disk without a vertical gradient in the rotation velocity can reproduce the observed $\Dv$ (Figures \ref{fig:SimTrends}a,b). The observed $\Dv$ can be reproduced with a vertical gradient in rotation velocity. For a linear vertical rotation velocity gradient, our results favor $\eta\equiv1-\frac{\vrot(z=h)}{V_0}\lesssim0.3$ (Figure \ref{fig:SimTrends}c). For a more realistic vertical rotation velocity gradient, our results can be reproduced with an eDIG scale height $\lesssim 1.5$\,kpc {\comment corresponding to a velocity dispersion $\lesssim 60$\,\kms\ } (Figure \ref{fig:SimTrends}d).
\end{enumerate}

An ideal way to test for eDIG in these galaxies would be to directly measure the CO and \ha\ scale heights in a sample of edge-on disks and to correlate the \ha\ scale height with $\Dv$. Additionally, a systematic decrease in the \ha\ rotation velocity with distance from the midplane would be compelling evidence for our proposed model of a thick disk with a vertical velocity gradient. We plan to carry out this analysis using the edge-on galaxies in the EDGE-CALIFA survey in a forthcoming paper (Levy et al. 2018, in preparation).

\acknowledgments
R.C.L. would like to thank Filippo Fraternali and Federico Lelli for useful discussions and advice. {\comment The authors would also like to thank the anonymous referee for their comments.} R.C.L. and A.D.B. acknowledge support from the National Science Foundation (NSF) grants AST-1412419 and AST-1615960. A.D.B. also acknowledges visiting support by the Alexander von Humboldt Foundation. P.T. and S.N.V. acknowledge support from NSF AST-1615960. S.F.S. acknowledges the PAPIIT-DGAPA-IA101217 project and CONACYT-IA-180125. R.G.B. acknowledges support from grant AYA2016-77846-P. L.B. and D.U. are supported by the NSF under grants AST-1140063 and AST-1616924. D.C. acknowledges support by the Deutsche Forschungsgemeinschaft, DFG through project number SFB956C. T.W. acknowledges support from the NSF through grants AST-1139950 and AST-1616199.  This study makes use of data from the EDGE (\url{http://www.astro.umd.edu/EDGE/}) and CALIFA (\url{http://califa.caha.es/}) surveys and numerical values from the HyperLeda database (\url{http://leda.univ-lyon1.fr}). Support for CARMA construction was derived from the Gordon and Betty Moore Foundation, the Kenneth T. and Eileen L. Norris Foundation, the James S. McDonnell Foundation, the Associates of the California Institute of Technology, the University of Chicago, the states of California, Illinois, and Maryland, and the NSF. CARMA development and operations were supported by the NSF under a cooperative agreement and by the CARMA partner universities. This research is based on observations collected at the Centro Astron\'{o}mico Hispano-Alem\'{a}n (CAHA) at Calar Alto, operated jointly by the Max-Planck Institut f\"{u}r Astronomie (MPA) and the Instituto de Astrofisica de Andalucia (CSIC). The National Radio Astronomy Observatory is a facility of the National Science Foundation operated under cooperative agreement by Associated Universities, Inc.

\facilities{CARMA, CAO:3.5, GBT, GB:300ft, Arecibo}
\software{Miriad \citep{miriad}, NEMO \citep{nemo}, \pipetd\ \citep{pipe3DI,pipe3DII}}

\bibliographystyle{yahapj}

\appendix
\section{Velocity Conventions}
\label{app:velconv}
When converting from frequency (or wavelength) to a velocity via the Doppler formula, the radio and optical communities use different conventions as approximations to the full relativistic conversion. The optical convention is
\begin{equation}
\frac{V_{\rm opt}}{c} = \frac{\lambda-\lambda_o}{\lambda_o} \equiv z,
\label{eq:opt}
\end{equation}
the radio convention is
\begin{equation}
\frac{V_{\rm radio}}{c} = \frac{\nu_o-\nu}{\nu_o} = \frac{z}{1+z},
\label{eq:radio}
\end{equation}
and the relativistic convention is
\begin{equation}
\frac{V_{\rm rel}}{c} = \frac{\nu_o^2-\nu^2}{\nu_o^2+\nu^2}.
\label{eq:rel}
\end{equation}
Therefore, combining Equations \ref{eq:opt} and \ref{eq:radio}, a velocity measured in the optical convention ($V_{\rm opt}$) can be converted to the radio convention ($V_{\rm radio}$) where
\begin{equation}
V_{\rm radio} = \frac{V_{\rm opt}}{1+\frac{V_{\rm opt}}{c}}.
\label{eq:opt2radio}
\end{equation}
A galaxy with a rotation speed measured in the optical convention ($V^{\rm rot}_{\rm opt}$) will result in a radio rotation speed ($V^{\rm rot}_{\rm radio}$), where
\begin{equation}
V^{\rm rot}_{\rm radio} = \frac{V_{\rm opt}+V^{\rm rot}_{\rm opt}}{1+\frac{V_{\rm opt}+V^{\rm rot}_{\rm opt}}{c}}-\frac{V_{\rm opt}}{1+\frac{V_{\rm opt}}{c}} \approx \frac{V^{\rm rot}_{\rm opt}}{1+\frac{2V_{\rm opt}}{c}}
\label{eq:vrotradio}
\end{equation}
to first order. The rotation velocity in the optical convention is larger than in the radio convention. As an example, for a galaxy with $V_{\rm opt} = 4500$\,\kms\ and $V^{\rm rot}_{\rm opt} = 250$\,\kms\ (similar to the CALIFA galaxies), $V^{\rm rot}_{\rm radio} = 243$\,\kms, so the rotation velocity in the radio convention is smaller than in the optical convention. For the velocity differences measured from the CO and \ha\ rotation curves, this discrepancy is not negligible. To avoid this, maps should be converted to the same convention.

Both the radio and optical convention scales become increasingly compressed at higher redshifts (large systemic velocities), but the relativistic convention does not suffer from this compression. Both EDGE and CALIFA reference to zero velocity, so typical velocities are a few thousand \kms. Conversions to relativistic from radio or optical by combining Equations \ref{eq:radio} or \ref{eq:opt}, respectively, with Equation \ref{eq:rel}, where
\begin{equation}
\frac{V_{\rm rel}}{c} = \frac{(\frac{V_{\rm opt}}{c}+1)^2-1}{(\frac{V_{\rm opt}}{c}+1)^2+1}
\label{eq:opt2rel}
\end{equation}
and
\begin{equation}
\frac{V_{\rm rel}}{c} = \frac{1-(1-\frac{V_{\rm radio}}{c})^2}{1+(1-\frac{V_{\rm radio}}{c})^2}.
\label{eq:radio2rel}
\end{equation}
For a rotational velocity measured in the relativistic frame ($V^{\rm rot}_{\rm rel}$), the corresponding $V^{\rm rot}_{\rm opt}$ and and $V^{\rm rot}_{\rm radio}$ are
\begin{equation}
V^{\rm rot}_{\rm opt} = \frac{V^{\rm rot}_{\rm rel}}{1-\frac{V_{\rm rel}}{c}}
\label{eq:rotrel2opt}
\end{equation}
\begin{equation}
V^{\rm rot}_{\rm radio} = \frac{V^{\rm rot}_{\rm rel}}{1+\frac{V_{\rm rel}}{c}}.
\label{eq:rotrel2radio}
\end{equation}
For a galaxy with $V_{\rm opt} = 4500$\,\kms, this results in $V_{\rm rel}=4466$\,\kms\ (Equation \ref{eq:opt2rel}) and $V_{\rm radio} = 4433$\,\kms\ (Equation \ref{eq:opt2radio}). For a $V^{\rm rot}_{\rm opt} = 250$\,\kms, this results in $V^{\rm rot}_{\rm rel} = 246$\,\kms\ (Equation \ref{eq:rotrel2opt}) and $V^{\rm rot}_{\rm radio} = 242$\,\kms\ (Equation \ref{eq:rotrel2radio}).

Therefore, to avoid the effects of different velocity conventions and compression, all EDGE and CALIFA velocity fields are converted to the relativistic convention using Equations \ref{eq:radio2rel} and \ref{eq:opt2rel} respectively. All velocities are reported in this convention unless otherwise noted.

{\comment
\section{Beam Smearing Correction}
\label{app:BS}

In order to accurately measure the CO and \hg\ velocity dispersions, a beam smearing correction must be applied to the data cubes. As the beam size increases, rotation velocities from different radii can be blended into the linewidth. To correct for this effect, we fit a rotation-only model to the CO and \ha\ rotation curves and construct model rotation velocity fields. The smooth model rotation curve for each galaxy was constructed using the ``Universal Rotation Curve'' (hereafter referred to as a Persic Profile) from \citet{persic96} which is given by
\begin{equation}
V_{\rm Persic}(x) = V(R_{\rm opt})\Bigg \{\Big[0.72+0.44\log(L_B/L_{B*})\Big]\frac{1.97x^{1.22}}{(x^2+0.78^2)^{1.43}}\\
 +1.6\exp{\Big[-0.6(L_B/L_{B*})\Big]\frac{x^2}{x^2+1.5^2(L_B/L_{B*})^{0.4}}} \Bigg\}^{1/2} {\rm km\ s^{-1}}
\label{eq:persic}
\end{equation}
where $x = R/R_{\rm opt}$. \citet{persic96} use $\log L_{B*} = 10.4$, where $L_{B*}=6\times10^{10} h_{50}^{-2} \ L_{B\odot}$ \citep{persic91} with $h=0.75$; we also adopt this value of $L_{B*}$. $R_{\rm opt}$ is the optical radius. The Persic Profile was fit to $\vrot$ using a non-linear least squares fit with $V(R_{\rm opt}),\ L_B,$ and $R_{\rm opt}$ left as parameters to be fit which are listed in Table \ref{tab:persicparams}. Model velocity fields were derived from the Persic Profile fits using a linear interpolation and rotated and inclined based on the PA and inclination of the corresponding galaxy.

\setcounter{table}{1}
\begin{deluxetable}{ccccccc}
\tablecaption{Parameters for the Persic Profile Fits \label{tab:persicparams}}
\tabletypesize\footnotesize
\tablehead{
\colhead{Name} &\colhead{CO V(R$_{\rm opt}$)} & \colhead{CO $L_B$} & \colhead{CO R$_{\rm opt}$} & \colhead{H$\alpha$ V(R$_{\rm opt}$)} & \colhead{H$\alpha$ $L_B$}  & \colhead{H$\alpha$ R$_{\rm opt}$} \\
& (km/s) &\colhead{($\times10^8 \ L_{B\odot}$ )} & (arcsec) & (km/s) & \colhead{($\times10^8 \ L_{B\odot}$)} & (arcsec)}
\startdata
IC1199    &  168.7 & 2.8 & 7.2 & 181.7 & 8.8 & 14.4 \\
NGC2253    &  145.9 & 1.7 & 6.3 & 146.6 & 2.0 & 10.5 \\
NGC2347    &  233.5 & 4.5 & 4.9 & 211.0 & 7.5 & 5.4 \\
NGC2410    &  198.8 & 2.5 & 10.9 & 175.1 & 9.8 & 4.3 \\
NGC3815    &  167.0 & 10.0 & 10.7 & 147.3 & 2.1 & 9.3 \\
NGC4047    &  174.8 & 10.0 & 6.4 & 177.9 & 10.0 & 10.1 \\
NGC4644    &  164.7 & 1.7 & 10.0 & 144.6 & 3.3 & 9.9 \\
NGC4711    &  131.1 & 9.4 & 12.6 & 138.4 & 10.0 & 16.1 \\
NGC5016    &  161.9 & 9.5 & 13.5 & 166.1 & 10.0 & 18.0 \\
NGC5480    &  90.7 & 9.9 & 6.7 & 110.3 & 10.0 & 18.6 \\
NGC5520    &  133.7 & 10.0 & 5.7 & 117.7 & 5.9 & 5.8 \\
NGC5633    &  165.2 & 10.0 & 11.3 & 148.4 & 2.9 & 9.6 \\
NGC5980    &  181.6 & 2.9 & 7.8 & 184.1 & 3.4 & 13.1 \\
UGC04132    &  199.8 & 10.0 & 10.7 & 207.2 & 10.0 & 15.5 \\
UGC05111    &  197.1 & 3.8 & 15.5 & 203.1 & 5.4 & 19.9 \\
UGC09067    &  185.8 & 2.1 & 8.0 & 169.9 & 7.4 & 7.7 \\
UGC10384    &  171.0 & 3.2 & 12.7 & 147.6 & 2.8 & 10.8 \\
\enddata
\tablecomments{V(R$_{opt}$), $L_B$, and R$_{opt}$ for CO and \ha\ are the parameters from the Persic Profile fitting (Equation \ref{eq:persic}).}
\end{deluxetable}

Using the rotation-only model velocity fields, we construct data cubes to simulate the effects of beam smearing. For the CO data, the channel width and spectral resolution are both 20\,\kms. For each spaxel in the simulated CO data cube, a line is placed at the channel corresponding to the model rotation velocity of that pixel. The amplitude of the line is given by the value of the CO data cube at that voxel. The linewidth is a single 20\,\kms\ channel. All other channels in that spaxel have zero amplitude. For the \hg\ however, the channel width is 0.7\,\AA\ (48.4\,\kms) and the spectral resolution (FWHM) is 2.3\,\AA\ (159.0\,\kms). Therefore, at each spaxel in the simulated \hg\ cube, a Gaussian line is placed at the channel corresponding to the model rotation velocity at that pixel. The FWHM of the Gaussian line is 2.3\,\AA. The amplitude of the line is given by the value of the \hg\ data cube at that voxel. These simulated data cubes are then convolved to the corresponding CO or \hg\ beam size using the \convol\ task in \miriad. Model velocity dispersion maps (sigmas, not FWHM) are created using the \moment\ task in \miriad, which quantify the velocity dispersion due to beam smearing. These model velocity dispersion maps are removed in quadrature from the corresponding CO or \hg\ velocity dispersion maps, yielding a beam smearing corrected velocity dispersion map. Since the simulated data cubes incorporate the instrumental linewidth, removing the simulated velocity dispersion map also corrects for the instrumental linewidth. The beam smearing corrected velocity dispersion maps are then masked to the same region where $\Dv$ is calculated, which excludes the centers where the beam smearing corrections are large. Figure \ref{fig:BScorr} shows the stages of the beam smearing correction for both CO (a) and \hg\ (b) for NGC\,2347. We note that after this masking, the difference between the beam smearing correction applied here and simply removing the instrumental linewidth in quadrature is \~2\,\kms over the whole KSS sample. Each CO and \hg\ velocity dispersion map has a corresponding error map. The weighted average CO or \hg\ velocity dispersion for each galaxy is 
\begin{equation}
\langle\sigma\rangle = \frac{\Sigma(\sigma_i/\delta_{\sigma_{i}}^2)}{\Sigma(1/\delta_{\sigma_{i}}^2)} 
\end{equation}
where $\sigma_i$ is the velocity dispersion and $\delta_{\sigma_{i}}$ is the error on the velocity dispersion at each pixel. The error on $\langle\sigma\rangle$ is the weighted standard deviation given by
\begin{equation}
\sigma_{\langle\sigma\rangle} = \sqrt{\frac{N}{\Sigma(1/\delta_{\sigma_{i}}^2)}}.
\end{equation}
The resulting beam smearing corrected average CO and \hg\ velocity dispersions are listed in Table \ref{tab:KSSparams}. A comparison of the CO and \hg\ velocity dispersions is shown in Figure \ref{fig:sigmaCOHg} for the KSS galaxies. We note that previous measurements of CO velocity dispersions from HERACLES in bright GMCs are $\sim7$\,\kms\ but increase to $\sim12$\,\kms\ if a larger beam is used which encompasses more diffuse CO \citep{mogotsi16}. The EDGE data are at \~kpc resolution and would encompass diffuse CO as well as denser GMCs, and the values we derive are consistent with these results.

\begin{figure}
\centering
\label{fig:BScorr}
\gridline{\fig{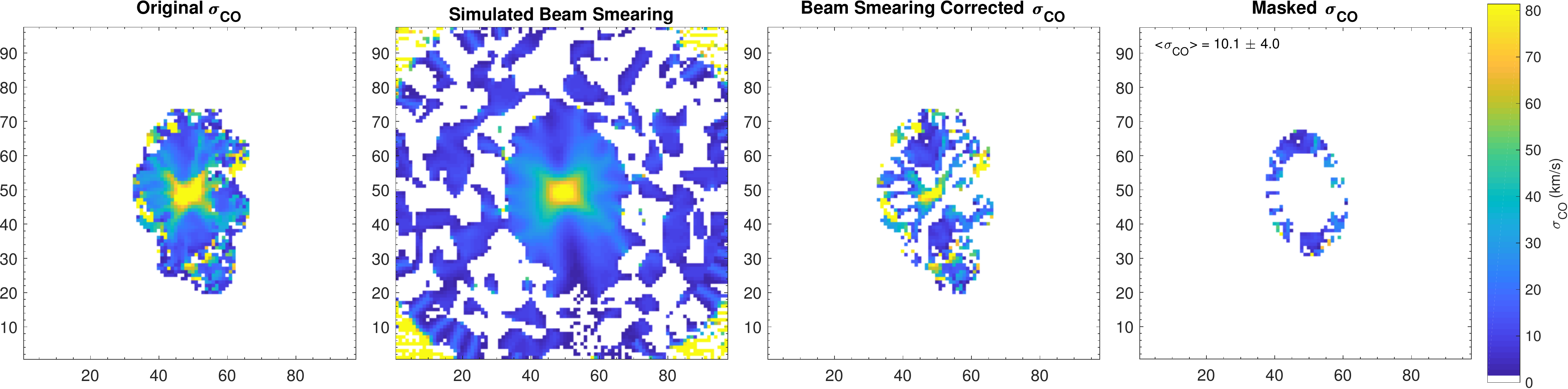}{\columnwidth}{(a)}}
\gridline{\fig{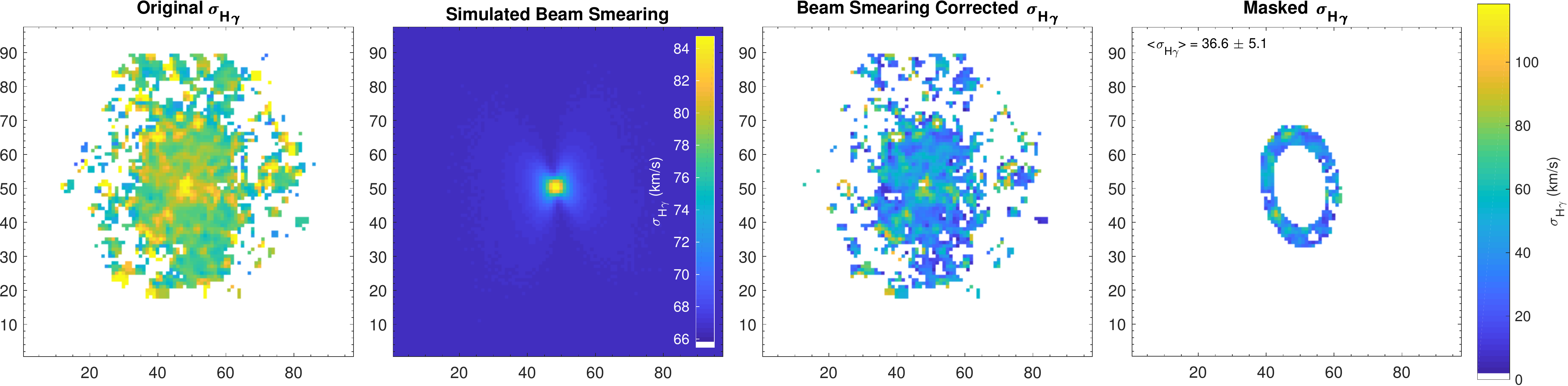}{\columnwidth}{(b)}}
\caption{\comment The progression of the beam smearing correction for (a) CO  and (b) \hg\ for NGC\,2347. From left to right, the panels show the original velocity dispersion map, the simulated velocity dispersion due to beam smearing, the beam smearing corrected velocity dispersion, and the masked beam smearing corrected velocity dispersion. All panels are normalized to the color scales shown to the right, except for the \hg\ simulated beam smearing whose color scale is shown in the inset. White patches in the right two columns are where the simulated beam smearing is larger than the original velocity dispersion and are imaginary when removed in quadrature. These points are discarded for the analysis.}
\end{figure}

\begin{figure}
\label{fig:sigmaCOHg}
\centering
\includegraphics[width=0.6\columnwidth]{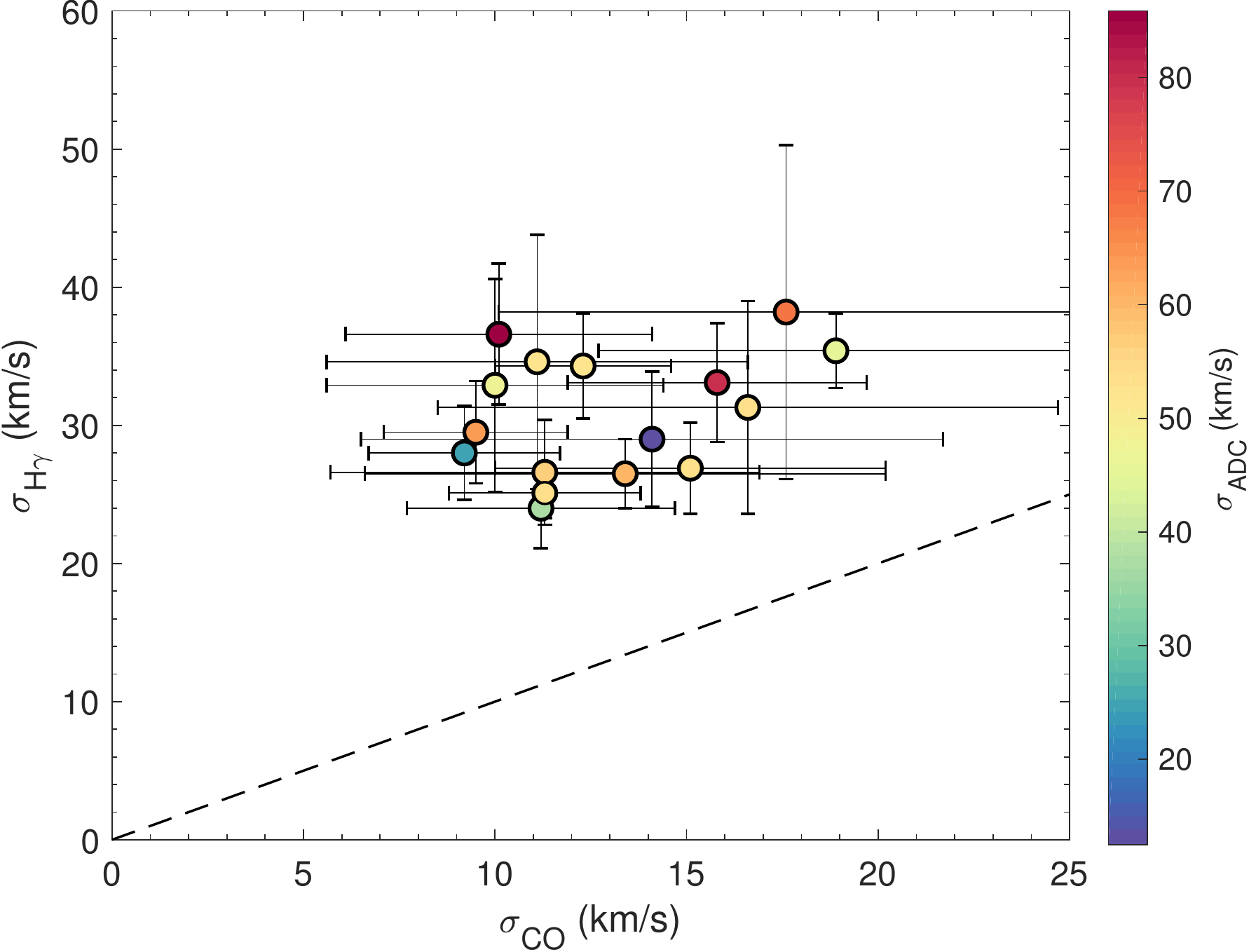}
\caption{The beam smearing corrected CO and \hg\ velocity dispersions for the KSS galaxies. Points are color-coded by the velocity dispersion inferred from the ADC. The dashed line is one-to-one.}
\end{figure}
}

\section{Galaxy-by-Galaxy Data, Figures, and Descriptions}
\label{app:galbygal}
\begin{figure*}[p]
\label{fig:Dvtrends}
\centering
	\gridline{\fig{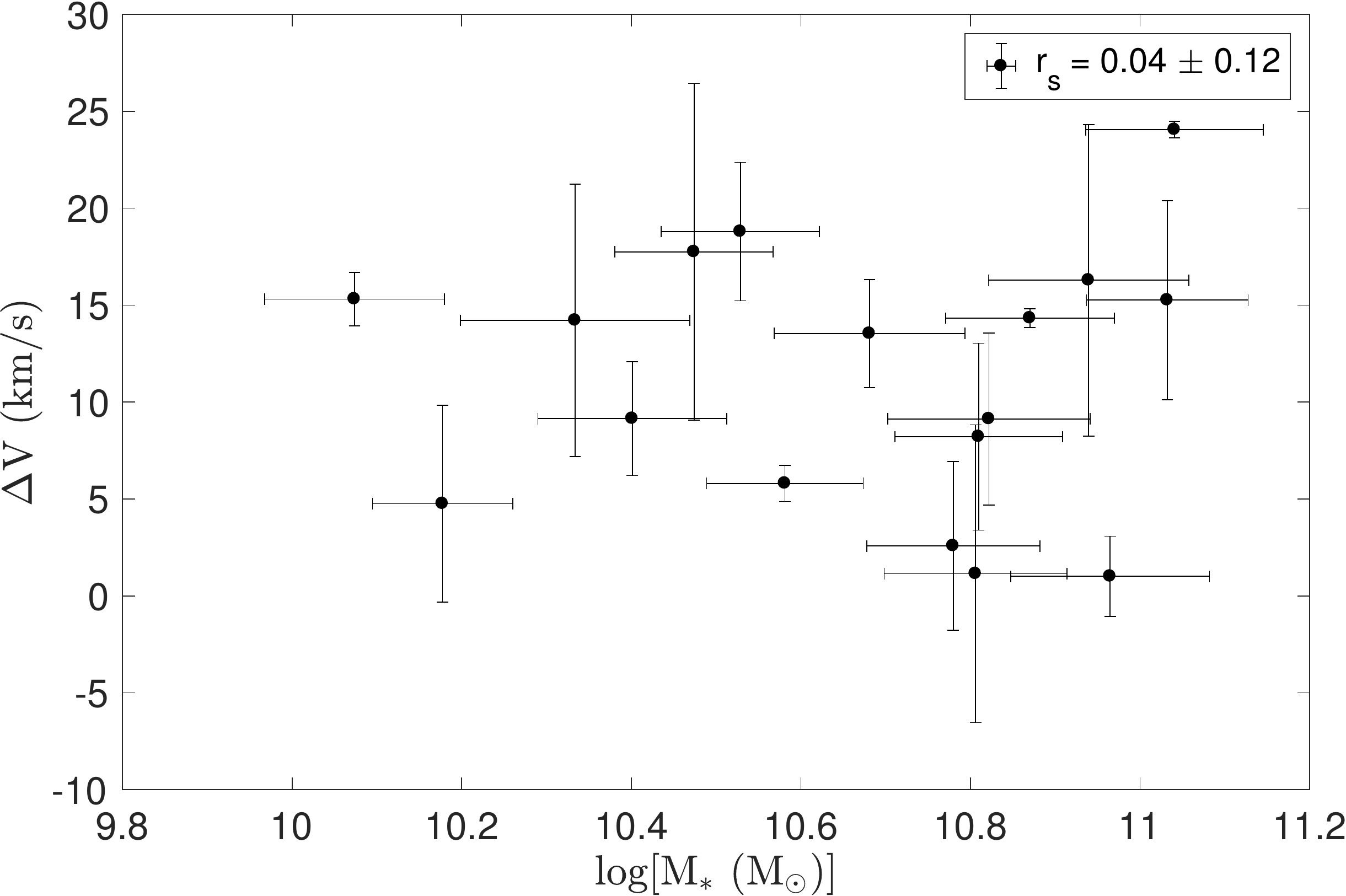}{0.45\columnwidth}{(a)}
    		\fig{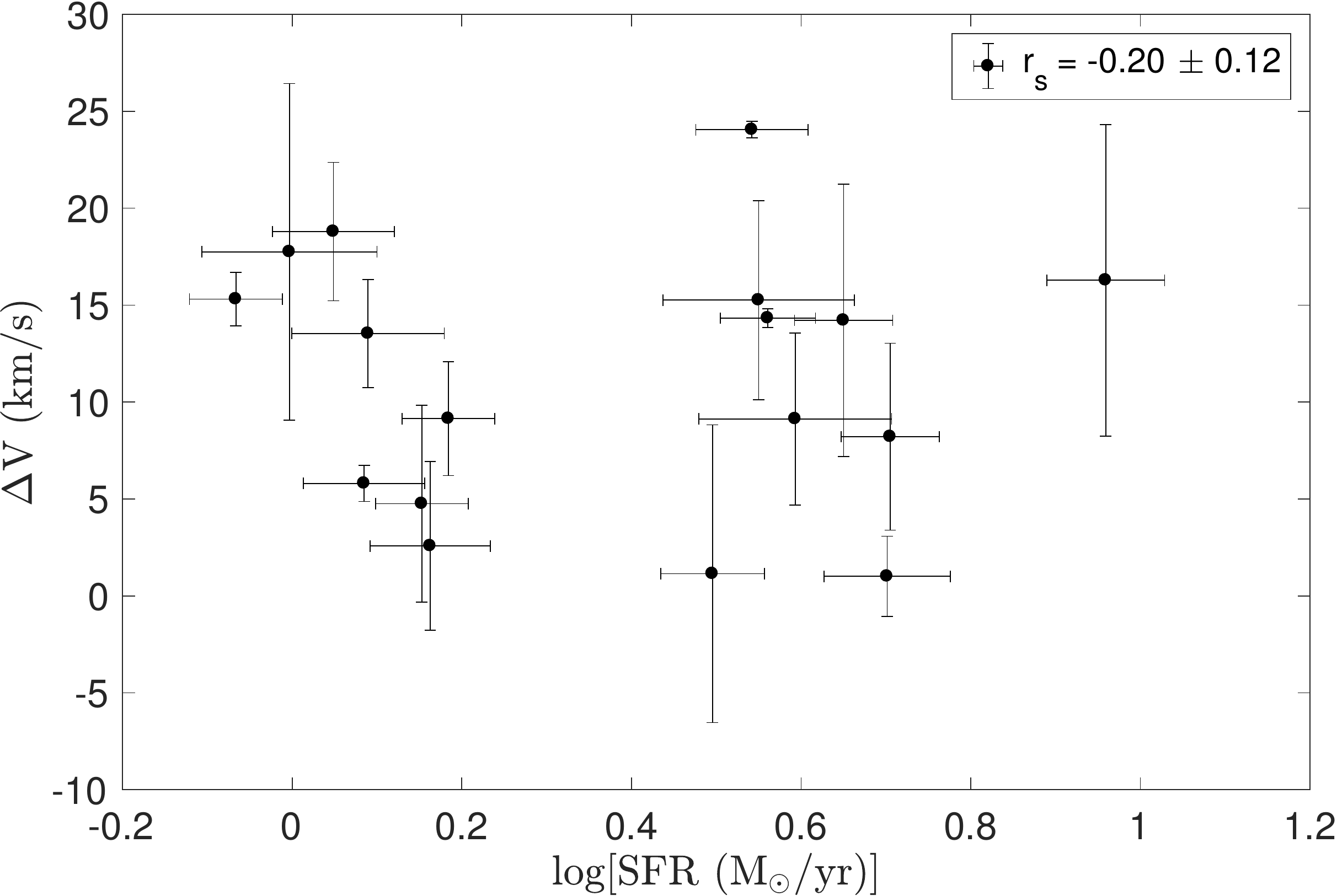}{0.45\columnwidth}{(b)}}
    \gridline{\fig{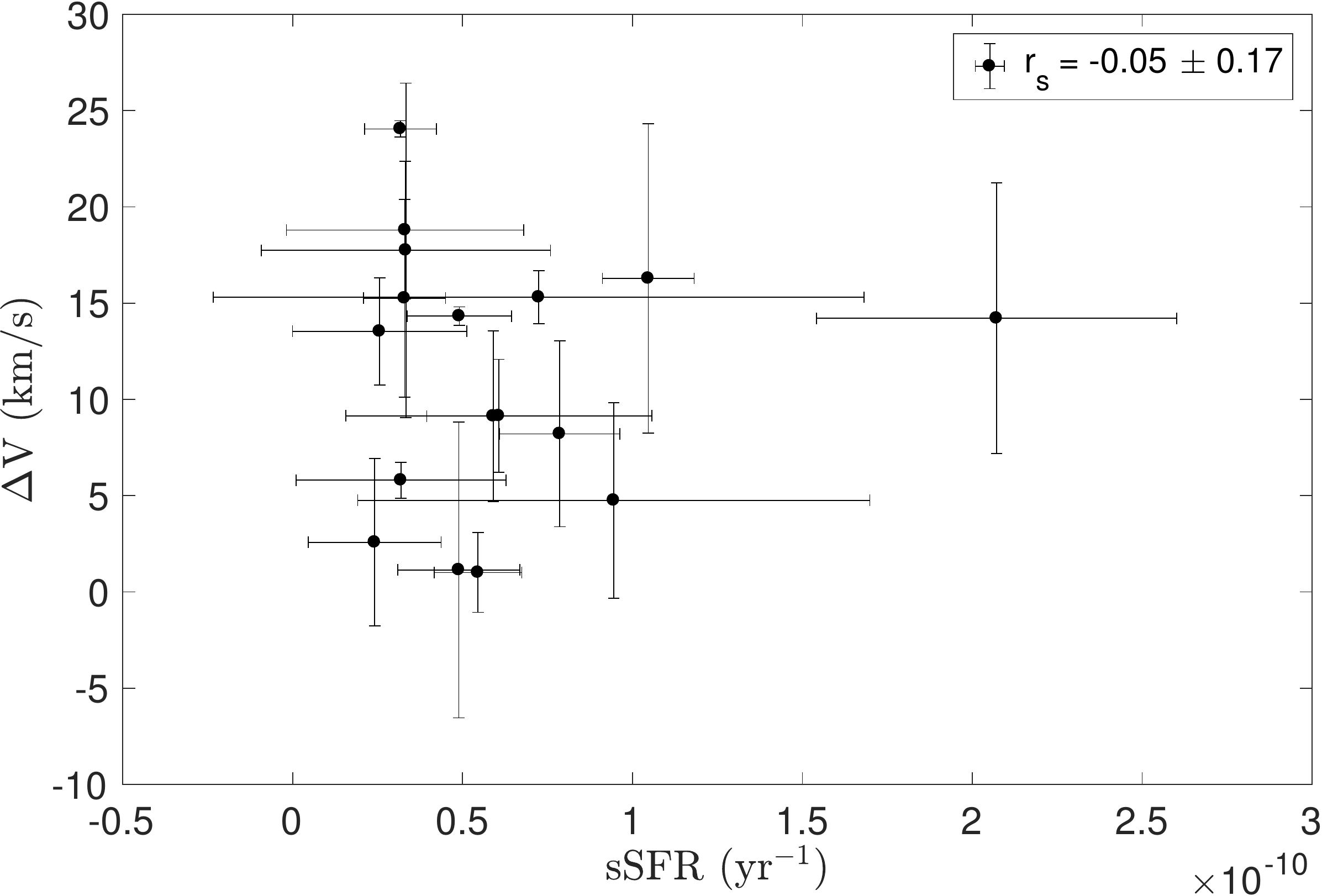}{0.45\columnwidth}{(c)}
    		\fig{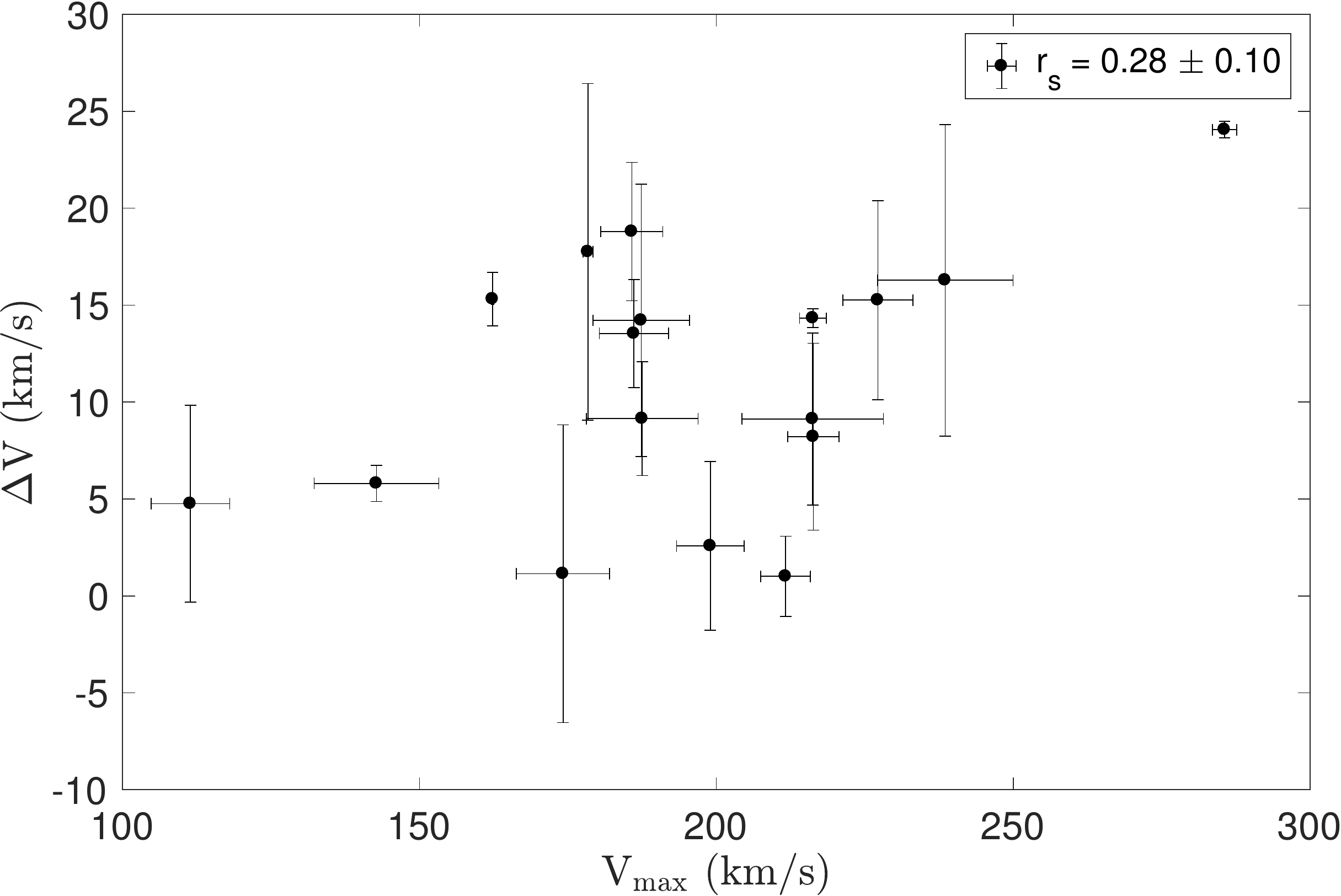}{0.45\columnwidth}{(d)}}
  	\gridline{\fig{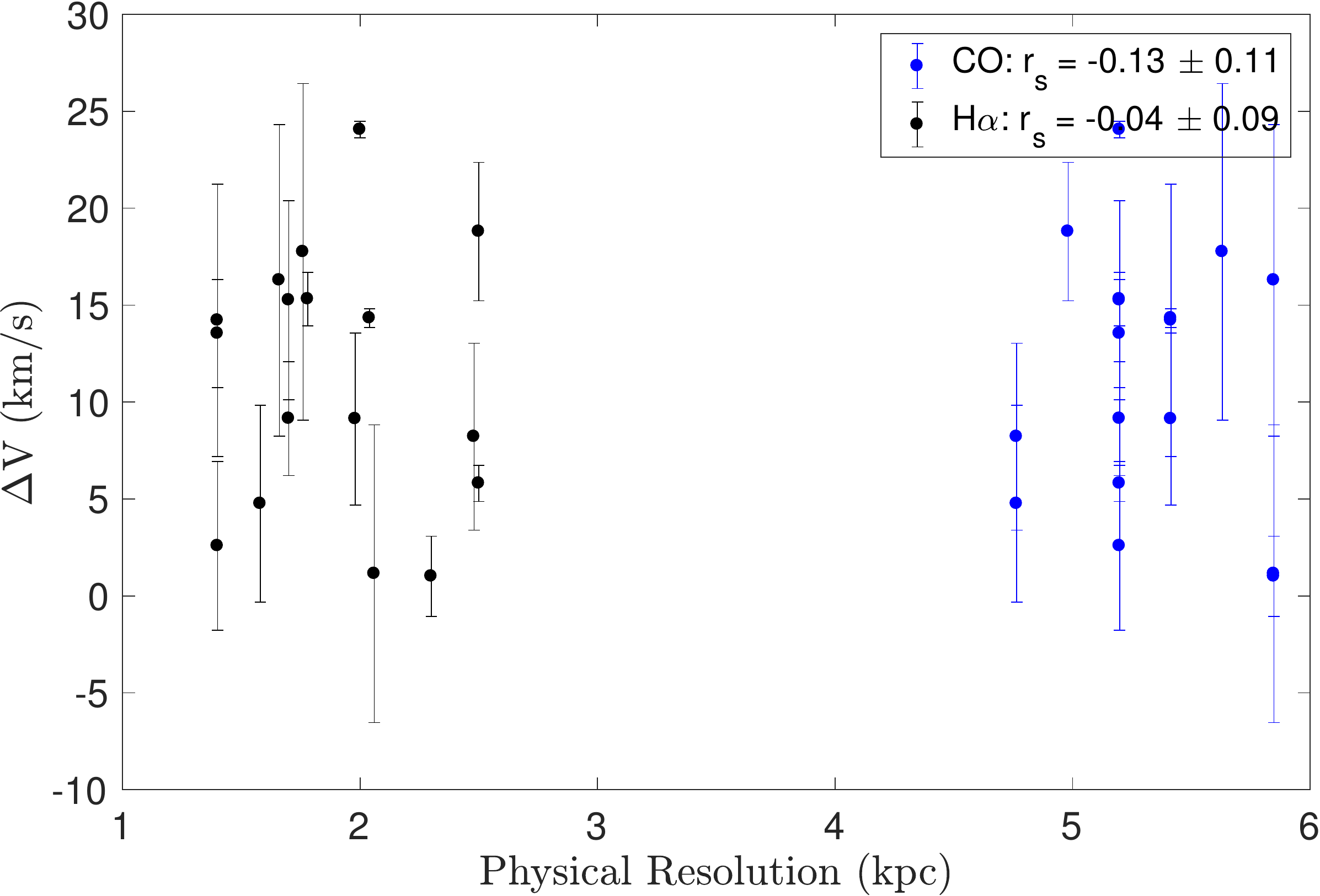}{0.45\columnwidth}{(e)}
    		\fig{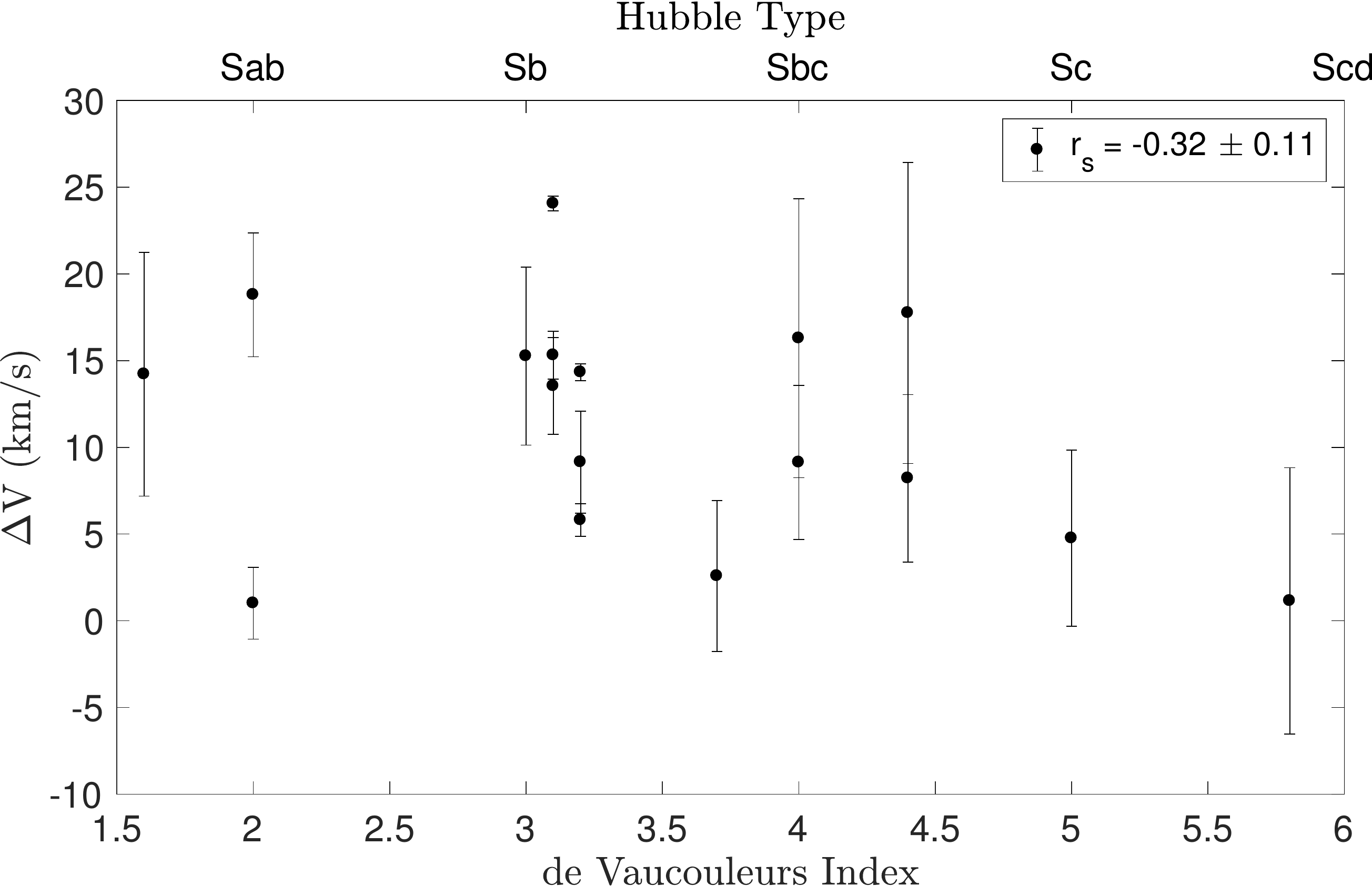}{0.45\columnwidth}{(f)}}
\caption{There are no trends between $\Dv$ and (a) $M_*$, (b) SFR, (c) sSFR, (d) CO V$_{\rm{max}}$, (e) physical resolution of the native (unconvolved) CO (blue) and \ha\ (black) data, and (f) morphology. The Spearman rank correlation coefficients ($r_s$) are shown in the top right of each panel. de Vaucouleurs indexes are taken from HyperLeda. The lack of trend with physical resolution in particular justifies our choice to convolve to a common angular resolution rather than to a common physical resolution.}
\end{figure*}

Kinematic parameters used for the EDGE and CALIFA data for all 126 galaxies are listed in Tables \ref{tab:EDGEparameters} and \ref{tab:CALIFAparameters} respectively. If kinematic parameters could not be fit, values were taken from outer isophote photometry \citep{falconbarroso17} or from HyperLeda. Previous work to determine kinematic position angles in a subset of the CALIFA galaxies has also been carried out by \citet{garcialorenzo15} and \citet{barreraballesteros14,barreraballesteros15b}. More details are in the table captions. The lack of trends with global parameters discussed in Section \ref{ssec:incdisk} are shown in Figure \ref{fig:Dvtrends}. CO and \ha\ velocity fields and rotation curves are shown for each galaxy in the KSS in Figure \ref{fig:KSS}. Below we provide brief comments on each galaxy in the KSS. 
\paragraph{IC 1199}
This galaxy's CO velocity field is somewhat patchy due to the SN masking; however, its CO rotation curve is excellent, having radial and systemic components near zero. The \ha\ rotation curve is interesting, as it flattens until \~13" then rises to meet the CO rotation curve. The \ha\ rotation curve generally has small radial and systemic components, although there is some non-zero radial component where the \ha\ rotation curve flattens which may explain the disagreement between the CO and \ha\ rotation curves in this region. The \ha\ systemic velocity is non-zero in the center, but since that region is excluded this will not affect the results. Neither the \hi\ W50 nor W90 agree with the CO or \ha\ rotation curves well.

\paragraph{NGC\,2253}
This galaxy's CO rotation curve is excellent, although there is a small trend in the radial component over the region of comparison. Its \ha\ rotation curve has large radial and systemic components in the center, but they are near zero over the comparison region. The CO and \ha\ rotation components cross, and the CO $\vrot$ continues to rise despite the flattened \ha. Both the \hi\ W50 and W90 are larger than the CO and \ha\ $\vrot$.
\paragraph{NGC\,2347}
This galaxy is used as an example throughout this work. Overall, its CO and \ha\ rotation curves are excellent, and the $\Dv$ is large and much greater than the errors on the rotation curves. Although there is a bump in the CO radial component in the comparison region, it likely cannot account for the $\Dv$. The \hi\ W50 and W90 straddle the CO $\vrot$, potentially indicating that the \ha\ scale height in this galaxy is larger than the \hi, which is closer to the CO scale height. This galaxy has a reported ring structure \citep{bolatto17}. It is also potentially an AGN candidate, based on its \OIII/\hb\ and \NII/\ha\ ratios. However, since the central 12" (\~4\,kpc) are excluded from analysis, any AGN contamination should be minimal. This galaxy has a large bulge at redder optical wavelengths, with a bulge-to-disk ratio of 0.3 in g-band, 0.7 in r-band, and 1.1 in i-band and an average bulge effective radius of $3.5\pm0.5$\,kpc \citep{mendezabreu17}, so bulge contamination is possible.
\paragraph{NGC\,2410}
Although this galaxy does exhibit non-zero CO and \ha\ radial and systemic components over the comparison region, the components all track one another so their effects on $\Dv$ should be minimal. There is a kinematic twist visible in the \ha\ velocity field, which is likely responsible for the varying \ha\ radial component. This feature is less obvious in the CO velocity field. This galaxy has a bar, as listed in \citet{bolatto17}, which is the likely source of this twist. Both the \hi\ W50 and W90 are higher than the CO $\vrot$.
\paragraph{NGC\,3815} 
Although this galaxy has non-zero radial and systemic components over the comparison region, they track each other so the effect on $\Dv$ should be minimal. The CO and \ha\ systemic components are near zero. At large radii, the \ha\ $\vrot$ is consistent with the \hi\ W50, whereas the W90 is much larger than both the CO and \ha\ $\vrot$.
\paragraph{NGC\,4047} 
This galaxy has excellent CO and \ha\ rotation curves and velocity fields. At large radii, the \hi\ W50 appears to be consistent with the \ha\ $\vrot$ and potentially the CO $\vrot$. The W90 is much larger.
\paragraph{NGC\,4644} 
The galaxy has excellent CO and \ha\ rotation curves, with small radial and systemic components overall. Interestingly, this galaxy is listed as having a bar and being a merger \citep{bolatto17}, although there are no obvious signs of either from the velocity fields or rotation curves. The \hi\ W50 tends to agree with the \ha\ $\vrot$. 
\paragraph{NGC\,4711} 
This galaxy has a warp in the center, as seen in the \ha\ velocity field and rotation component of the rotation curve. It is listed as having a bar \citep{bolatto17}, which likely accounts for these features. These are not seen in the CO velocity field or rotation curve. Although both the CO and \ha\ have non-zero radial components in the comparison region, they track each other closely so their effect on $\Dv$ is likely small. Neither the CO nor \ha\ rotation curve flattens over the radii probed, so it is unclear whether the \hi\ W50 or W90 agree with either.
\paragraph{NGC\,5016}
The $\vrad$ and $\vsys$ components for CO and \ha\ are nearly zero over the comparison region, although they deviate at small and large radii. This galaxy has a bar \citep{bolatto17}, although it is not obvious from either velocity field. The \hi\ W50 and W90 are more consistent with $\vrot$(CO), at least over the comparison region. 
\paragraph{NGC\,5480} Neither the CO nor \ha\ rotation curve for this galaxy flattens out to large radii. It has small radial and systemic components over the comparison region, although there is a decreasing trend in the both of these components in \ha. \hi\ data is taken from \citet{springob05} using the Green Bank 300 ft telescope.
\paragraph{NGC\,5520} The difference between the CO and \ha\ rotation velocity is striking in this galaxy. The radial and systemic components are all consistent with zero. The \ha\ rotation curve flattens quickly. The \hi\ data agree well with the CO rotation velocity.  
\paragraph{NGC\,5633} The CO and \ha\ rotation velocities are remarkably similar and only begin to deviate in the comparison region. Although there are trends in the radial and systemic components in the center, they are near zero over the radii of interest. This galaxy in known to have a ring structure \citep{bolatto17}. \hi\ data is taken from \citet{springob05} using the Green Bank 300 ft telescope.
\paragraph{NGC\,5980} This galaxy's CO rotation curve flattens, whereas the \ha\ rotation curve continues to rise slightly. Although the radial and systemic components dip in the center, they are near zero over the radii of interest. There is a kinematic twist visible in both the CO and \ha\ velocity fields, which likely causes the dip in the radial and systemic components. The \hi\ W50 is close to $\vrot$(CO), but the W90 is higher. 
\paragraph{UGC\,4132} There is an interesting dip in the \ha\ rotation curve that is not visible in the CO rotation curve. There is a slight trend in the \ha\ radial and systemic components. The \hi\ W50 is close to the CO rotational velocity, but the W90 is substantially larger. 
\paragraph{UGC\,5111} The CO and \ha\ velocity fields look different near the major axis; in the \ha\ velocity fields there are thin regions of high velocities whereas these larger velocities are not as concentrated along the major axis in the CO velocity field. The radial and systemic components are small over all radii. There is not \hi\ data available for this galaxy from either our GBT observations or from \citet{springob05}. 
\paragraph{UGC\,9067} The galaxy's CO and \ha\ rotation velocities are nearly identical. Both the CO and \ha\ radial components are increasing over the comparison region but have the same values. \hi\ data is taken from \citet{springob05} using the Arecibo telescope (line feed system).
\paragraph{UGC\,10384} This galaxy has small radial and systemic components over the region of interest, although there are small deviations at small and large radii. The rotation components are identical at small radii, but diverge as the \ha\ rotation curve flattens while the CO rotation curve continues to rise. The \hi\ W50 and W90 are more consistent with the CO rotation velocity than the \ha. 

\clearpage
\setcounter{table}{\value{figure}}
\begin{longtable*}{c}
\label{fig:KSS}
\endfirsthead
\endhead
\endfoot
\endhead
\endfoot
\endhead
\endfoot
\endhead
\endfoot
\endhead
\endfoot
\endlastfoot
\includegraphics[width=0.85\textwidth]{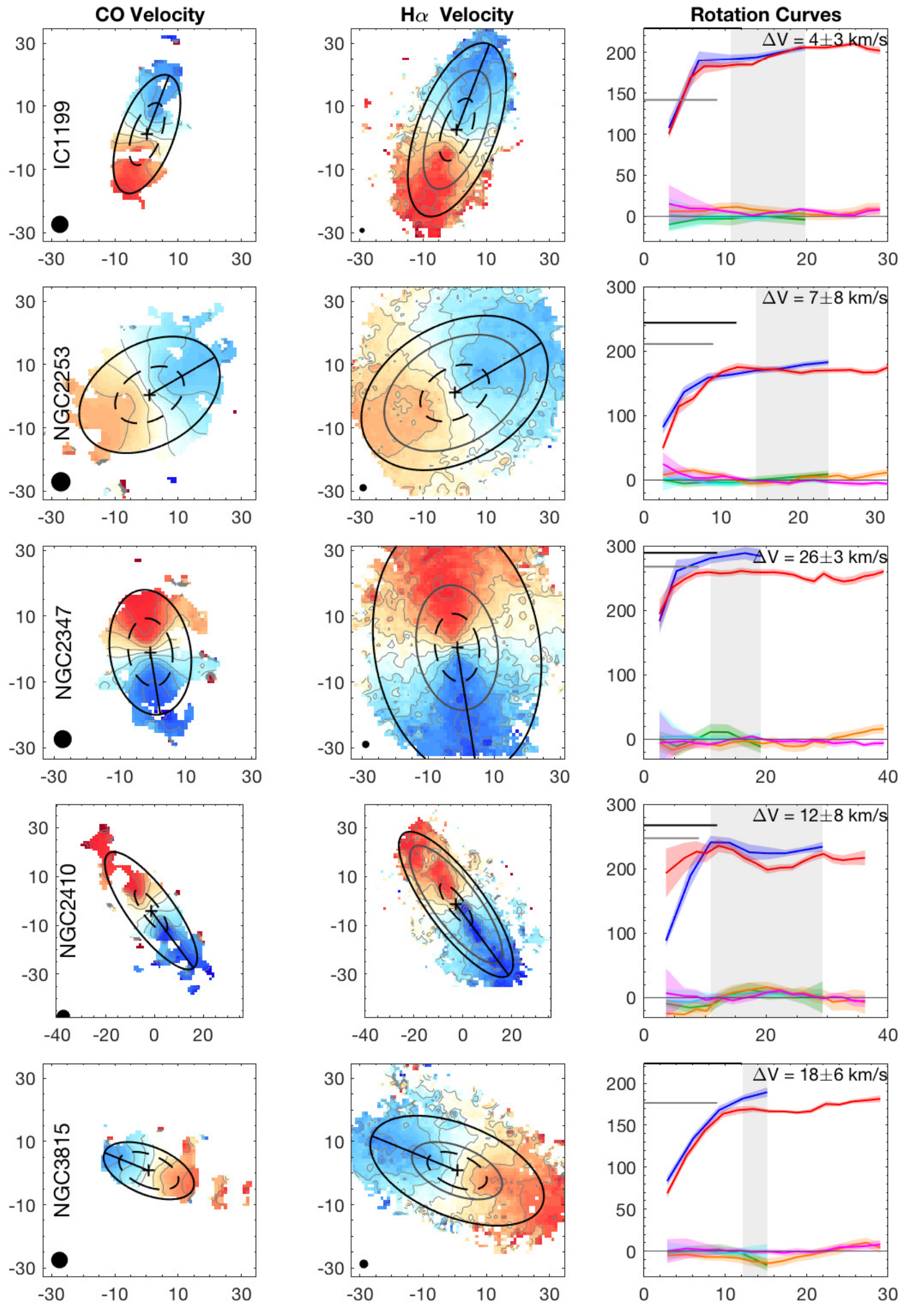}\\
\includegraphics[width=0.85\textwidth]{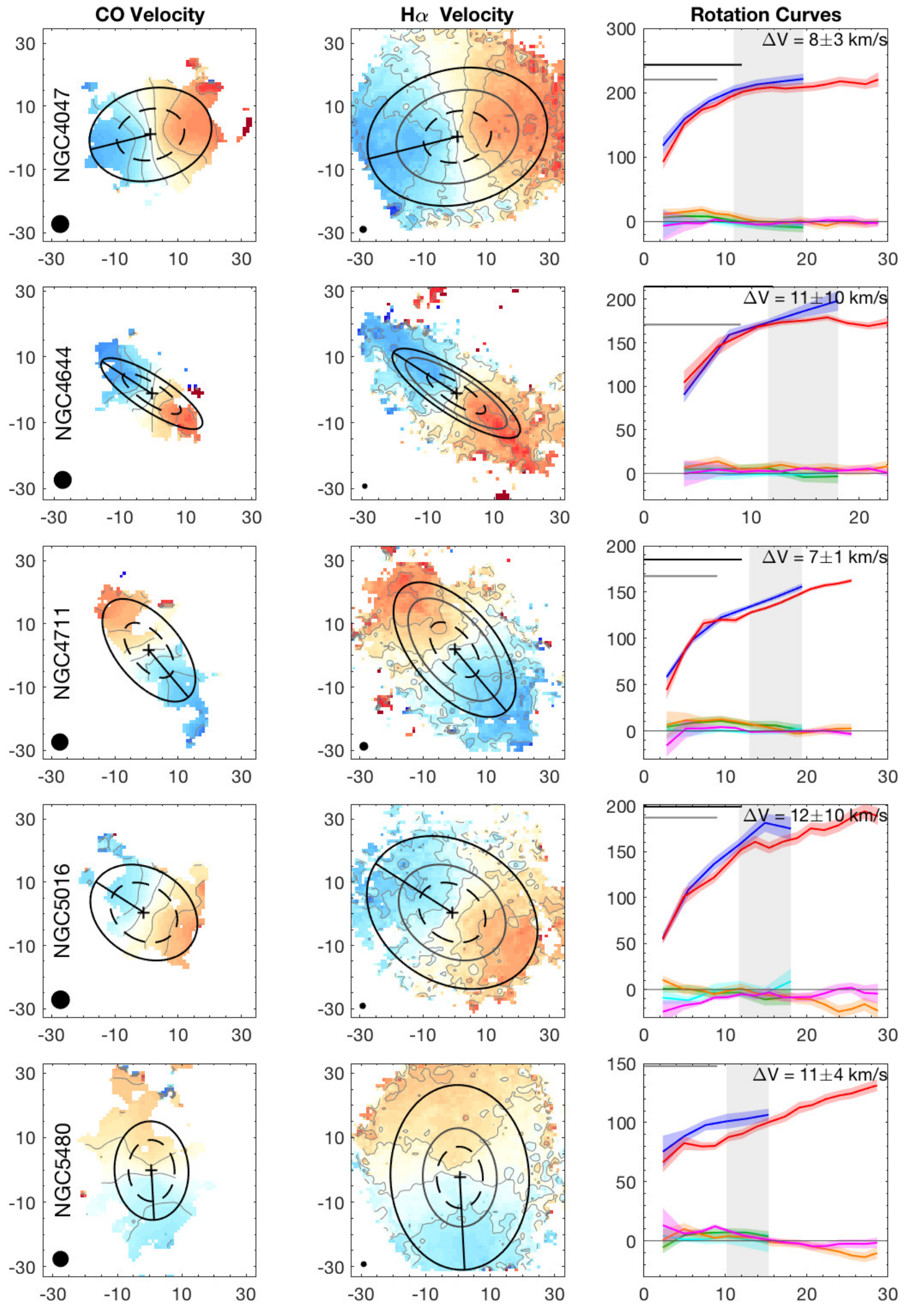}\\
\includegraphics[width=0.85\textwidth]{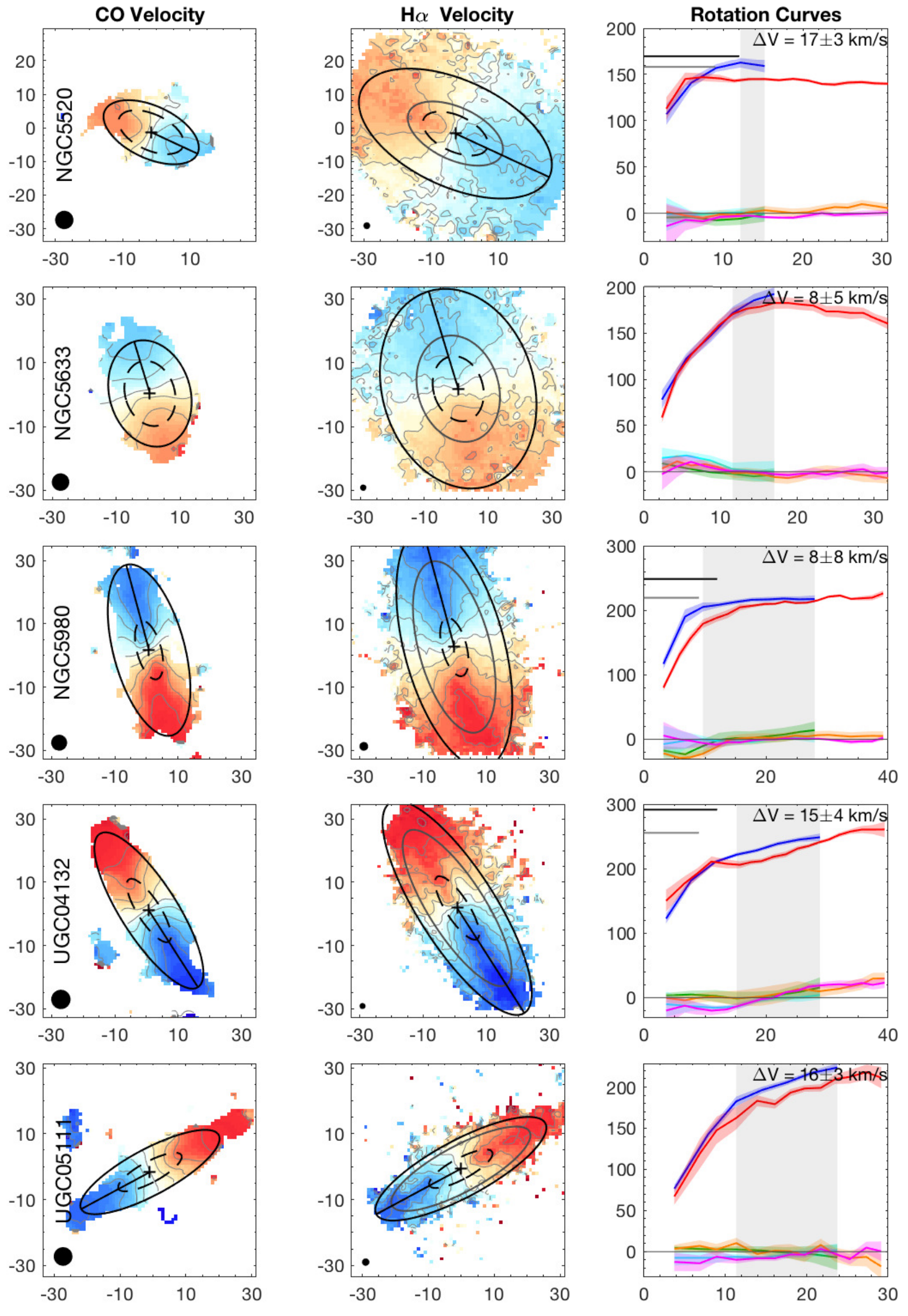}\\
\includegraphics[width=0.85\textwidth]{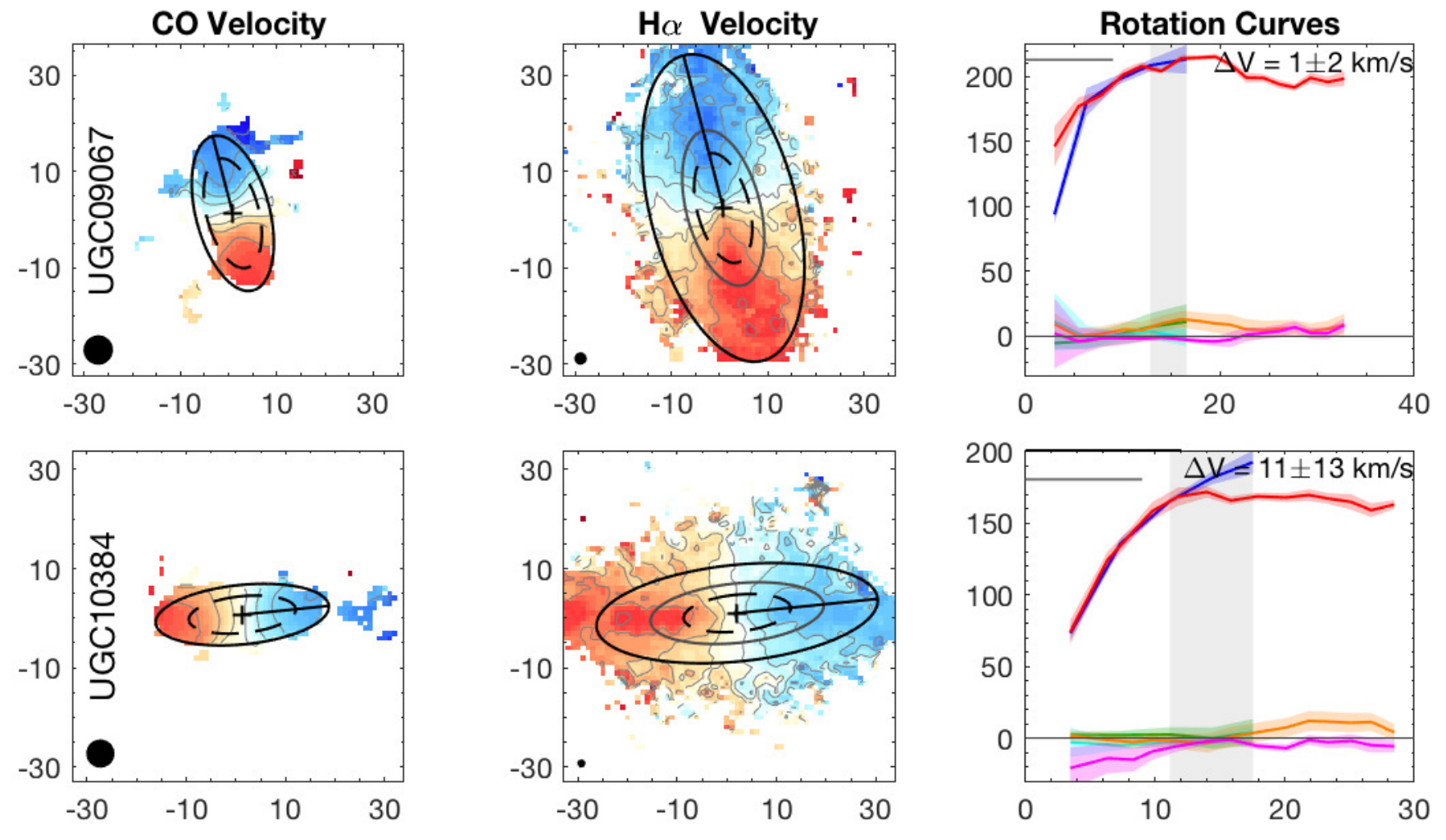}\\
\end{longtable*}
\noindent {\bf Figure \arabic{table}.} The left column shows the CO velocity fields for the KSS galaxies. The color scale spans $\pm$300\,\kms, with negative velocities showing the approaching side in blue. {\comments Isovelocity contours are shown in gray spanning $\pm$300\,\kms\ in 50\,\kms\ increments.} The values on the x- and y-axes show the offset in arcsec from the center position. The solid black circle shows the 6" beam size. The cross marks the kinematic center of the galaxy. The black line shows the semimajor axis. The black solid ellipse shows the orientation, and the size corresponds to the outermost ring in the rotation curve. The dashed ellipse indicates twice the CO beam; points within this ellipse are excluded from the rotation curve comparison due to possible beam smearing. The center column shows the \ha\ velocity fields. The color scale also spans $\pm$300\,\kms. The solid black circle shows the 6" beam size, the cross marks the kinematic center of the galaxy, and the solid black line shows the semimajor axis. The solid black ellipse shows the orientation, and the size corresponds to the outermost ring in the \ha\ rotation curve. The gray ellipse shows the outermost CO ring (the same as the black ellipse in the CO velocity panels). The right column shows the CO and \ha\ rotation curves. The x-axis is galactocentric radius (") and the y-axis is velocity (\kms). The colored curves show CO $\vrot$ (blue), $\vrad$ (green), and $\vsys$ (cyan), and \ha\ $\vrot$ (red), $\vrad$ (orange), and $\vsys$ (magenta). The colored shaded regions around the curves show the errors on the rotation curve components from the Monte Carlo analysis. The gray shaded region shows the radii over which the CO and \ha\ rotation curves are compared, where the inner radius is twice the CO beam size and the outer radius is the furthest CO extent. The $\Dv$ in the upper right corner is the median difference between the CO and \ha\ $\vrot$ over the gray region, and the error is the standard deviation of the differences at each radius. The thin solid gray line marks $V=0$. The gray and black horizontal line segments show the \hi\ rotation velocity from W50 and W90, respectively. All HI data have been corrected for inclination, using the values in Table \ref{tab:EDGEparameters}.

\clearpage
\setcounter{table}{2}
\startlongtable
\begin{deluxetable*}{ccccccccccc}
\tabletypesize{\footnotesize}
\tablecaption{Geometric Parameters for the EDGE Data \label{tab:EDGEparameters}}
\tablehead{\colhead{Name} & \colhead{RA} & \colhead{Dec}  & \colhead{PA}  & \colhead{Inc} & \colhead{V$_{\rm sys}$} & \colhead{X$_{\rm off}$} & \colhead{Y$_{\rm off}$} & \colhead{PA Flag} & \colhead{Inc Flag} & \colhead{V$_{\rm sys}$ Flag}\\
 & \colhead{(J2000)} & \colhead{(J2000)} & \colhead{(\D)} & \colhead{(\D)} & \colhead{(km\,s$^{-1}$)} & \colhead{('')} & \colhead{('')} &  &  &  }
\startdata
ARP220 & 233.73900 & 23.50270 & 337.7 & 29.7 & 5247 & 0.0 & 0.0 & P & P & C\\
IC0480 & 118.84665 & 26.74280 & 167.9 & 76.6 & 4595 & 0.0 & 0.0 & P & P & C\\
IC0540 & 142.54290 & 7.90259 & 350.0 & 68.3 & 2022 & 0.0 & 0.0 & P & P & C\\
IC0944 & 207.87900 & 14.09200 & 105.7 & 75.0 & 6907 & 0.0 & 0.0 & P & C & C\\
IC1151 & 239.63550 & 17.44150 & 203.9 & 68.0 & 2192 & 0.0 & 0.0 & C & P & C\\
IC1199 & 242.64300 & 10.04010 & 339.3 & 64.5 & 4686 & 0.6 & 1.1 & C & P & C\\
IC1683 & 20.66190 & 34.43700 & 15.6 & 54.8 & 4820 & 0.0 & 0.0 & P & C & C\\
IC2247 & 123.99615 & 23.19960 & 328.5 & 77.7 & 4254 & 0.0 & 0.0 & P & P & C\\
IC2487 & 142.53840 & 20.09090 & 162.9 & 77.9 & 4310 & 0.0 & 0.0 & P & P & C\\
IC4566 & 234.17550 & 43.53940 & 145.0 & 53.9 & 5537 & 0.0 & 0.0 & C & P & C\\
IC5376 & 0.33233 & 34.52570 & 3.4 & 71.6 & 4979 & 0.0 & 0.0 & P & P & C\\
NGC0444 & 18.95685 & 31.08020 & 158.7 & 74.9 & 4776 & 0.0 & 0.0 & P & P & H\\
NGC0447 & 18.90660 & 33.06760 & 227.0 & 29.1 & 5552 & 0.8 & 0.8 & C & P & C\\
NGC0477 & 20.33475 & 40.48820 & 140.0 & 60.0 & 5796 & 0.0 & 0.0 & C & P & C\\
NGC0496 & 20.79810 & 33.52890 & 36.5 & 57.0 & 5958 & -0.4 & 1.6 & C & P & C\\
NGC0523 & 21.33660 & 34.02500 & 277.2 & 71.6 & 4760 & 0.0 & 0.0 & C & P & C\\
NGC0528 & 21.38985 & 33.67150 & 57.7 & 61.1 & 4638 & -3.0 & 1.8 & P & P & H\\
NGC0551 & 21.91935 & 37.18290 & 315.0 & 64.2 & 5141 & 0.0 & 0.0 & C & P & C\\
NGC1167 & 45.42660 & 35.20570 & 87.5 & 39.5 & 4797 & -6.5 & 3.5 & C & P & C\\
NGC2253 & 100.92420 & 65.20620 & 300.0 & 47.4 & 3545 & 0.0 & 0.0 & C & P & C\\
NGC2347 & 109.01625 & 64.71080 & 189.1 & 50.2 & 4387 & 2.5 & 1.9 & P & P & C\\
NGC2410 & 113.75940 & 32.82210 & 216.6 & 71.6 & 4642 & 0.0 & 0.0 & C & P & C\\
NGC2480 & 119.29350 & 23.77980 & 343.1 & 55.4 & 2287 & 0.0 & 0.0 & C & P & C\\
NGC2486 & 119.48535 & 25.16080 & 92.9 & 55.6 & 4569 & 0.0 & 0.0 & P & P & H\\
NGC2487 & 119.58540 & 25.14920 & 117.5 & 31.4 & 4795 & 0.0 & 0.0 & C & P & C\\
NGC2623 & 129.60000 & 25.75410 & 255.0 & 45.6 & 5454 & 0.0 & 0.0 & C & P & C\\
NGC2639 & 130.90845 & 50.20540 & 314.3 & 49.5 & 3162 & -1.4 & 0.6 & C & P & C\\
NGC2730 & 135.56580 & 16.83830 & 260.8 & 27.7 & 3802 & 0.0 & 0.0 & P & P & C\\
NGC2880 & 142.39410 & 62.49060 & 322.9 & 49.9 & 1530 & 0.0 & 0.0 & P & P & H\\
NGC2906 & 143.02605 & 8.44159 & 262.0 & 55.7 & 2133 & 0.0 & 0.0 & C & P & C\\
NGC2916 & 143.73990 & 21.70520 & 199.9 & 49.9 & 3620 & 0.0 & 0.0 & P & P & C\\
NGC2918 & 143.93355 & 31.70550 & 75.1 & 46.1 & 6569 & 0.0 & 0.0 & P & P & H\\
NGC3303 & 159.25050 & 18.13570 & 159.6 & 60.5 & 6040 & 0.0 & 0.0 & P & P & H\\
NGC3381 & 162.10350 & 34.71140 & 333.1 & 30.8 & 1625 & 0.0 & 0.0 & C & P & H\\
NGC3687 & 172.00200 & 29.51100 & 326.0 & 19.6 & 2497 & 0.0 & 0.0 & C & P & H\\
NGC3811 & 175.32000 & 47.69080 & 351.5 & 39.9 & 3073 & -2.6 & 0.1 & P & P & C\\
NGC3815 & 175.41300 & 24.80040 & 67.8 & 59.9 & 3686 & 3.0 & 0.1 & P & P & C\\
NGC3994 & 179.40300 & 32.27730 & 188.1 & 59.5 & 3097 & 2.4 & 1.6 & P & P & C\\
NGC4047 & 180.71100 & 48.63620 & 104.0 & 42.1 & 3419 & 0.5 & 0.0 & C & P & C\\
NGC4149 & 182.63700 & 58.30410 & 85.0 & 66.2 & 3050 & -0.5 & -0.9 & C & P & C\\
NGC4185 & 183.34200 & 28.51100 & 344.4 & 48.2 & 3874 & 0.0 & 0.0 & P & P & C\\
NGC4210 & 183.81600 & 65.98540 & 277.7 & 40.9 & 2714 & 0.0 & 0.0 & P & P & C\\
NGC4211NED02 & 183.90600 & 28.16960 & 25.0 & 30.0 & 6605 & 0.0 & 0.0 & C & C & C\\
NGC4470 & 187.40700 & 7.82390 & 349.5 & 47.5 & 2338 & 0.0 & 0.0 & C & P & C\\
NGC4644 & 190.67850 & 55.14550 & 57.0 & 72.9 & 4915 & -3.2 & -0.1 & P & P & C\\
NGC4676A & 191.54250 & 30.73210 & 185.3 & 50.0 & 6541 & -2.1 & -0.8 & C & C & C\\
NGC4711 & 192.19050 & 35.33270 & 220.0 & 58.3 & 4044 & 3.0 & 0.6 & C & P & C\\
NGC4961 & 196.44900 & 27.73390 & 100.0 & 46.6 & 2521 & -2.2 & 0.4 & C & P & C\\
NGC5000 & 197.44800 & 28.90680 & 31.3 & 20.0 & 5557 & 0.0 & 0.0 & C & C & C\\
NGC5016 & 198.02850 & 24.09500 & 57.4 & 39.9 & 2588 & -0.1 & -0.6 & P & P & C\\
NGC5056 & 199.05150 & 30.95020 & 178.0 & 61.4 & 5544 & -1.0 & 1.0 & C & L & C\\
NGC5205 & 202.51500 & 62.51150 & 169.0 & 49.8 & 1762 & 0.0 & 0.0 & P & P & C\\
NGC5218 & 203.04300 & 62.76780 & 236.4 & 30.1 & 2888 & 0.0 & 0.0 & C & P & C\\
NGC5394 & 209.64000 & 37.45350 & 189.3 & 70.2 & 3431 & 0.0 & 0.0 & C & P & C\\
NGC5406 & 210.08400 & 38.91540 & 111.4 & 45.0 & 5350 & 0.0 & 0.0 & P & P & C\\
NGC5480 & 211.59000 & 50.72510 & 183.0 & 41.5 & 1879 & -0.9 & -0.4 & C & L & C\\
NGC5485 & 211.79700 & 55.00160 & 74.5 & 47.2 & 1893 & 2.0 & -0.6 & C & P & H\\
NGC5520 & 213.09450 & 50.34850 & 245.1 & 59.1 & 1870 & 1.1 & 0.4 & C & P & C\\
NGC5614 & 216.03150 & 34.85890 & 270.0 & 35.9 & 3859 & 0.0 & 0.0 & C & P & C\\
NGC5633 & 216.86850 & 46.14640 & 16.9 & 41.9 & 2319 & 0.0 & 0.0 & P & P & C\\
NGC5657 & 217.68150 & 29.18070 & 349.0 & 68.3 & 3860 & 2.1 & 0.2 & C & P & C\\
NGC5682 & 218.68800 & 48.66950 & 310.6 & 76.3 & 2242 & -0.9 & 1.7 & C & P & H\\
NGC5732 & 220.16250 & 38.63770 & 43.2 & 58.4 & 3723 & -2.0 & -0.2 & P & P & C\\
NGC5784 & 223.56900 & 42.55780 & 255.0 & 45.0 & 5427 & -2.1 & 0.6 & C & C & C\\
NGC5876 & 227.38200 & 54.50650 & 51.4 & 65.9 & 3240 & -0.1 & 0.6 & P & P & H\\
NGC5908 & 229.18050 & 55.40940 & 153.0 & 77.0 & 3294 & 0.0 & 0.0 & C & C & C\\
NGC5930 & 231.53250 & 41.67610 & 155.0 & 45.0 & 2637 & 2.2 & 0.1 & C & C & C\\
NGC5934 & 232.05300 & 42.92990 & 5.0 & 55.0 & 5566 & 0.0 & 0.0 & C & C & C\\
NGC5947 & 232.65300 & 42.71720 & 248.6 & 32.2 & 5898 & -0.9 & 0.1 & C & P & C\\
NGC5953 & 233.63550 & 15.19380 & 48.3 & 26.0 & 1988 & -1.0 & 0.4 & C & C & C\\
NGC5980 & 235.37700 & 15.78760 & 15.0 & 66.2 & 4060 & -1.0 & 1.0 & L & P & C\\
NGC6004 & 237.59400 & 18.93920 & 272.3 & 37.3 & 3818 & 3.2 & 1.1 & C & P & C\\
NGC6021 & 239.37750 & 15.95600 & 157.1 & 43.4 & 4673 & 1.7 & -0.1 & P & P & H\\
NGC6027 & 239.80200 & 20.76330 & 231.4 & 30.9 & 4338 & 0.0 & 0.0 & P & P & H\\
NGC6060 & 241.46700 & 21.48490 & 102.0 & 64.3 & 4416 & -1.2 & -0.2 & P & P & C\\
NGC6063 & 241.80450 & 7.97887 & 331.6 & 56.2 & 2807 & 0.0 & 0.0 & C & P & H\\
NGC6081 & 243.23700 & 9.86703 & 308.2 & 65.6 & 4978 & 0.0 & 0.0 & P & P & H\\
NGC6125 & 244.79850 & 57.98410 & 4.8 & 16.9 & 4522 & 0.0 & 0.0 & P & P & H\\
NGC6146 & 246.29250 & 40.89260 & 78.3 & 40.7 & 8693 & 0.0 & 0.0 & C & P & H\\
NGC6155 & 246.53400 & 48.36680 & 130.0 & 44.7 & 2418 & 3.0 & 3.0 & C & P & C\\
NGC6168 & 247.83750 & 20.18550 & 111.1 & 76.6 & 2540 & 0.0 & 0.0 & L & P & C\\
NGC6186 & 248.60700 & 21.54090 & 64.6 & 71.2 & 2940 & -3.0 & 0.5 & C & L & C\\
NGC6301 & 257.13600 & 42.33900 & 288.5 & 52.8 & 8222 & 0.0 & 0.0 & P & P & C\\
NGC6310 & 256.98900 & 60.99010 & 69.9 & 73.7 & 3459 & 0.0 & 0.0 & P & P & C\\
NGC6314 & 258.16200 & 23.27020 & 356.0 & 57.7 & 6551 & -2.2 & 0.0 & C & P & C\\
NGC6361 & 259.67100 & 60.60810 & 46.8 & 75.0 & 3791 & 0.0 & 0.0 & C & C & C\\
NGC6394 & 262.59000 & 59.63990 & 237.4 & 60.0 & 8453 & 0.0 & 0.0 & C & C & C\\
NGC6478 & 267.15900 & 51.15720 & 29.2 & 73.4 & 6756 & 0.0 & 0.0 & C & C & C\\
NGC7738 & 356.00850 & 0.51671 & 234.7 & 65.6 & 6682 & 0.0 & 0.0 & C & P & C\\
NGC7819 & 1.10206 & 31.47200 & 270.3 & 54.0 & 4918 & 0.0 & 0.0 & C & P & C\\
UGC00809 & 18.96615 & 33.81070 & 18.6 & 78.9 & 4171 & 0.0 & 0.0 & C & P & C\\
UGC03253 & 79.92345 & 84.05250 & 267.7 & 58.3 & 4040 & 0.0 & 0.0 & P & P & H\\
UGC03539 & 102.22470 & 66.26130 & 302.9 & 72.1 & 3278 & 0.0 & 0.0 & C & P & C\\
UGC03969 & 115.30965 & 27.61410 & 134.3 & 70.0 & 8029 & 0.0 & 0.0 & P & C & C\\
UGC03973 & 115.63560 & 49.80980 & 143.8 & 39.1 & 6594 & 0.5 & -1.8 & C & P & C\\
UGC04029 & 117.07920 & 34.33220 & 63.5 & 77.6 & 4389 & 0.0 & 0.0 & C & P & C\\
UGC04132 & 119.80425 & 32.91490 & 212.6 & 72.0 & 5151 & 0.0 & -0.8 & C & P & C\\
UGC04280 & 123.63885 & 54.79950 & 183.7 & 71.5 & 3500 & 0.0 & 0.0 & P & P & C\\
UGC04461 & 128.34450 & 52.53230 & 222.8 & 70.1 & 4941 & 0.0 & 0.0 & C & P & C\\
UGC05108 & 143.85960 & 29.81260 & 136.1 & 66.1 & 8015 & 0.0 & 0.0 & C & P & C\\
UGC05111 & 144.21855 & 66.78840 & 118.3 & 72.9 & 6660 & 2.0 & -1.0 & C & P & C\\
UGC05244 & 147.20070 & 64.16800 & 32.8 & 77.9 & 2974 & 0.0 & 0.0 & P & P & H\\
UGC05359 & 149.71530 & 19.21500 & 94.5 & 72.3 & 8344 & -0.9 & 0.8 & C & P & C\\
UGC05498NED01 & 153.01500 & 23.08540 & 61.8 & 81.0 & 6250 & 0.0 & 0.0 & C & C & H\\
UGC05598 & 155.55900 & 20.58940 & 215.6 & 74.8 & 5591 & 0.0 & 0.0 & P & P & C\\
UGC06312 & 169.50000 & 7.84466 & 224.6 & 68.7 & 6266 & 0.0 & 0.0 & C & P & H\\
UGC07012 & 180.51300 & 29.84810 & 184.1 & 60.5 & 3052 & 0.0 & 0.0 & C & P & H\\
UGC08107 & 194.91600 & 53.34130 & 228.2 & 71.4 & 8201 & 0.0 & 0.0 & C & P & C\\
UGC08250 & 197.58450 & 32.48260 & 11.7 & 76.2 & 5169 & 0.0 & 0.0 & C & P & H\\
UGC08267 & 197.79750 & 43.72650 & 223.0 & 75.4 & 7159 & -1.0 & 0.0 & C & P & C\\
UGC09067 & 212.68950 & 15.20920 & 14.6 & 62.4 & 7740 & 0.0 & 0.0 & C & P & C\\
UGC09476 & 220.38300 & 44.51270 & 307.0 & 48.5 & 3243 & 0.0 & 0.0 & C & P & C\\
UGC09537 & 222.11100 & 34.99800 & 135.7 & 72.0 & 8662 & 0.0 & 0.0 & C & C & C\\
UGC09542 & 222.25500 & 42.46400 & 214.3 & 72.7 & 5417 & 2.1 & -1.9 & P & P & C\\
UGC09665 & 225.38550 & 48.31980 & 138.2 & 74.0 & 2561 & 0.0 & 0.0 & P & P & C\\
UGC09759 & 227.67000 & 55.35040 & 49.7 & 66.8 & 3394 & 3.6 & -2.1 & P & P & C\\
UGC09873 & 232.46100 & 42.62900 & 129.0 & 75.3 & 5575 & 0.0 & 0.0 & C & P & C\\
UGC09892 & 233.21700 & 41.19140 & 101.0 & 72.2 & 5591 & 0.0 & 0.0 & P & P & C\\
UGC09919 & 233.91450 & 12.60630 & 349.2 & 77.9 & 3160 & 0.0 & 0.0 & P & P & C\\
UGC10043 & 237.17250 & 21.86950 & 327.8 & 90.0 & 2154 & -2.4 & -0.6 & C & L & C\\
UGC10123 & 239.76150 & 51.30460 & 231.6 & 70.0 & 3738 & 3.6 & -0.4 & C & C & C\\
UGC10205 & 241.66800 & 30.09900 & 128.6 & 51.7 & 6491 & 0.0 & 0.0 & C & P & C\\
UGC10331 & 244.33800 & 59.32010 & 140.8 & 76.2 & 4415 & 0.0 & 0.0 & P & P & H\\
UGC10380 & 246.45750 & 16.57610 & 288.2 & 77.9 & 8624 & 0.0 & 0.0 & P & P & C\\
UGC10384 & 246.69450 & 11.58020 & 275.8 & 70.0 & 4927 & 0.0 & 0.0 & C & C & C\\
UGC10710 & 256.71900 & 43.12210 & 329.5 & 69.6 & 8228 & 0.0 & 0.0 & L & P & C\\
\enddata
\tablecomments{The table lists the geometric parameters used for the EDGE data for each galaxy. RA and Dec values are taken from \citet{bolatto17}. The V$_{\rm sys}$ values are reported in the relativistic convention. PA Flag, Inc Flag, and V$_{\rm sys}$ Flag indicate whether those respective values were derived from CO kinematic fits (C) or  \ha\ kinematic fits (H) done in this work, from photometric fits to the outer isophotes \citep[P, ][]{falconbarroso17}, or from HyperLeda (L).}
\end{deluxetable*}

\startlongtable
\begin{deluxetable*}{cccccccc}
\tablecaption{Geometric Parameters for the CALIFA Data \label{tab:CALIFAparameters}}
\tablehead{\colhead{Name} & \colhead{$R_e$} & \colhead{V500 V$_{\rm sys}$} & \colhead{V500 X$_{\rm off}$} & \colhead{V500 Y$_{\rm off}$} & \colhead{V1200 V$_{\rm sys}$} & \colhead{V1200 X$_{\rm off}$} & \colhead{V1200 Y$_{\rm off}$}\\
 & \colhead{('')} & \colhead{(km\,s$^{-1}$)} &  \colhead{('')} & \colhead{('')} & \colhead{(km\,s$^{-1}$)} & \colhead{('')} & \colhead{('')}}
\startdata
ARP220 & 12.7 & 5294 & 0.0 & 0.0 & 5247 & 0.0 & 0.0\\
IC0480 & 11.5 & 4553 & 0.0 & 0.0 & 4518 & 0.0 & 0.0\\
IC0540 & 14.9 & 2050 & 0.0 & 0.0 & 2043 & 0.0 & 0.0\\
IC0944 & 9.8 & 6854 & 0.2 & 2.5 & 6805 & -2.5 & 3.4\\
IC1151 & 19.3 & 2122 & 0.0 & 0.0 & 2114 & 0.0 & 0.0\\
IC1199 & 18.8 & 4636 & 0.0 & 2.5 & 4625 & 0.3 & 3.0\\
IC1683 & 10.0 & 4787 & -3.4 & 1.4 & 4704 & -3.0 & 3.4\\
IC2247 & 16.3 & 4218 & 0.4 & 0.6 & 4188 & 0.0 & 0.0\\
IC2487 & 16.8 & 4281 & 0.0 & 0.0 & 4250 & 0.0 & 0.0\\
IC4566 & 13.2 & 5504 & 1.2 & 0.1 & 5453 & 1.2 & 0.1\\
IC5376 & 11.6 & 4944 & 0.0 & 0.0 & 4903 & 0.0 & 0.0\\
NGC0444 & 17.4 & 4776 & -4.0 & 2.6 & 4738 & -4.0 & 2.6\\
NGC0447 & 18.6 & 5489 & 0.8 & 0.8 & 5439 & 0.8 & 0.8\\
NGC0477 & 18.6 & 5794 & 0.0 & 0.0 & 5738 & 0.0 & 0.0\\
NGC0496 & 16.5 & 5966 & -3.0 & 3.0 & 5898 & 0.4 & 2.1\\
NGC0523 & 8.1 & 4719 & 0.0 & 0.0 & 4682 & 0.0 & 0.0\\
NGC0528 & 9.0 & 4638 & -3.7 & 0.7 & 4602 & -3.7 & 0.7\\
NGC0551 & 14.4 & 5106 & 0.0 & 0.0 & 5091 & -0.9 & 1.6\\
NGC1167 & 21.6 & 4875 & -6.6 & 2.3 & 4835 & -6.6 & 2.3\\
NGC2253 & 4.1 & 3530 & 2.4 & 0.6 & 3540 & 1.1 & 2.1\\
NGC2347 & 13.8 & 4383 & 3.6 & 3.3 & 4373 & 1.9 & 3.6\\
NGC2410 & 17.9 & 4650 & 1.6 & 2.6 & 4648 & 0.9 & 0.9\\
NGC2480 & 10.8 & 2305 & 0.0 & 0.0 & 2296 & 0.0 & 0.0\\
NGC2486 & 13.0 & 4569 & 0.0 & 0.0 & 4534 & 0.0 & 0.0\\
NGC2487 & 18.8 & 4808 & 0.0 & 0.0 & 4769 & 0.0 & 0.0\\
NGC2623 & 11.9 & 5440 & 2.5 & 0.5 & 5391 & 2.5 & 0.5\\
NGC2639 & 13.4 & 3157 & -1.1 & -2.6 & 3142 & 1.9 & 2.3\\
NGC2730 & 14.6 & 3773 & 0.0 & 0.0 & 3749 & 0.0 & 0.0\\
NGC2880 & 13.7 & 1530 & 0.0 & 0.0 & 1526 & 0.0 & 0.0\\
NGC2906 & 15.2 & 2142 & 1.6 & 0.5 & 2140 & -0.1 & 2.1\\
NGC2916 & 20.6 & 3664 & 0.0 & 0.0 & 3642 & 0.0 & 0.0\\
NGC2918 & 9.3 & 6569 & -0.6 & 2.4 & 6497 & -0.6 & 2.4\\
NGC3303 & 9.2 & 6040 & 0.0 & 0.0 & 5979 & 0.0 & 0.0\\
NGC3381 & 14.8 & 1625 & 0.0 & 0.0 & 1621 & 0.0 & 0.0\\
NGC3687 & 15.4 & 2497 & 3.2 & 0.1 & 2487 & 3.2 & 0.1\\
NGC3811 & 14.7 & 3061 & -2.6 & 0.1 & 3071 & -1.0 & 0.0\\
NGC3815 & 8.8 & 3690 & 3.0 & 0.1 & 3648 & 1.6 & 2.1\\
NGC3994 & 7.1 & 3089 & 3.2 & 1.0 & 3055 & 1.6 & 1.5\\
NGC4047 & 14.8 & 3376 & 1.2 & -0.7 & 3396 & 0.9 & 2.0\\
NGC4149 & 11.5 & 3042 & -0.5 & -0.9 & 3027 & -0.5 & -0.9\\
NGC4185 & 22.6 & 3831 & 0.0 & 0.0 & 3807 & 0.0 & 0.0\\
NGC4210 & 16.9 & 2689 & 2.6 & 3.1 & 2687 & -1.4 & 2.1\\
NGC4211NED02 & 14.0 & 6555 & 0.0 & 0.0 & 6483 & 0.0 & 0.0\\
NGC4470 & 11.5 & 2319 & 3.2 & 0.5 & 2310 & 3.2 & 0.5\\
NGC4644 & 14.3 & 4889 & -1.2 & 0.0 & 4885 & -2.9 & -0.9\\
NGC4676A & 13.5 & 6518 & -2.1 & -0.8 & 6447 & -2.1 & -0.8\\
NGC4711 & 12.3 & 4044 & 4.0 & 0.8 & 4007 & 2.5 & 0.0\\
NGC4961 & 9.7 & 2528 & -2.2 & 0.4 & 2517 & -2.2 & 0.4\\
NGC5000 & 10.2 & 5505 & 0.0 & 0.0 & 5454 & 0.0 & 0.0\\
NGC5016 & 15.3 & 2539 & -0.1 & -0.6 & 2586 & -1.9 & 0.5\\
NGC5056 & 13.8 & 5530 & -1.5 & 1.0 & 5453 & -0.4 & 0.9\\
NGC5205 & 16.4 & 1743 & 0.0 & 0.0 & 1738 & 0.0 & 0.0\\
NGC5218 & 12.3 & 2855 & 0.0 & 0.0 & 2841 & 0.0 & 0.0\\
NGC5394 & 16.8 & 3404 & 0.0 & 0.0 & 3385 & 0.0 & 0.0\\
NGC5406 & 14.9 & 5313 & 0.0 & 0.0 & 5331 & -0.8 & 0.9\\
NGC5480 & 17.4 & 1851 & -0.9 & -2.5 & 1908 & -3.5 & -0.3\\
NGC5485 & 21.8 & 1893 & 2.0 & -0.6 & 1887 & 2.0 & -0.6\\
NGC5520 & 11.9 & 1852 & 3.0 & 0.0 & 1893 & 1.8 & 1.1\\
NGC5614 & 15.7 & 3824 & 3.4 & -0.6 & 3800 & 3.4 & -0.6\\
NGC5633 & 12.9 & 2295 & 0.0 & 1.5 & 2290 & 3.1 & 1.8\\
NGC5657 & 11.6 & 3861 & 3.1 & -0.5 & 3836 & 3.1 & -0.5\\
NGC5682 & 19.6 & 2242 & 0.0 & 1.1 & 2234 & 0.0 & 1.1\\
NGC5732 & 12.3 & 3703 & -1.0 & -0.6 & 3680 & -1.0 & -0.6\\
NGC5784 & 11.9 & 5420 & -0.4 & 2.5 & 5307 & -2.4 & -0.4\\
NGC5876 & 15.1 & 3240 & 1.0 & 0.5 & 3222 & 1.0 & 0.5\\
NGC5908 & 14.6 & 3258 & 1.0 & -0.3 & 3240 & 1.0 & -0.3\\
NGC5930 & 14.4 & 2590 & 3.4 & -0.1 & 2579 & 3.4 & -0.1\\
NGC5934 & 6.7 & 5556 & 2.5 & -1.2 & 5505 & 2.5 & -1.2\\
NGC5947 & 10.5 & 5863 & -3.2 & 2.4 & 5811 & -3.2 & -0.1\\
NGC5953 & 9.1 & 1967 & 0.1 & 0.0 & 1961 & 0.1 & 0.0\\
NGC5980 & 12.6 & 4049 & 0.0 & 2.0 & 4021 & -0.9 & 0.3\\
NGC6004 & 20.4 & 3781 & 4.4 & 0.8 & 3757 & 4.4 & 0.8\\
NGC6021 & 8.5 & 4673 & 2.6 & -0.4 & 4637 & 2.6 & -0.4\\
NGC6027 & 10.8 & 4338 & 0.0 & 0.0 & 4307 & 0.0 & 0.0\\
NGC6060 & 20.2 & 4337 & 0.0 & 0.0 & 4306 & 0.0 & 0.0\\
NGC6063 & 17.8 & 2807 & 0.0 & 0.0 & 2794 & 0.0 & 0.0\\
NGC6081 & 10.4 & 4978 & 0.0 & 0.0 & 4937 & 0.0 & 0.0\\
NGC6125 & 15.4 & 4522 & 0.0 & 0.0 & 4488 & 0.0 & 0.0\\
NGC6146 & 11.0 & 8693 & 0.0 & 0.0 & 8567 & 0.0 & 0.0\\
NGC6155 & 13.5 & 2381 & 4.0 & 1.0 & 2396 & -1.4 & 0.4\\
NGC6168 & 16.3 & 2505 & 0.0 & 0.0 & 2495 & 0.0 & 0.0\\
NGC6186 & 12.7 & 2910 & -3.0 & 0.5 & 2896 & -3.0 & 0.5\\
NGC6301 & 20.0 & 8221 & 3.5 & -0.8 & 8118 & 2.0 & -0.5\\
NGC6310 & 15.8 & 3377 & 0.0 & 0.0 & 3358 & 0.0 & 0.0\\
NGC6314 & 8.7 & 6493 & -2.2 & 0.0 & 6423 & -2.2 & 0.0\\
NGC6361 & 15.4 & 3759 & 0.0 & 0.0 & 3714 & 3.3 & -2.4\\
NGC6394 & 9.0 & 8387 & 0.0 & 0.0 & 8261 & -0.6 & 0.5\\
NGC6478 & 17.4 & 6735 & 0.0 & 0.0 & 6659 & 0.0 & 0.0\\
NGC7738 & 11.5 & 6642 & 0.0 & 0.0 & 6568 & 0.0 & 0.0\\
NGC7819 & 15.0 & 4898 & 0.0 & 0.0 & 4858 & 0.0 & 0.0\\
UGC00809 & 11.0 & 4143 & -3.0 & 1.0 & 4114 & 0.0 & 0.0\\
UGC03253 & 12.7 & 4040 & 0.0 & 0.0 & 4013 & 0.0 & 0.0\\
UGC03539 & 13.7 & 3244 & 2.8 & 4.6 & 3250 & -1.4 & 3.3\\
UGC03969 & 11.2 & 8001 & 0.0 & 0.0 & 7896 & -2.4 & -0.5\\
UGC03973 & 9.9 & 6551 & 0.5 & -1.8 & 6479 & 0.5 & -1.8\\
UGC04029 & 15.0 & 4367 & 0.0 & 0.0 & 4335 & 0.0 & 0.0\\
UGC04132 & 13.2 & 5158 & 0.0 & 0.0 & 5105 & 0.2 & 0.6\\
UGC04280 & 11.2 & 3485 & 0.0 & 0.0 & 3465 & 0.0 & 0.0\\
UGC04461 & 11.9 & 4954 & 0.0 & 0.0 & 4913 & 0.0 & 0.0\\
UGC05108 & 9.6 & 7987 & -1.0 & 1.0 & 7881 & 0.0 & 0.0\\
UGC05111 & 12.7 & 6657 & 0.0 & 0.0 & 6557 & 3.4 & -1.3\\
UGC05244 & 12.6 & 2974 & -0.5 & -5.0 & 2959 & -2.0 & -4.0\\
UGC05359 & 12.4 & 8332 & 0.0 & 0.0 & 8226 & -0.4 & 1.5\\
UGC05498NED01 & 10.5 & 6250 & 0.0 & 0.0 & 6185 & 0.0 & 0.0\\
UGC05598 & 11.4 & 5601 & 0.0 & 0.0 & 5549 & 0.0 & 0.0\\
UGC06312 & 12.8 & 6266 & 0.0 & 0.0 & 6201 & 0.0 & 0.0\\
UGC07012 & 11.9 & 3052 & 0.0 & 0.0 & 3036 & 0.0 & 0.0\\
UGC08107 & 17.7 & 8199 & 0.0 & 0.0 & 8139 & -0.4 & 2.5\\
UGC08250 & 11.8 & 5169 & 0.0 & 0.0 & 5124 & 0.0 & 0.0\\
UGC08267 & 10.5 & 7102 & -0.4 & 0.3 & 7018 & -0.4 & 0.3\\
UGC09067 & 11.3 & 7733 & 0.1 & 1.1 & 7661 & 1.0 & 0.0\\
UGC09476 & 15.5 & 3201 & 0.0 & 0.0 & 3184 & 0.0 & 0.0\\
UGC09537 & 15.8 & 8653 & 0.0 & 0.0 & 8528 & 0.0 & 0.0\\
UGC09542 & 12.9 & 5399 & 1.2 & -0.4 & 5350 & 1.2 & -0.4\\
UGC09665 & 11.6 & 2511 & 0.0 & 0.0 & 2500 & 0.0 & 0.0\\
UGC09759 & 13.5 & 3397 & 3.6 & -2.1 & 3378 & 3.6 & -2.1\\
UGC09873 & 14.8 & 5533 & 0.0 & 0.0 & 5482 & 0.0 & 0.0\\
UGC09892 & 13.7 & 5582 & 0.0 & 0.0 & 5530 & 0.0 & 0.0\\
UGC09919 & 13.1 & 3167 & 0.0 & 0.0 & 3150 & 0.0 & 0.0\\
UGC10043 & 24.6 & 2128 & -0.5 & -1.0 & 2120 & -1.5 & -1.0\\
UGC10123 & 11.0 & 3701 & 4.8 & -0.5 & 3709 & 3.6 & -0.4\\
UGC10205 & 14.0 & 6445 & 0.0 & 0.0 & 6376 & 0.0 & 0.0\\
UGC10331 & 15.4 & 4415 & 0.0 & 0.0 & 4382 & 0.0 & 0.0\\
UGC10380 & 12.8 & 8592 & 0.0 & 0.0 & 8469 & 0.0 & 0.0\\
UGC10384 & 9.3 & 4891 & -0.8 & 0.3 & 4886 & 0.4 & -0.6\\
UGC10710 & 12.0 & 8184 & 0.2 & -0.2 & 8072 & 0.2 & -0.2\\
\enddata
\tablecomments{The table lists the geometric parameters used for the CALIFA data for each galaxy, which are not already reported in Table \ref{tab:EDGEparameters}. $R_e$ values are derived from growth curves fit to SDSS r-band images and are provided by the CALIFA team. V500 parameters were determined from kinematic fits to \ha, and V1200 parameters were determined from kinematic fits to \hg. The V$_{\rm sys}$ values are reported in the relativistic convention.}
\end{deluxetable*}

\end{document}